\documentclass{kluwer}    

\usepackage{graphicx}

\newdisplay{guess}{Conjecture}

\def\gtsima{$\; \buildrel > \over \sim \;$}
\def\ltsima{$\; \buildrel < \over \sim \;$}
\def\simgt{\lower.5ex\hbox{\gtsima}}
\def\simlt{\lower.5ex\hbox{\ltsima}}

\def\CIV{\hbox{C~$\scriptstyle\rm IV\ $}}
\def\SiIV{\hbox{Si~$\scriptstyle\rm IV\ $}}
\def\OVI{\hbox{O~$\scriptstyle\rm VI\ $}}

\def\HI{\hbox{H~$\scriptstyle\rm I\ $}}
\def\HII{\hbox{H~$\scriptstyle\rm II\ $}}
\def\HeI{\hbox{He~$\scriptstyle\rm I\ $}}
\def\HeII{\hbox{He~$\scriptstyle\rm II\ $}}
\def\HeIII{\hbox{He~$\scriptstyle\rm III\ $}}
\def\simlt{\mathrel{\rlap{\lower 3pt\hbox{$\sim$}}\raise 2.0pt\hbox{$<$}}}
\def\simgt{\mathrel{\rlap{\lower 3pt\hbox{$\sim$}} \raise 2.0pt\hbox{$>$}}}
\def\be{\begin{equation}}
\def\ee{\end{equation}}

\def\cm3{\;\mbox{cm}^{-3}}

\begin{document}                                                                                   
\begin{article}
\begin{opening}
\title{The First Cosmic Structures and their Effects}               
\author{Benedetta \surname{Ciardi}} 
\institute{Max-Planck-Institut f\"{u}r Astrophysik, Garching, Germany}
\author{Andrea \surname{Ferrara}}
\institute{SISSA/International School for Advanced Studies, Trieste, Italy}
\date{April, 2004}  

\end{opening}

{\footnotesize\tableofcontents}


\section{Introduction}

Our present understanding of the evolution of the universe is based on the
Standard Hot Big Bang Model. The expansion of the universe, the synthesis of
the light elements and the Cosmic Microwave Background (CMB) radiation are
the pillars of this Model. Its definite observational confirmation was the
CMB accidental detection in 1965 by Penzias \& Wilson, who received the Nobel
Prize for this discovery.
In the 1990s the {\tt COBE}\footnote{http://lambda.gsfc.nasa.gov/product/cobe}
(COsmic Background Explorer)
satellite improved on the previous observations and
measured an almost isotropic blackbody radiation with temperature
$T_{\rm CMB}=2.726  \pm 0.010$ K (Mather et al. 1994), together with temperature
anisotropies on angular scales $\approx 90^\circ$ (Smoot et al. 1992).
These discoveries warranted yet another Nobel Prize (in 2006)
to the principal investigators of the mission, 
Smoot and Mather.
The isotropy of the microwave background
indicates that on large scales ($\ge 200$~Mpc) the universe is indeed
very smooth, as postulated by Einstein in his ``cosmological principle''.
On small scales, in contrast, it presents inhomogeneities,
from planets and stars, to galaxies, clusters and super-clusters of
galaxies. These structures are not uniformly distributed, but
rather show some spatial correlation, and regions of space almost
totally devoid of galaxies are alternated to high density regions.
The commonly adopted theory for the formation of these structures is the
gravitational instability scenario, in which primordial density
perturbations grow through gravitational Jeans instability to form
all the structures we observe today.
The evidence for the existence of such perturbations is based on 
combined CMB observations by {\tt BOOMERanG}\footnote{http://cmb.phys.cwru.edu/boomerang}
(Balloon Observations Of Millimetric Extragalactic Radiation and Geomagnetics),
{\tt MAXIMA}\footnote{http://cosmology.berkeley.edu/group/cmb} (Millimeter
Anisotropy eXperiment IMaging Array), {\tt DASI}\footnote{http://astro.uchicago.edu/dasi}
(Degree Angular Scale Interferometer) and, more recently, {\tt WMAP}\footnote
{http://map.gsfc.nasa.gov} (Wilkinson Microwave Anisotropy
Probe) which provided the power spectrum of temperature anisotropies
on a wide range of angular scales (Hu et al. 2001; Sievers et al. 2003;
Spergel et al. 2003; Spergel et al. 2007).  The anisotropies at scales $< 90^\circ$
are interpreted as the result of perturbations in the
energy density at the decoupling epoch and hence are directly related
to the primordial density fluctuations from which all present
structures originate.

While the observation of the
very high-redshift universe, $z\approx 1000$,
is possible through the CMB, there is a lack of observational data in the 
redshift interval
$7 < z < 1000$, with only a few objects with $z \approx 6.5-8$
(Hu et al. 2002; Kodaira et al. 2003; Kneib et al. 2004; Bouwens et al. 2004;
Malhotra et al. 2005; Mobasher et al. 2005; Iye et al. 2006; Schaerer \& Pell\'o
2005; Ota et al. 2007). A deep, 
narrow J-band search for Ly$\alpha$
emitting galaxies at $z\approx 9$ by Willis \& Courbin (2005) has given no
result, but the NICMOS field at $z \approx 10$ has revealed 3 possible galaxies
(Bouwens et al. 2005). 
Another tentative detection of a galaxy at $z\approx 10$ has been reported 
(Pell\'o et al. 2004a,b), but this has not been confirmed by an analysis of the
same data from a different group (Weatherley, Warren \& Babbedge 2004).
Other two lensed candidates at $z\approx 10$ have been reported by Stark et al. (2007b).

 The state of the art of the
telescopes observing in the NIR/optical/UV range is represented by 
{\tt HST}\footnote{http://www.stsci.edu/hst}
(Hubble Space Telescope), that has allowed to take high resolution
images ranging from supernova (SN) explosions, to nebulae with ongoing star
formation, to the Hubble Deep Field and Ultra Deep Field.
Additional important contributions are brought to the field by the {\tt Spitzer}\footnote{http://www.spitzer.caltech.edu/spitzer/index.shtml}
satellite, a space-based infrared telescope launched in 2003 working in the 3-180~$\mu$m range. 
Thanks to its exquisite sensitivity in this band, it has allowed
to better understand the evolution of galaxies up to $z \approx 5$  through a variety 
of mid- and far-infrared imaging surveys and to constrain the
nature  and amplitude of the Cosmic Infrared Background (CIB).   
In order to observe radiation from objects at even higher redshift,
telescopes with exceptional sensitivity in the IR and radio bands are needed. 
{\tt JWST}\footnote{http://ngst.gsfc.nasa.gov}
(James Webb Space Telescope), for example, with its nJy
sensitivity in the $1-10$~$\mu$m infrared regime, is ideally suited
for probing optical-UV emission from sources at $z>10$. Similarly, the
planned generation of radio telescopes as 
{\tt SKA}\footnote{http://www.skatelescope.org} (Square Kilometer Array),
{\tt LOFAR}\footnote{http://www.lofar.org} (LOw Frequency ARray),
{\tt 21cmA}\footnote{http://astrophysics.phys.cmu.edu/~jbp} (21cm Array;       
Pen, Wu \& Peterson 2004; Peterson, Pen \& Wu 2005) and
{\tt MWA}\footnote{http://web.haystack.mit.edu/arrays/MWA/} (Murchison Widefield Array)
will open a new observational window on the high redshift universe.

Thus, in
the near future, we should be able to image the first sources of light
that had formed in the universe, and unveil their nature. Nowadays,
observational evidence suggests that they are stellar type objects,
although the existence of primordial mini-quasars is not
completely ruled out.

In the standard cosmological hierarchical scenario for structure
formation, the objects which form first are predicted to have masses
corresponding to virial temperatures $T_{vir}<10^4$ K. Once the gas has
virialized in the potential wells of pre-existing dark matter halos, additional
cooling is required to further collapse the gas and form
stars. For a gas of primordial composition at such low temperatures the
main coolant is molecular hydrogen. The typical primordial H$_2$ fraction
is usually lower than the one required for the formation of such objects,
but during the collapse phase the H$_2$ content can reach suitable values,
depending on the mass of the object itself.
Once the gas has collapsed and cooled, star formation (SF) is ignited.
The study of the star formation process in primordial
objects is still in its early stages, and although large improvements have
been done in both analytical and numerical approaches, 
a solid physical scenario requires substantial further investigation.

Once the first sources have formed, their mass deposition, energy injection
and emitted radiation can deeply affect the subsequent
galaxy formation process and influence the evolution of the Intergalactic
Medium (IGM) via a number of so-called ``feedback'' effects. 
Although a rigorous classification of the various effects is not feasible,
they can be divided into three broad classes: radiative,
mechanical and chemical feedback. Into the first class fall all those
effects associated, in particular, with ionization/dissociation of hydrogen
atoms/molecules; the second class is produced by the mechanical energy injection of 
massive stars in form of winds or supernova explosions; the chemical feedback is, 
in contrast, related to the postulated existence of a 
critical metallicity governing the
cosmic transition from very massive stars to ``normal'' stars.

In this review we will summarize the recent achievements in understanding
the first structure formation process.

\section{First Structure Formation: Overview}    

In this Section we will describe the main processes leading from the evolution of
primordial density perturbations to the formation of the first structures
in the universe.

\subsection{The Cosmological Model}
\label{cosmomod}

Throughout this review we will usually refer to the popular $\Lambda$CDM model, in which dark matter
is composed of cold, weakly interacting, massive particles. The cosmological model
is completely defined once the value of the following parameters is specified:
the adimensional density of the universe, $\Omega_0=\Omega_m+\Omega_\Lambda$
(where $\Omega_m$ and $\Omega_\Lambda$ are the contribution from matter and vacuum,
respectively), the Hubble constant, $H_0=100 h$~km~s$^{-1}$~Mpc$^{-1}$, the
adimensional baryon density, $\Omega_b$, the rms mass fluctuations on $8 h^{-1}$~Mpc scale,
$\sigma_8$, and the spectral index of the primordial density fluctuation, $n$.
Recent observations have set stringent constraints on the values allowed for
these parameters. Studies of the temperature anisotropies in the CMB
strongly suggest that the universe is flat,
$\Omega_0 \approx 1$ (Balbi et al. 2000, 2001; O'Meara et al. 2001;
Pryke et al. 2002), with a
matter contribution of $\Omega_m \approx 0.10-0.36$. The latter values are derived
from the local abundance of galaxy clusters (Pen 1998; Viana \& Liddle 1999;
Schuecker et al. 2003),
their redshift evolution (Bahcall \& Fan 1998; Eke et al. 1998; Borgani et al. 2001),
mass-to-light ratio of groups of galaxies (Hoekstra et al. 2001), X-ray gas mass
fraction of clusters of galaxies (Allen, Schmidt \& Fabian 2002), clustering 
measurements from large galaxy surveys (e.g. Blake et al. 2007)
and high-$z$ supernovae
(Schmidt et al. 1998; Perlmutter et al. 1999; but see also Rowan-Robinson 2002).
The baryon contribution to matter,
$\Omega_b h^2 \approx 0.020-0.030$, is derived through measurements of
CMB anisotropies (Balbi et al. 2000, 2001;
Pryke et al. 2002), clustering measurements from large galaxy surveys (e.g. Blake et al. 2007)
 and D/H abundance in quasars spectra (Tytler et al. 2000;
Pettini \& Bowen 2001; Levshakov et al. 2002; Crighton et al. 2004). All the estimates
of the Hubble constant are based on secondary distance indicators, such as
the Tully-Fisher relation, the fundamental plane of elliptical galaxies, Type Ia
supernovae and surface brightness fluctuations, and give $h \approx
0.5-0.8$ (Theureau et al. 1997;
Tonry et al. 1997; Jha et al. 1999; Mould et al. 2000; Freedman et al. 2001;
Gibson \& Stetson 2001;
Liu \& Graham 2001; Tutui et al. 2001; Willick \& Batra 2001). Local and high-$z$
evolution of galaxy clusters is used to set a limit to $\sigma_8 \approx 0.7-1.2$
(Bahcall \& Fan 1998; Eke et al. 1998; Pen 1998; Viana \& Liddle 1999; Borgani et al. 2001;
Hoekstra, Yee \& Gladders 2002). Finally, studies of the temperature anisotropies in the CMB
give $n \approx 1.01-1.08$ (Balbi et al. 2000, 2001; Pryke et al. 2002).
\begin{figure}
\centerline{\includegraphics[width=28pc]{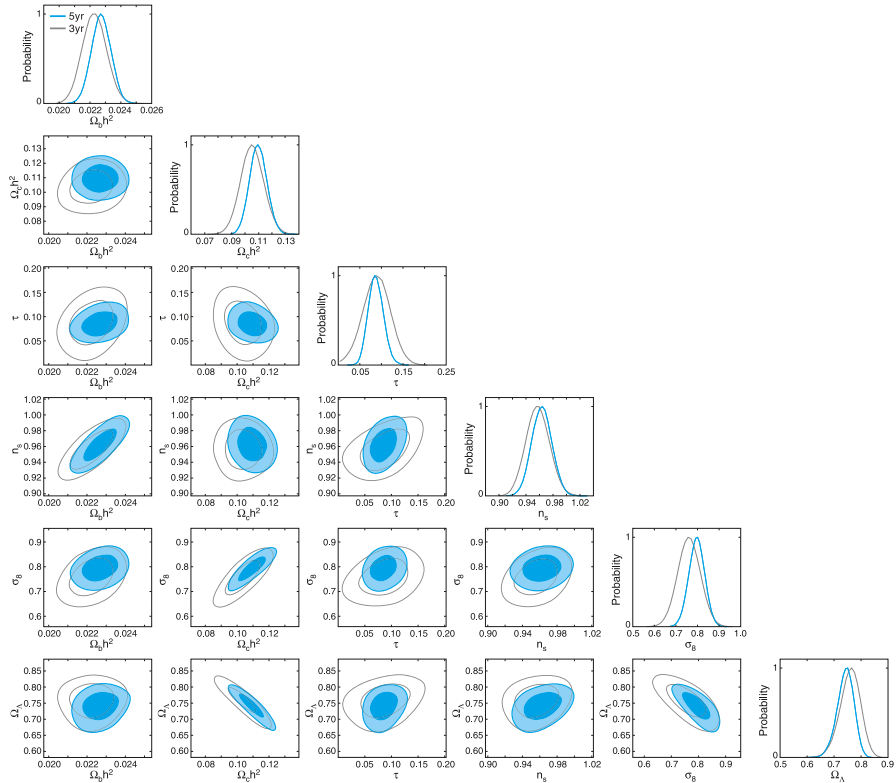}}
\caption{Constraints from the 5-yr {\tt WMAP} data on $\Lambda$CDM parameters (blue),
showing marginalized 1-D distributions and 2-D 68\% and 95\% limits (Dunkley et al. 2008).
Parameters are consistent with the 3-yr limits (grey) from Spergel et al. (2007).}
\label{dunkley08}
\end{figure}
These parameters have been confirmed by the CMB data from {\tt WMAP}
(e.g. Spergel et al. 2003, 2006; Dunkley et al. 2008).
But, although the excess power at $l \approx 2000-6000$ reported
by {\tt CBI}\footnote{http://www.astro.caltech.edu/~tjp/CBI}
(Cosmic Background Imager) is roughly consistent with a
secondary contribution resulting from the Sunyaev-Zeldovich effect, this requires
a normalization for the matter power spectrum higher than measured by other means
(Sievers et al. 2003).

\begin{figure}
\centerline{\includegraphics[width=25pc]{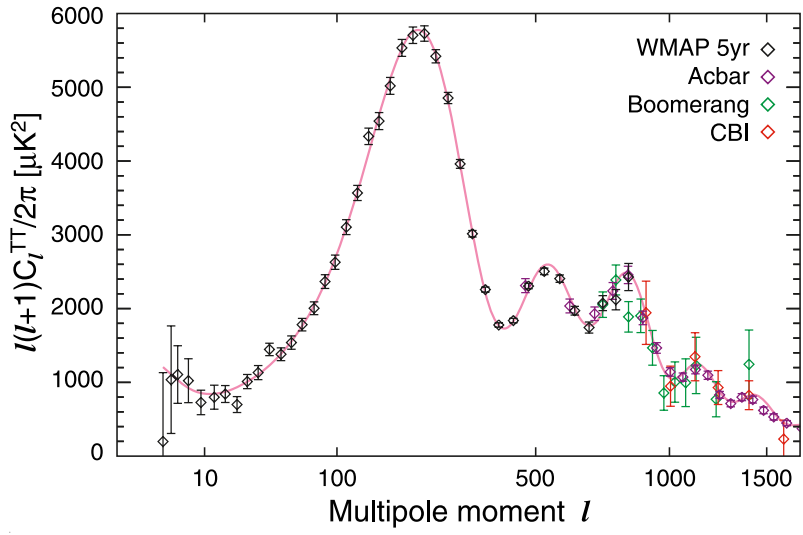}}
\caption{CMB temperature fluctuations as measured by different missions (symbols). The 
curve is the best-fit $\Lambda$CDM model to the {\tt WMAP} data  (see Nolta et al. 2008 for details).}
\label{cmbps}
\end{figure}

{\it The most popular, or concordance, model is a} $\Lambda$CDM with $\Omega_0=1.02 \pm
0.02$, $\Omega_\Lambda=0.73 \pm 0.04$, $\Omega_m=0.27 \pm 0.04$,
$\Omega_b=0.044 \pm 0.004$,  $h=0.71^{+0.04}_{-0.03}$, $\sigma_8=0.84 \pm 0.04$ and 
$n=0.93 \pm 0.03$ (Bennett et al. 2003). The above parameter set,
determined from CMB experiments, is in impressively good agreement with that obtained
independently by experiments based on high-redshift Type Ia supernovae (Knop et al. 2003) 
and galaxy cluster evolution data. Throughout the review
we will adopt these values. Slight changes to the above values are constantly being
made as {\tt WMAP} is collecting additional data. The results after 5 years of mission have
been released very recently (see e.g. Hinshaw et al. 2008). The reader is then referred to 
the {\tt WMAP} web page for further updates.  In Fig.~\ref{dunkley08} expected values for the
cosmological parameters as measured by {\tt WMAP} are plotted, while Fig.~\ref{cmbps} shows 
the most recent comparison 
(i.e. after the 5-th year of {\tt WMAP} observations) between the CMB temperature fluctuations 
predicted by such $\Lambda$CDM model and those measured by CMB experiments.

\subsection{Linear Growth of Fluctuations}

The commonly adopted theory for structure formation is the {\it
gravitational instability
scenario}, in which primordial density perturbations grow through gravitational
Jeans instability to form all the structures we observe today.
The most favored model for the origin of these perturbations is {\it inflation}. 
In this scenario, the universe expands exponentially for a brief period of time
at early epochs ($t \approx 10^{-35}-10^{-33}$~s), and, at the end of this inflationary
period, it is highly homogeneous on large scales, but is locally
perturbed as a consequence of quantistic fluctuations. To derive the evolution
of these primordial density fluctuations into bound objects, we can proceed as
follows. 

Let us describe the universe in terms of a fluid made of collisionless dark
matter and baryons, with a mean mass density $\bar{\rho}$. At any time and location,
the mass density can be written as $\rho({\bf x},t)=\bar{\rho}(t)[1+\delta({\bf x},t)]$,
where ${\bf x}$ indicates the comoving spatial coordinates and $\delta({\bf x},t)$
is the mass density contrast. The time evolution equation for $\delta$ during the
linear regime ($\delta \ll 1$) reads (Peebles 1993):
\begin{equation}
\ddot{\delta}({\bf x},t)+2H(t)\dot{\delta}({\bf x},t)=4\pi G \bar{\rho}(t)
\delta({\bf x},t)+ \frac {c_s^2}{a(t)^2} \nabla^2 \delta({\bf x},t).
\label{deltaev}
\end{equation}
Here, $c_s$ is the sound speed, $a \equiv (1+z)^{-1}$ is the scale factor which describes the
expansion and $H(t)=H_0 [\Omega_m (1+z)^3+\Omega_\Lambda]^{1/2}$.
The second term on the left hand side of the above equation represents
the effect of cosmological expansion. This, together with the pressure
support (second term on the right hand side) acts against the growth of
the perturbation due to the gravitational collapse (first term on the
right hand side). The pressure in the baryonic gas is essentially provided
by collisions, while in the collisionless dark matter component the pressure
support arises from the readjustment of the particle orbits. The above equation can
be used also to follow separately the evolution of the different components
of a multi-component medium. In this case, $c_s$ would be the velocity of the
perturbed component (which provides the pressure support) and $\bar{\rho}$
would be the density of the component which is most dominant gravitationally
(as it drives the collapse of the perturbation).
The equation has two independent solutions, one of which
grows in time and governs the formation of structures.
The total density contrast at any spatial location can be described in
the Fourier space as a superposition of modes with different wavelengths:
\begin{equation}
\delta({\bf x},t)=\int \frac{d^3 {\bf k}}{(2 \pi)^3} \delta_{\bf k}(t)
{\rm exp}(i {\bf k} \cdot {\bf x}),
\end{equation}
where ${\bf k}$ is the comoving wave number of the Fourier series.
The evolution of the single Fourier components is then given by:
\begin{equation}
\ddot{\delta}_{\bf k}+2H\dot{\delta}_{\bf k}=\left (4\pi G \bar{\rho}-
\frac{k^2 c_s^2}{a^2} \right )\delta_{\bf k}.
\label{deltakev}
\end{equation}
This sets a critical wavelength, the Jeans length, at which the
competing pressure and gravitational forces cancel (Jeans 1928):
\begin{equation}
\lambda_J=\frac{2\pi a}{k_J}=\left (\frac{\pi c_s^2}{G \bar{\rho}} \right )^{1/2}.
\label{jlength}
\end{equation}
For $\lambda \gg \lambda_J$ the pressure term is negligible because
the response time for
the pressure wave is long compared to the growth time for the density contrast,
and the zero pressure solutions apply. On the contrary, at $\lambda < \lambda_J$
the pressure force is able to counteract gravity and 
the density contrast oscillates as a sound wave.
It is conventional to introduce also the Jeans mass as the mass within a sphere
of radius $\lambda_J/2$:
\begin{equation}
M_J=\frac{4 \pi}{3} \bar{\rho} \left (\frac{\lambda_J}{2} \right )^3.
\label{jmass}
\end{equation}
In a perturbation with mass greater than $M_J$ the pressure force is not counteracted
by gravity and the structure collapses.
This sets a limit on the scales that are able to collapse at each epoch and has a
different value according to the component under consideration, reflecting the
differences in the velocity of the perturbed component.

Given the initial power spectrum of the perturbations, $P(k)\equiv
\vert \delta_k \vert ^2$, the evolution of each mode can be followed
through eq.~\ref{deltakev} and then integrated to recover the global
spectrum at any time. The inflationary model predicts that
$P(k)\propto k^n$, with $n=1$. This value corresponds to a scale invariant
spectrum, in which neither small nor large scales dominate. Although the
initial power spectrum is a pure power-law, perturbation growth
results in a modified final power spectrum. In fact, while on
large scales the power spectrum follows a simple linear evolution,
on small scales it changes shape due to the additional non-linear
gravitational growth of perturbations and it results in a bended
spectrum, $P(k) \propto k^{n-4}$. The amplitude of the power spectrum,
however, is not specified by current models of inflation and must be
determined observationally. Note that most of the power of the  
fluctuation spectrum of the standard CDM model is on           
small scales; therefore these are the first to
become non-linear. In the following Section we will discuss the
non-linear evolution of the dark matter halos.

\subsection{Non Linear Evolution: Dark Matter Halos}

Because dark matter is made of collisionless particles that interact very
weakly with the rest of matter and with the radiation field, density contrast
in this component can start to grow at early times.
Once a scale has become non-linear, the linear perturbation theory
described in the previous Section does not apply anymore.
In overdense regions ($\delta >0$)
the self-gravity of the local mass concentration will work against the expansion
of the universe. This region will expand at a progressively slower rate compared
to the background universe and, eventually, will collapse and form
a bound object. The details of this process depend on the initial density
profile. 
A simple and elegant approximation to describe the non-linear stage of gravitational
evolution has been developed by Zel'dovich (1970). In this approach, sheetlike structures
(``pancakes'') are the first non-linear structures to form from collapse
along one of the principal axes, when gravitational instability amplifies density
perturbations. Other structures, like filaments and knots, would result from
simultaneous contractions along two and three axes, respectively. As the probability
distribution for simultaneous contraction along more axes is small, in this
scenario pancakes are the dominant features arising from the first stages of
non-linear evolution.
Numerical simulations have been employed to test the Zeldovich approximation, finding
that it works remarkably well at the beginning of the non-linear evolution. At later
times, however, its predictions are not as accurate.
Hence, although the perturbations in Gaussian density fields are inherently 
triaxial (e.g. Bardeen et al. 1986) and an ellipsoidal collapse
would be more suitable to follow their evolution (e.g. Eisenstein \& Loeb
1995; Sheth, Mo \& Tormen 2001; Sheth \& Tormen 2002), the simplest model is the one of a spherically 
symmetric, constant
density region, for which the collapse can be followed analytically. At a
certain point the region reaches the maximum radius of expansion, then it turns
around and starts to contract. 
In the absence of any symmetry violation,
the mass would collapse into a point. However, long before this
happens, the dark matter experiences a violent relaxation process and
quickly reaches virial equilibrium. 
If we indicate with $z$ the redshift
at which such a condition is reached, the halo can be described in terms of its 
virial radius,
$r_{vir}$, circular velocity, $v_c = \sqrt{GM/r_{vir}}$, and virial temperature, 
$T_{vir}= \mu m_p v_c^2/2 k_B$, whose expressions are (Barkana \& Loeb 2001):
\begin{equation}
r_{vir}=0.784 \left ( \frac {M} {10^8 h^{-1} {\rm M}_\odot} \right )^{1/3}
\left [ \frac{\Omega_m}{\Omega_m^z} \frac {\Delta_c}{18 \pi^2} \right ]^{-1/3}
\left ( \frac{1+z}{10} \right )^{-1} h^{-1} {\rm kpc},
\end{equation}
\begin{equation}
v_c= 23.4
\left (\frac {M} {10^8 h^{-1} {\rm M}_\odot} \right )^{1/3}
\left [ \frac{\Omega_m}{\Omega_m^z} \frac {\Delta_c}{18 \pi^2} \right ]^{1/6}
\left ( \frac{1+z}{10} \right )^{1/2} {\rm km \; s}^{-1},
\end{equation}
\begin{equation}
T_{vir}= 2 \times 10^4 \left (\frac{\mu}{0.6} \right)
\left ( \frac {M} {10^8 h^{-1} {\rm M}_\odot} \right )^{2/3}
\left [ \frac{\Omega_m}{\Omega_m^z} \frac {\Delta_c}{18 \pi^2} \right ]^{1/3}
\left ( \frac{1+z}{10} \right ) {\rm K}.
\end{equation}
Here, $\mu$ is the mean molecular weight, $m_p$ is the proton mass, and
(Bryan \& Norman 1998):
\begin{equation}
\Delta_c=18 \pi^2+82 (\Omega_m^z -1)-39 (\Omega_m^z -1)^2,
\end{equation}
\begin{equation}
\Omega_m^z=\frac{\Omega_m (1+z)^3}{\Omega_m (1+z)^3+\Omega_\Lambda}.
\end{equation}

\begin{figure}
\centerline{\includegraphics[width=26pc]{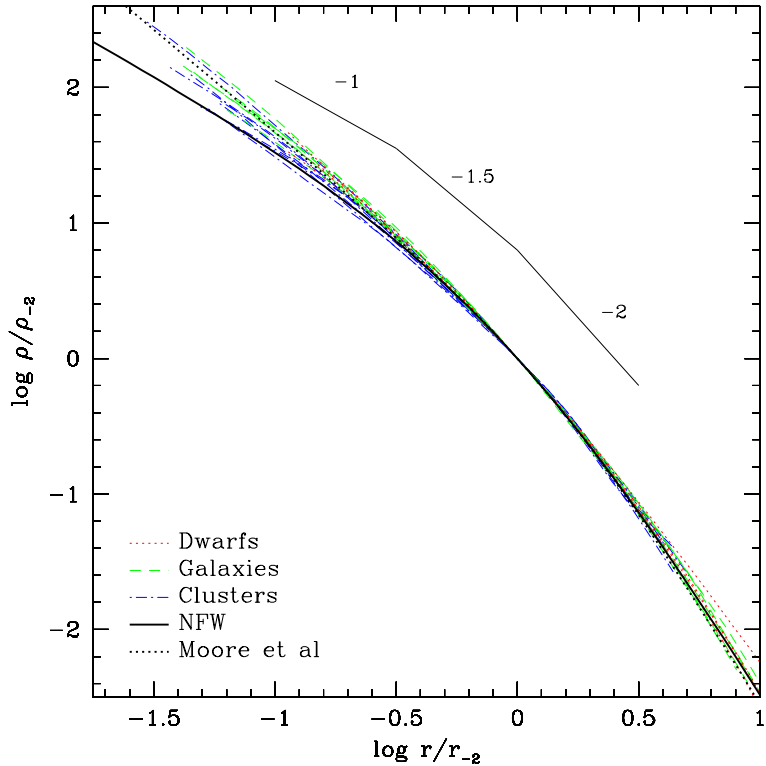}}
\caption{Density profiles of halos in a simulation by Navarro et al. (2004),
scaled to the radius, $r_{-2}$, where the local logarithmic slope of the
density profile takes the isothermal value of 2. Densities are scaled to
$\rho_{-2}=\rho(r_{-2})$.}
\label{navarro04}
\end{figure}
Although spherical collapse captures some of the physics governing the
formation of halos, their inner structure should be investigated
through numerical simulations. Navarro, Frenk \& White (1996, 1997 NFW)
have simulated the formation of dark matter halos of masses ranging from
dwarfs to rich clusters, finding that their density profile has a universal
shape, independent of the halo mass, the initial density fluctuation spectrum
and the cosmological parameters:
\begin{equation}
\rho(r)=\frac {\rho_s}{(r/r_s)(1+r/r_s)^2},
\end{equation}
where $\rho_s$ and $r_s$ are a characteristic density and radius. Usually,
the quantity $c \equiv r_{vir}/r_{s}$, the concentration parameter, is
introduced. As $\rho_s$ can be written in terms of $c$, the above
equation is a one-parameter form. An improved fitting formula (which reproduces
the more gradual shallowing of the density profile towards the centre, the 
apparent lack of evidence for convergence to a well-defined central power law
and the scatter in profile for different halos) has been developed by Navarro 
et al. (2004; eqs. 4-5).
Fig.~\ref{navarro04} shows that, with proper scaling, there is little difference
in the shape of density profile of halos of different mass.                  
The halo profiles are approximately isothermal over a large range of radii, but they
are shallower ($\rho \propto r^{-1}$) near the center and steeper
($\rho \propto r^{-3}$) near the virial radius.
The halo mass and the characteristic density are strongly correlated:
low-mass halos are denser than more massive systems, and this reflects the
higher collapse redshift of small halos.
The existence of a universal shape is explained as the result of the violent
relaxation process producing an equilibrium largely independent of
initial conditions. On the other hand, simulations which accurately
follow the mass accretion histories of dark matter halos (Wechsler et al. 2002;
Zhao et al. 2003), find an early phase of fast accretion, during which the
inner structure of halos
is mainly built, followed by a late phase of gentle accretion due to
the secondary infall, which gives rise to an outer profile matching
the universal form. Violent relaxation would be an active process only during
the first phase.

A universal shape of the density profile with an internal cusp, has 
been confirmed by other authors (Moore et al. 1999; Del Popolo et al. 2000;
Ghigna et al. 2000; Jing 2000; Jing \& Suto 2000; Power et al. 2003;
Fukushige, Kawai \& Makino 2004),
although it has been noted that the halos simulated by NFW
are intentionally selected to be in an equilibrium
state, and thus the shape would not be, strictly speaking, universal.
Moreover, compared with the original NFW simulations, some of these 
studies find a steeper central profile ($\rho \propto r^{-\beta}, 1.2<\beta<1.5$).
This result could be related to an increased resolution
of the numerical simulations (Ghigna et al. 2000), which nowadays 
can resolve down to less than 1\% of the virial radius. 
The simulations by Reed at al. (2005a), which
resolve 0.5\% of the virial radius, confirm the existence of an internal cusp
with $1<\beta<1.5$, depending on the mass of the halo (smaller halos have steeper
cusps).

{\it Not all the studies converge towards a universal shape of the halo
density profile.}
Rather, some authors (e.g. Kravtsov et al. 1998; Subramanian, Cen \& Ostriker 2000;
Ricotti 2003; Cen et al. 2004) 
find that the scatter in the halo profile is large,
depending on a different number of items. For example, the outer slope of the
density profile strongly depends on the environment (Avila-Reese et
al. 1999); the slope at 1\% of the virial radius increases with decreasing
halo mass (Jing \& Suto 2000); the slope depends on the initial fluctuation
field and cosmology (Nusser \& Sheth 1999); the concentration parameter
is not constant but decreases with increasing redshift and mass and halos in dense
environment are more concentrated than isolated ones (Bullock et al. 2001; Macci\`o
et al. 2007);
clusters (dwarf galaxies) have steeper (shallower) cusps than Milky 
Way type galaxies with a concentration parameter that is a
universal constant at the virialization redshift (Ricotti 2003). Some of the 
apparent tensions with works that find a universal shape of the density profile
might be due to the different method used to analyse the simulations
(Ricotti, Pontzen \& Viel 2007).

Also on the presence of a central cusp there is no general agreement
(e.g. Stoehr 2006). For
example, if the effect of gas is included in the calculation a flat core
is more plausible than a central cusp. In fact, if
the gas is not initially smoothly distributed, but is concentrated in
clumps, dynamical friction
acting on these clumps dissipates their orbital energy and deposits it
in the dark matter, resulting in a heating of the halo and the
formation of a finite, non-divergent core (El-Zant, Shlosman \& Hoffman 2001;
Mashchenko, Couchman \& Wadsley 2006).
A core would be also expected if the standard description of the post-collapse
object were replaced by a more realistic truncated, non-singular, isothermal 
sphere (TIS) in virial and hydrostatic equilibrium (Shapiro, Iliev \& Raga 1999).
An alternative mechanism that could
prevent a cusp forming is the following. The build up of dark matter
halos by merging satellites inevitably leads to an inner cusp with
$\beta >1$.    A flatter core with $\beta <1$   exerts on each satellite
tidal compression which prevents deposit of stripped satellite material
in this region. An inner cusp is expected as long as enough satellite
material is captured by the inner halo.
If, instead,  satellites are disrupted in the outer halo, a flatter
core is maintained (Dekel, Devor \& Hetzroni 2003).
A flat core could be also explained with the existence of a dark spheroid of
baryons of mass comparable with that of the cold dark matter
component (Burkert \& Silk 1997). 

\begin{figure}
\centerline{\includegraphics[width=26pc]{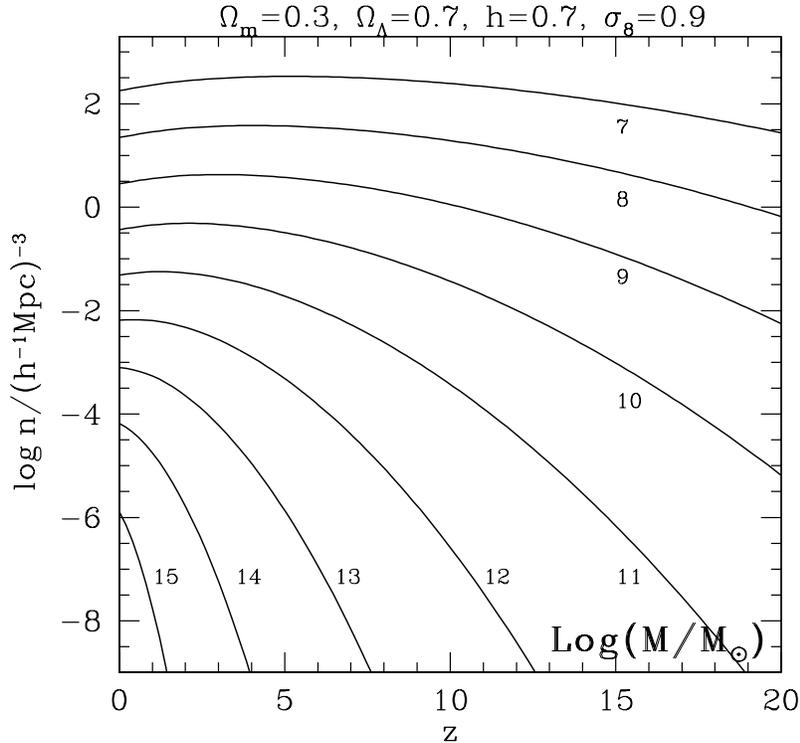}}
\caption{Each curve indicates the variation with redshift of the comoving number
density of dark matter halos with masses exceeding a specific value $M$ in the
standard $\Lambda$CDM model. The label on each curve indicates the corresponding
value of log($M/$M$_\odot$) (see Mo \& White 2002 for details).}
\label{halodens}
\end{figure}
The density of the dark matter halos and their spatial distribution are better
known. Two main methods have emerged to evaluate
them: numerical computations that solve the equations of gravitational
collapse, and analytical techniques that approximate these
results with simple  one-dimensional functions. While only numerical
simulations give the spatial distribution of halos, analytical techniques
are extremely useful as they are much faster and allow the analysis of a wide
range of parameters. The most commonly applied method of this type was
first developed by Press \& Schechter (1974). In their approach, the
abundance of halos at a redshift $z$ is determined from the linear density
field by applying a model of spherical collapse to associate peaks
in the field with virialized objects in a full non-linear treatment. This
simple model, later refined by Bond et al. (1991) and Lacey \& Cole (1994),
has had great success in describing the formation of structures and reproducing
the numerical results. The method provides the comoving number
density of halos, $dn$, with mass between $M$ and $M+dM$ as (Peebles 1993):
\begin{equation}
M \frac{dn}{dM}=\left (\frac{2}{\pi} \right )^{1/2} \frac{-d ({\rm ln}\sigma)}
{d ({\rm ln}M)} \; \frac{\rho_0}{M} \; \nu_c \; {\rm e}^{-\nu_c^2/2},
\end{equation}
where $\rho_0$ is the present mean mass density, $\sigma$ is the standard
deviation of the density contrast smoothed through a certain window and
$\nu_c$ is the minimum number of standard deviations of a collapsed fluctuation;
Fig.~\ref{halodens} shows  an application of the above equation.

While the model accurately reproduces the number density of the dark matter
halos, it does not predict their spatial distribution,
which is important to study structure
formation. For this reason, Porciani et al. (1998) and 
Scannapieco \& Barkana (2002) have developed analytical
techniques to derive halo correlation functions and non-linear biasing, 
by studying the joint statistics of dark matter halos forming at two points.
An alternative approach to the analytical determination of the abundance
and spatial distribution of dark matter halos has been developed by Sheth 
\& Tormen (2002). This is known as the excursion set approach and can be applied
in the case of both a spherical (to reproduce the standard Press-Schechter results)
and an elliptical collapse, which seems to be in better agreement with numerical
simulations. 

\subsection{Formation of protogalaxies}

In contrast to dark matter,
as long as the gas is fully ionized, the radiation drag on free electrons
prevents the formation of gravitationally bound systems. When recombination
(defined as the time at which the electron fraction has dropped to 0.1) occurs 
at $z_{rec} \approx 1250$ the primeval plasma combines into neutral 
atomic hydrogen.
The residual ionization of the cosmic gas keeps
its temperature locked to the CMB temperature through Compton scattering,
down to a redshift $1+z_t\approx 1000 (\Omega_b h^2)^{2/5}$ (Peebles 1993).
Following recombination, the nearly neutral matter decouples from the
radiation and perturbations in this component are finally able to grow in
the pre-existing dark matter halo potential wells and eventually form the
first bound objects. The process leading to the virialization of the gaseous
component of matter is similar to the dark matter one. In this case, during
the contraction following the turnaround, the gas develops shocks and gets
reheated to a temperature at which pressure support can prevent further
collapse.

The mass of these first bound objects can be derived from eqs.~\ref{jlength}-
~\ref{jmass}, where $c_s$ is the velocity of the baryonic gas.
In particular, at $z>z_t$ the Jeans mass is time-independent,
while at $z<z_t$, when the gas temperature declines adiabatically,
$M_J$ decreases with decreasing redshift: 
\begin{equation}
M_J=3.08 \times 10^3 \left ( \frac{\Omega_m h^2}{0.13} \right )^{-1/2} 
\left ( \frac{\Omega_b h^2}{0.022} \right )^{-3/5} \left ( \frac{1+z}
{10}\right )^{3/2} \; {\rm M}_\odot.
\end{equation}
As the determination of the Jeans mass is based on a perturbative approach,
it can only describe the initial phase of the collapse. $M_J$ is simply a
limit given by the linear theory to the minimum mass that is able to
collapse. It is worth noticing that  the Jeans mass
represents only a necessary but not sufficient condition for collapse:
in addition, one has to require that the atomic/molecular cooling time
is shorter than the Hubble time. 
The dynamical and chemical evolution of collapse must then 
be derived in detail to assess if the object can actually 
become a protogalaxy.                   
In the following we will discuss the minimum mass of protogalaxies.

In the standard cosmological hierarchical scenario for structure
formation, the objects which collapse first are predicted to have masses
corresponding to virial temperatures $T_{vir}<10^4$ K. 
Once the gas has virialized in the potential wells of dark matter halos, 
additional cooling is required to further collapse the gas and form
stars. For a gas of primordial composition at such low temperatures the
main coolant is molecular hydrogen.
We define {\it Pop~III objects} those for which H$_2$ cooling is required
for collapse (note that in the literature Pop~III 
objects are sometimes also referred to as
{\it minihalos}).
After an H$_2$ molecule gets rotationally or vibrationally excited
through a collision with an H atom or another H$_2$ molecule,  a
radiative de-excitation leads to
cooling of the gas. This mechanism has been investigated in great detail by
several authors (Lepp \& Shull 1984; Hollenbach \& McKee 1989; Martin,
Schwarz \& Mandy 1996; Galli \& Palla 1998) and has been applied to the
study of the formation of primordial small mass objects (e.g. Haiman, Thoul
\& Loeb 1996; Haiman, Rees \& Loeb 1996; Abel et al. 1997;
Tegmark et al. 1997; Abel et al. 1998;
Omukai \& Nishi 1999; Abel, Bryan \& Norman 2000; Yoshida et al. 2003;
Maio et al. 2007).
Primordial H$_2$ forms with a fractional abundance of $\approx  10^{-7}$
at redshifts $\gsim 400$ via the H$_2^+$ formation channel. At redshifts
$\lsim 110$, when the CMB radiation intensity
becomes weak enough to allow for significant formation of H$^-$ ions, even
more H$_2$ molecules can be formed:
\begin{eqnarray}
{\rm H}  +  {\rm e}^- & \rightarrow & {\rm H}^-  +  h\nu \nonumber \\
{\rm H}^-  +  {\rm H} & \rightarrow & {\rm H}_2  +  {\rm e}^- \nonumber
\end{eqnarray}
Due to the lack of molecular data,
it has unfortunately not been possible to follow the details of the
H$_2^+$ chemistry as its level distribution decouples from the CMB.
Assuming the rotational and vibrational states of H$_2^+$ to be in
equilibrium with the CMB, the H$_2^+$ photo--dissociation rate is much
larger than the one obtained by considering only photo--dissociations
out of the ground state.  Conservatively, one concludes that these two
limits constrain the H$_2$ fraction to be in the range 
$10^{-6}-10^{-4}$. In fact, both limits have been used in the literature
(e.g. Lepp \& Shull 1984; Palla, Galli \& Silk 1995;
Haiman, Thoul \& Loeb 1996; Tegmark et al. 1997). If we assume that the H$^-$
channel for H$_2$ formation is the dominant mechanism, i.e. that the
H$_2^+$ photo--dissociation rate at high redshift is close to its
equilibrium value, this leads to a typical primordial H$_2$
fraction of $f_{H_{2}}\approx 2 \times 10^{-6}$ (Anninos \&
Norman 1996) which is found for model universes that satisfy the
constraint $\Omega_b h^2 = 0.019$.
Hirata \& Padmanabhan (2006) have recently revised the formation of primordial
H$_2$ including {\it (i)} the effect of spectral distortions due to radiation
emitted during hydrogen and helium recombination (which destroys H$^-$ at 
$z>70$) and {\it (ii)} a detailed treatment of the H$_2^+$ level population
(which suppresses the efficiency of the H$_2^+$ formation channel). As a result,
they find the slightly lower primordial value of $6 \times 10^{-7}$. It should
be noted that the above effects do not influence the H$_2$ formation and cooling 
at lower redshift and in overdense conditions.
In any case, this primordial fraction is usually lower than the one required for the
formation of Pop~III objects, but during the collapse 
the molecular hydrogen content can reach high enough values to
trigger star formation. Thus, the fate of a
virialized lump depends crucially on its ability to rapidly increase
its H$_2$ content during the collapse phase.
\begin{figure*}
\centerline{\includegraphics[width=26pc]{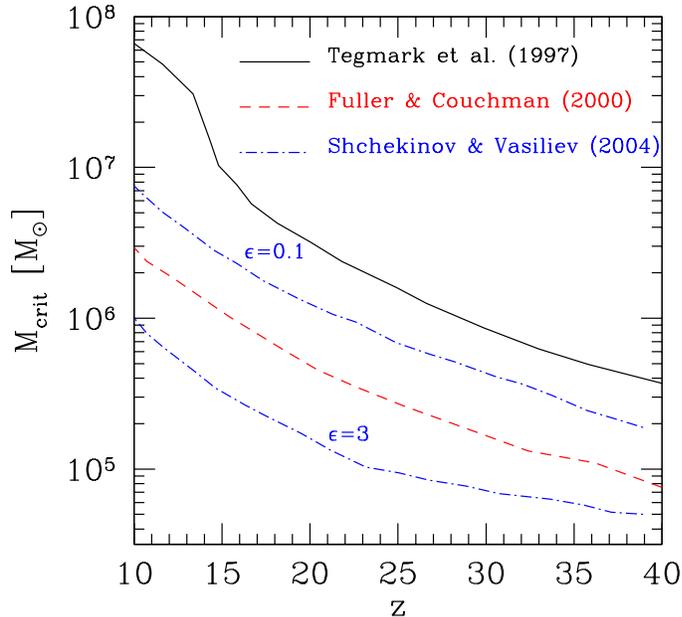}}
\caption{
Minimum mass able to cool and collapse as a function of redshift
in the formulation of Tegmark et al. (1997) (solid line), Fuller \&
Couchman (2000) (dashed) and Shchekinov \& Vasiliev (2004) (dashed-dotted).
The latter curves are derived
for two different values of the rate of ionizing photon production
by UHECR, $\epsilon=(0.1, 3)/(1+z)$ for the upper and lower line, respectively.
See original papers for details. }
\label{mass_min}
\end{figure*}

Haiman, Thoul \& Loeb (1996), have followed the growth of
spherical perturbations into the non-linear regime,
using a one-dimensional spherical Lagrangian
hydrodynamics code to follow the dynamical, thermal and non-equilibrium
chemical evolution of the gas. They
note that H$_2$ cooling becomes important only after virialization.
Interestingly, though, they find that
shell crossing by dark matter allows baryonic objects with masses well
below the Jeans mass to form.
A similar result is obtained by Haiman \& Loeb (1997).
Tegmark et al. (1997) have addressed the same question by
calculating the evolution of the H$_2$ abundance for different halo masses
and initial conditions for a standard CDM cosmology. They use an analytical
approach to the problem and find that the minimum baryonic mass $M_b$ is
strongly redshift dependent, dropping from $10^6$~M$_\odot$ at $z\approx 15$
to $5 \times 10^3$~M$_\odot$ at $z\approx 100$, as molecular hydrogen cooling
becomes effective (see Fig.~\ref{mass_min}). 
They conclude that if the prevailing conditions are such
that a molecular hydrogen fraction $\approx
5 \times 10^{-4}$ is produced, then the lump will cool, fragment and
eventually form stars.
This criterion is met only by larger halos, implying that for
each virialization redshift there is          
a critical mass, $M_{crit}$, such that protogalaxies with total mass
$M>M_{crit}$ will be able to collapse and form stars
and those with $M<M_{crit}$ will fail.
In a subsequent paper Nishi \& Susa (1999) claim that the primordial
number fraction of H$_2$ as calculated by Tegmark et al. (1997) is overestimated by about two orders of magnitude, because the destruction rate of
H$_2^+$ by CMB radiation at high redshift is underestimated (Galli \&
Palla 1998). Thus, since their primordial value $\approx 10^{-4}$
is comparable
to the one necessary to cool, their cooling criterion is not generally
reliable. Fuller \& Couchman (2000) have included
the chemical rate network responsible for the formation of H$_2$ in
an N-body hydrodynamical code, finding a value for $M_{crit}$
considerably lower than the one derived by Tegmark et al. (1997)
(see Fig.~\ref{mass_min}),
since they have used a different molecular hydrogen cooling function
and have included different chemical reactions, but they nevertheless find
the same {\it critical molecular hydrogen fraction for the collapse of $f_{H_2}\approx
5 \times 10^{-4}$.} Most studies (e.g. Abel, Bryan \& Norman 2000;
Machacek, Bryan \& Abel 2001; Reed et al. 2005b; O'Shea \& Norman 2007) agree that 
{\it the minimum mass 
allowed to collapse is as low as $10^5 M_\odot$}. 

\begin{figure*}
\centerline{\includegraphics[width=15pc]{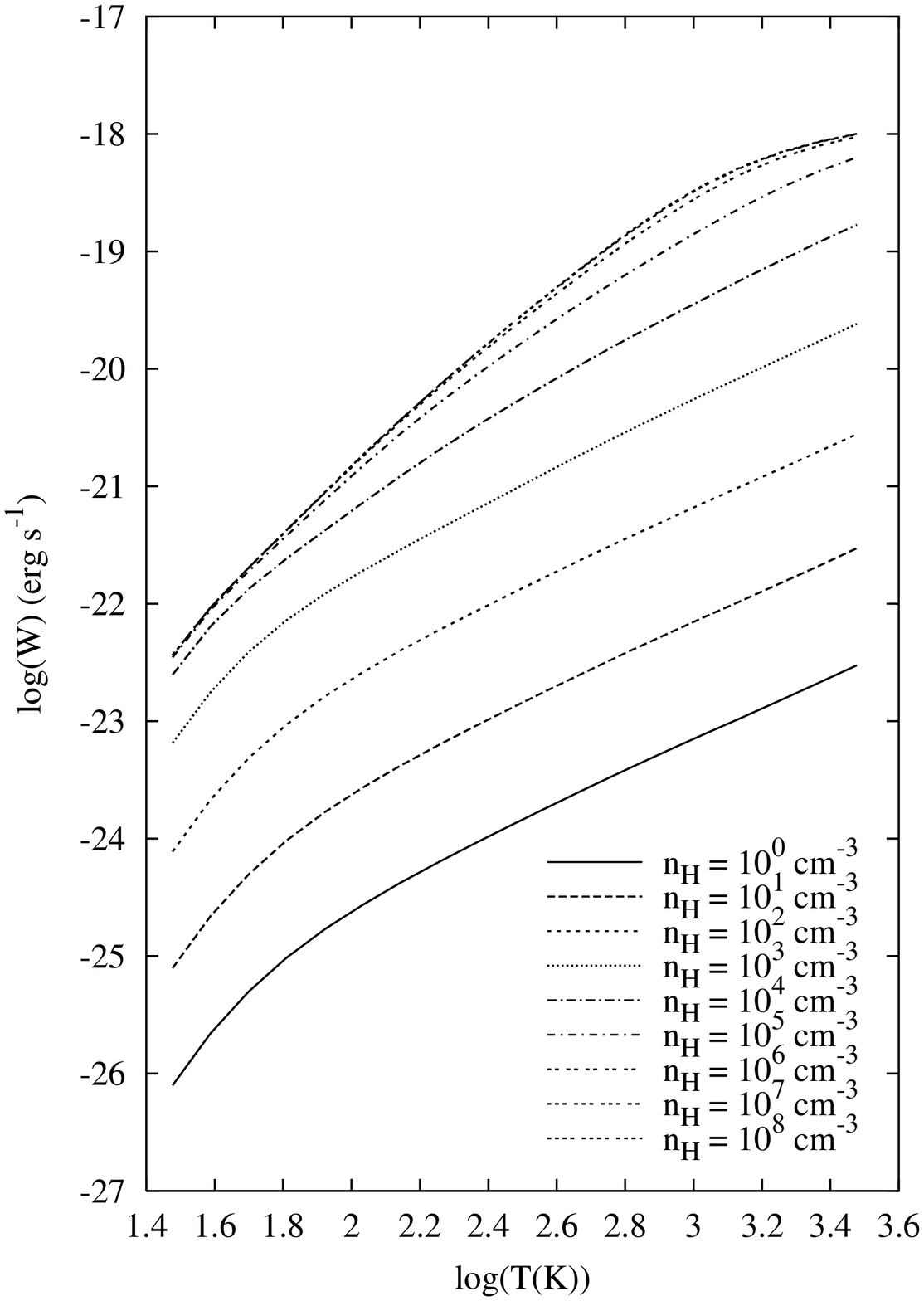}
\includegraphics[width=14.7pc]{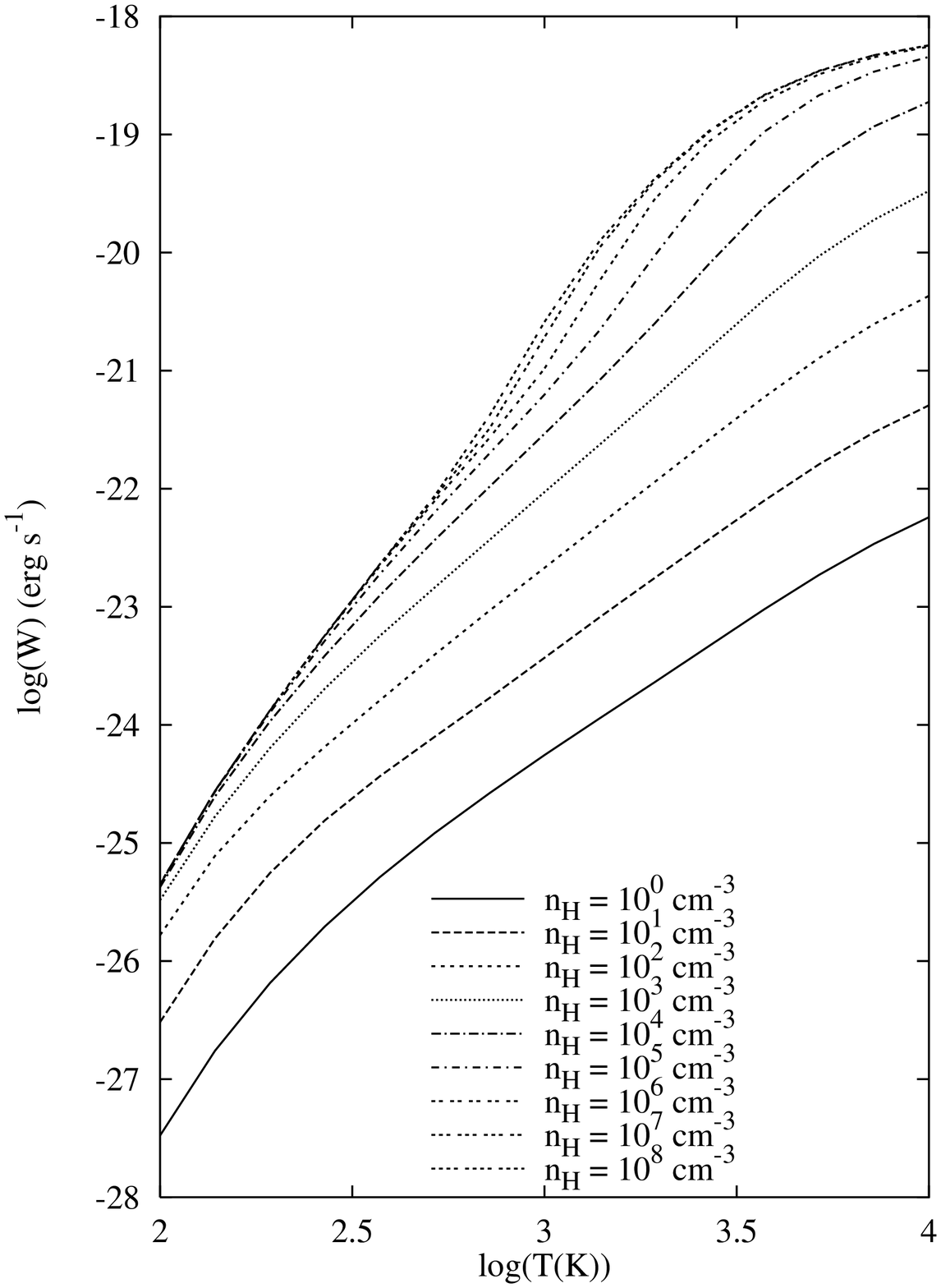}}
\caption{
{\it Left panel:} cooling rate per HD molecule calculated for an
ortho/para H$_2$ density ratio of 1 and an H/H$_2$ abundance ratio of 1.
{\it Right panel:} cooling rate per H$_2$ molecule. The parameters are
the same as in the left panel. For details refer to Flower et al. (2000)
and Le Bourlot, Pineau des For\^ets \& Flower (1999).}
\label{cool}
\end{figure*}

The results of recent quantum mechanical calculations of cross-sections
for rotational transitions within the vibrational ground state of HD
enabled the contribution of HD to the primordial cooling function to
be reliably calculated for the first time. In spite of its low abundance
($10^{-3}$) relative to H$_2$, HD can be as important in the
thermal balance of the primordial gas. In particular, the permanent
dipole moment and lower rotational constant of HD favor cooling by HD
at high densities and low kinetic temperatures (Flower 2000; Flower et
al. 2000; Galli \& Palla 2002), possibly allowing even smaller objects to collapse. If 
also vibrational transitions are taking into account, the cooling rate
is increased (Lipovka, N\'u\~nez-Lopez \& Avila-Reese 2005). In 
Fig.~\ref{cool} a comparison between the H$_2$ and HD cooling rates is
shown.

A considerably smaller 
characteristic  mass (a factor $\approx$ 10 below the one predicted by Tegmark et al. 1997)
of first luminous objects could  also
result from the presence of ultra-high energy cosmic rays (UHECRs), if they formed 
in the so-called top-to-bottom
scenario from decaying superheavy dark matter particles with masses $\simgt
10^{12}$~GeV. In fact, the small additional electron fraction produced by UHECR
photons results in more efficient H$_2$ and HD molecule formation  and cooling 
(Shchekinov \& Vasiliev 2004, 2006; see Fig.~\ref{mass_min}). Also
the presence of an early population of cosmic rays (that would produce free electrons;
Jasche, Ciardi \& En{\ss}lin 2007; Stacy \& Bromm 2007) or of sterile neutrinos with masses 
of several keV (that would decay in X-rays and produce free electrons; Biermann
\& Kusenko 2006) could boost the formation of the above molecules. 
On the other hand, Ripamonti, Mapelli \& Ferrara (2007b)
find that moderately massive dark matter particle (as sterile neutrinos and
light dark matter) decays and annihilations favors the formation of H$_2$
and HD, but at the same time induce heating of the gas. The net effect in this case is
a slight increase of M$_{crit}$.

It should be noted though that all the above estimates do not take into account
the large uncertainties on the reaction rates that govern the H$_2$ formation
and that can have a non negligible impact on the prediction for early structure
formation (e.g. Glover, Savin \& Jappsen 2006).

\subsection{Key Observations}

\subsubsection{Cosmological Model}

\begin{figure}
\centerline{\includegraphics[width=30pc]{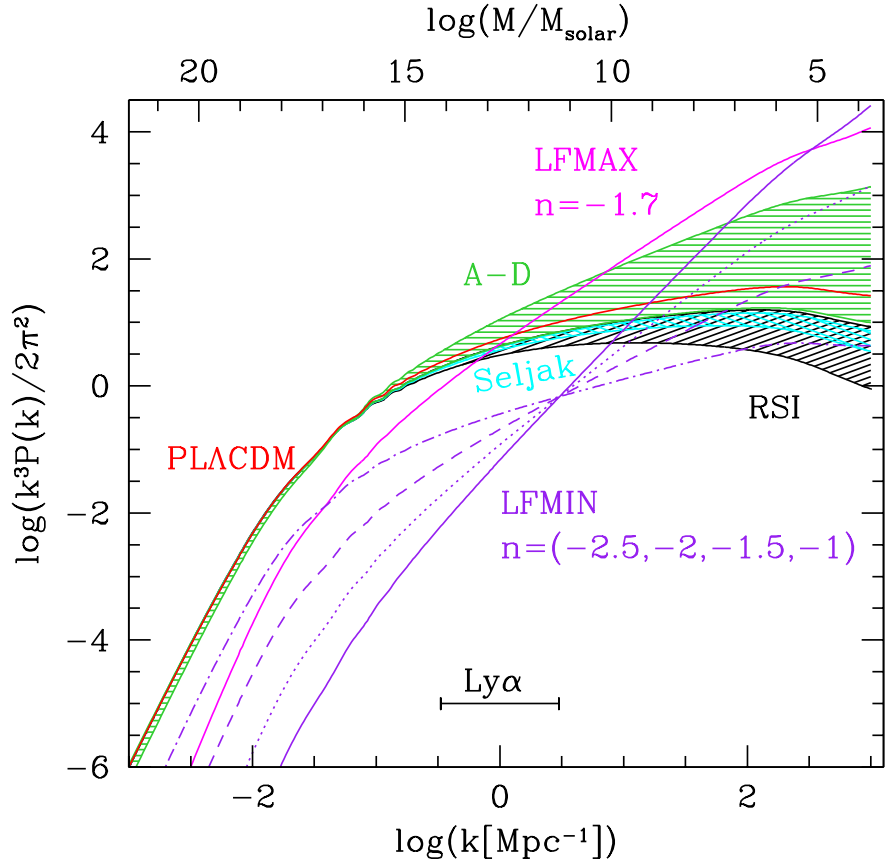}}
\caption{Matter power spectra (from Sugiyama, Zaroubi \& Silk 2003). The {\tt WMAP}
best fitted power law result is indicated by PL$\Lambda$CDM, and it is compared
with the {\tt WMAP} Running Spectral Index model (RSI) with errors combined
with different additional data
(hatched regions); the curves are isocurvature spectra with different prescriptions.
See original paper for details.
}
\label{massps}
\end{figure}

\begin{figure*}
\centerline{\includegraphics[width=14pc]{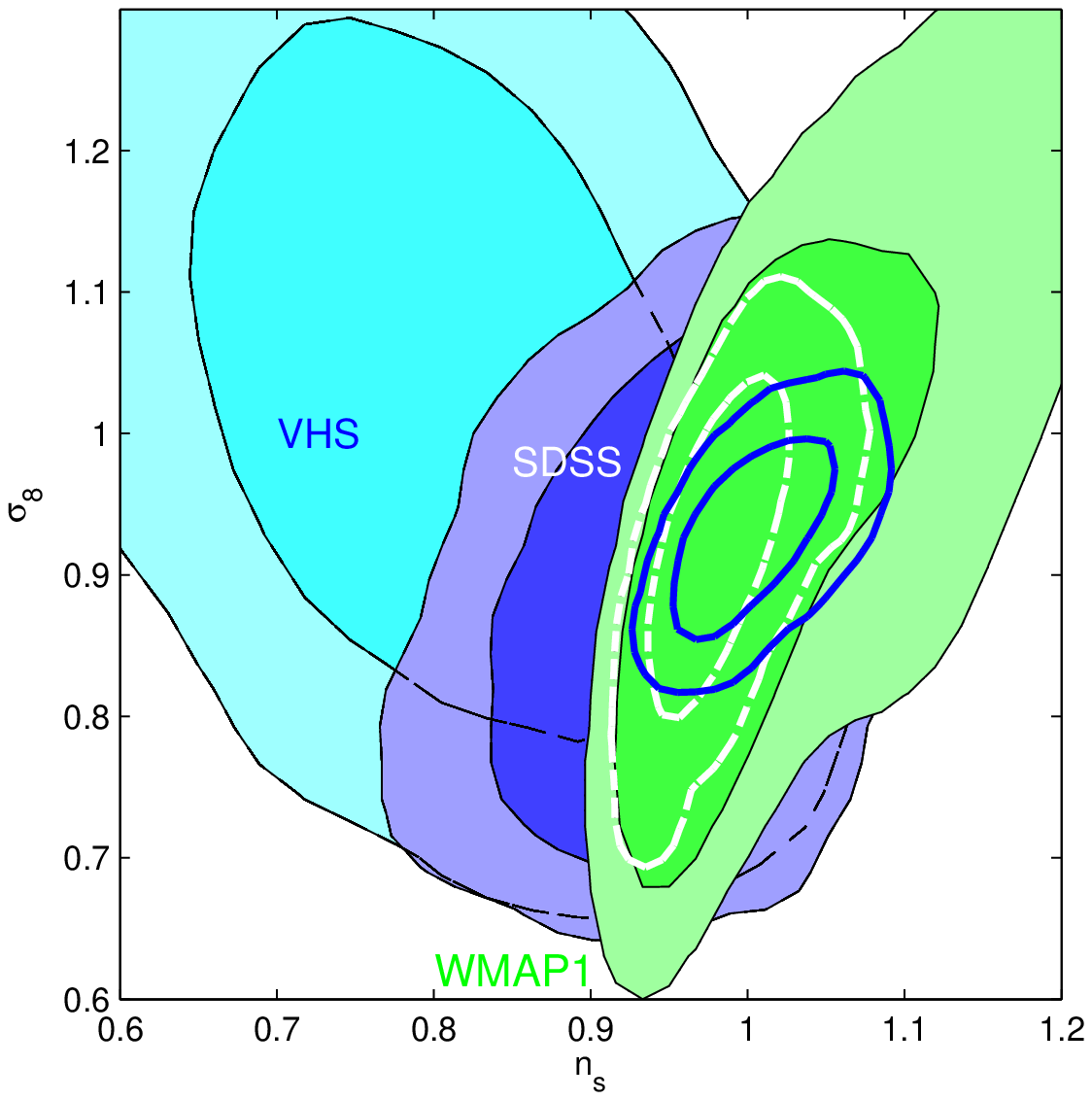}
\includegraphics[width=14pc]{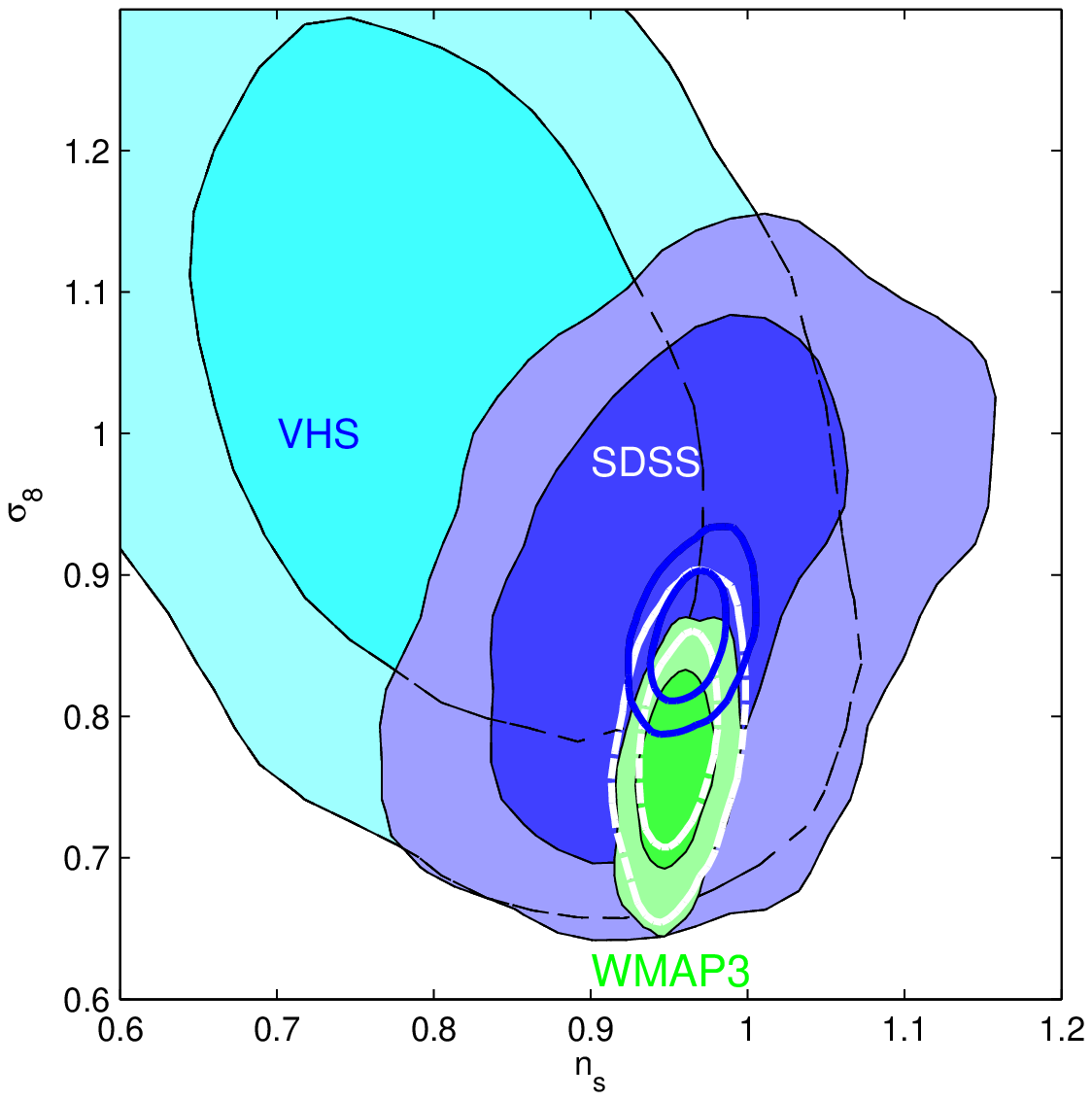}}
\caption{
1$\sigma$ and 2$\sigma$ likelihoods for $\sigma_8$ and the spectral index
(in this figure indicated with $n_s$), marginalized over all other parameters.
{\it Left panel:} constraints for 1-yr {\tt WMAP} only (green), the {\tt LUQAS}
(Kim et al. 2004)
+ Croft et al. (2002) data set as analyzed by Viel, Haehnelt \& Springel (2004, VHS) (cyan) and
the {\tt SDSS} Ly$\alpha$ forest data of McDonald et al. (2005) (blue).
The thick dashed white empty
contours refer to 1-yr {\tt WMAP} + VHS, while the solid blue contours are for
1-yr {\tt WMAP} + {\tt SDSS}.
{\it Right panel:} as in the left panel but for the 3-yr {\tt WMAP}.}
\label{viel06}
\end{figure*}

The main observational evidence in support of the inflationary scenario
comes from observations of the CMB, showing temperature anisotropies
on  large angular scales (e.g. Hu et al. 2001), which are interpreted as
the results of the primordial density fluctuations at the decoupling
epoch. The same measurements confirm that the fluctuations are adiabatic
and the value of the spectral index is the one
predicted by the inflationary model (Balbi et al. 2000, 2001; Pryke et al. 2002;
Sievers et al. 2003), although there is an indication for a possible
running spectral index (e.g. Spergel et al. 2003, 2007), with $n>1$ on large scales
and $n<1$ on small scales (see Fig.~\ref{massps} for possible power spectra).
This is not confirmed though when the constraints from the recent Ly$\alpha$
forest and galaxy bias analysis of the {\tt SDSS}\footnote{http://www.sdss.org/}
(Sloan Digital Sky Survey) are
combined with previous constraints from {\tt SDSS} galaxy clustering, the latest
supernovae and the {\tt WMAP} results (Seljak et al. 2005; Seljak, Slosar \& 
McDonald 2006). In fact, while CMB anisotropies, galaxy clustering from {\tt 2dF}
\footnote{http://magnum.anu.edu.au/~TDFgg/}
and {\tt SDSS}, supernovae and weak lensing trace large scale structures (scales
larger than 10~Mpc), tracers on smaller scale are needed to get correct information
on the running spectral index.
For the same reason, a tension is present also in the $\sigma_8$ value measured 
with or without the Ly$\alpha$ forest data, giving a higher value in the 
latter case (Seljak, Slosar \& McDonald 2006; Viel, Haehnelt \& Lewis 2006). 
More specifically, if only the 3-rd year
{\tt WMAP} data are used, $\sigma_8=0.76 \pm 0.05$, but the value increases with
additional constraints (Spergel et al. 2007), in particular those from the
Ly$\alpha$ forest (see Fig.~\ref{viel06};
Seljak, Slosar \& McDonald 2006; Viel, Haehnelt \& Lewis 2006).
In addition, analyses of the {\tt WMAP} observations
have reported evidence for a positive detection of non-Gaussian features (e.g.
Chiang et al. 2003; R\"ath, Schuecker \& Banday 2007; Eriksen et al. 2007;
McEwen et al. 2008).

As discussed in the previous Sections, the amplitude of the power spectrum is not
predicted by the model and has to be set observationally. Historically, this has been 
derived by comparing the observed CMB quadrupole anisotropies with theoretical ones 
(e.g. Wright et al. 1992; Efstathiou, Bond \& White 1992) or from the abundance of
galaxy clusters (see Sec. \ref{cosmomod}). More recently, this method
has been complemented by measuring the amplitude of the power spectrum from  
available large galaxy redshift surveys (e.g. Lahav et al. 2002).
The main problem related to the latter method is the attempt to estimate the
distribution of matter using galaxies as tracers. In fact, as
the galaxies form preferentially in high density regions of the mass distribution,
they are more strongly correlated than the underlying distribution and
an additional factor, the bias, should be specified.
It has been found from the {\tt 2dF} Galaxy Redshift
Survey that, at least on large scales ($5-30 h^{-1}$~Mpc), optically selected
galaxies do indeed trace the underlying mass distribution (Verde et al. 2002).

An alternative method to measure $P(k)$ is based on observations of the
high-redshift Ly$\alpha$ forest. The method is motivated by the physical picture
that has emerged from hydrodynamical cosmological simulations and related
semi-analytical models, in which typical Ly$\alpha$ forest lines arise
in a diffuse IGM. The thermal state of this low-density gas is governed
by simple physical processes, which lead to a tight correlation between the
Ly$\alpha$ optical depth, $\tau$, and the underlying matter density (e.g.
Bi \& Davidsen 1997; Hui, Gnedin \& Zhang 1997). Thus, as the transmitted
flux in a QSO spectrum is $F=e^{-\tau}$, one can extract information about
the underlying mass density field from the observed flux distribution
(Croft et al. 1998, 1999, 2002; Viel, Haehnelt \& Springel 2004;
McDonald et al. 2005). These works confirm a basic
prediction of the inflationary CDM scenario, an approximately scale invariant
spectrum of primeval fluctuations modulated by a transfer function that
bends it toward $k^{n-4}$ on small scales. A possible caveat on these applications
is that radiative transfer effects (Abel \& Haenhelt 1999; Bolton,
Meiksin \& White 2004) are not included in current
numerical simulations of the Ly$\alpha$ forest. Such effects might {\it (i)} blur the
polytropic temperature-density relation derived from the simulations
and usually adopted, hence introducing a non-negligible error on the 
predicted cosmological spectrum, and {\it (ii)} produce an inversion of the temperature-density
relation in regions of low density (voids), as pointed out by Bolton et al. (2008). 

It has been proposed (Loeb \& Zaldarriaga 2004; Ali, Bharadwaj \& Panday 2005;
Barkana \& Loeb 2005a; Bharadwaj \& Ali 2005; 
Pillepich, Porciani \& Matarrese 2007; Lewis \& Challinor 2007; Cooray, Li \& 
Melchiorri 2008; Mao et al. 2008) that a wealth of
information about the initial density fluctuations (e.g. the presence of a
running spectral index or deviations from gaussianity) could come from the 21~cm
line in absorption against the CMB, from $z\approx 200$ down to lower redshift
(as long as the effects of reionization are not dominant; Bowman, Morales \& Hewitt 2007)
or from the application of the Alcock-Paczy\'nski test to the 21~cm line (Nusser 2005b).
In these cases, measurements from upcoming radio telescopes in combination with CMB data,
would produce constraints on cosmological parameters tighter that those obtained by
CMB data alone (McQuinn et al. 2006). To make precision cosmology though, particular
care should be used in the modeling of the spin temperature (Hirata \& Sigurdson 2007). 

\begin{figure*}
\centerline{\includegraphics[angle=90, width=14.5pc]{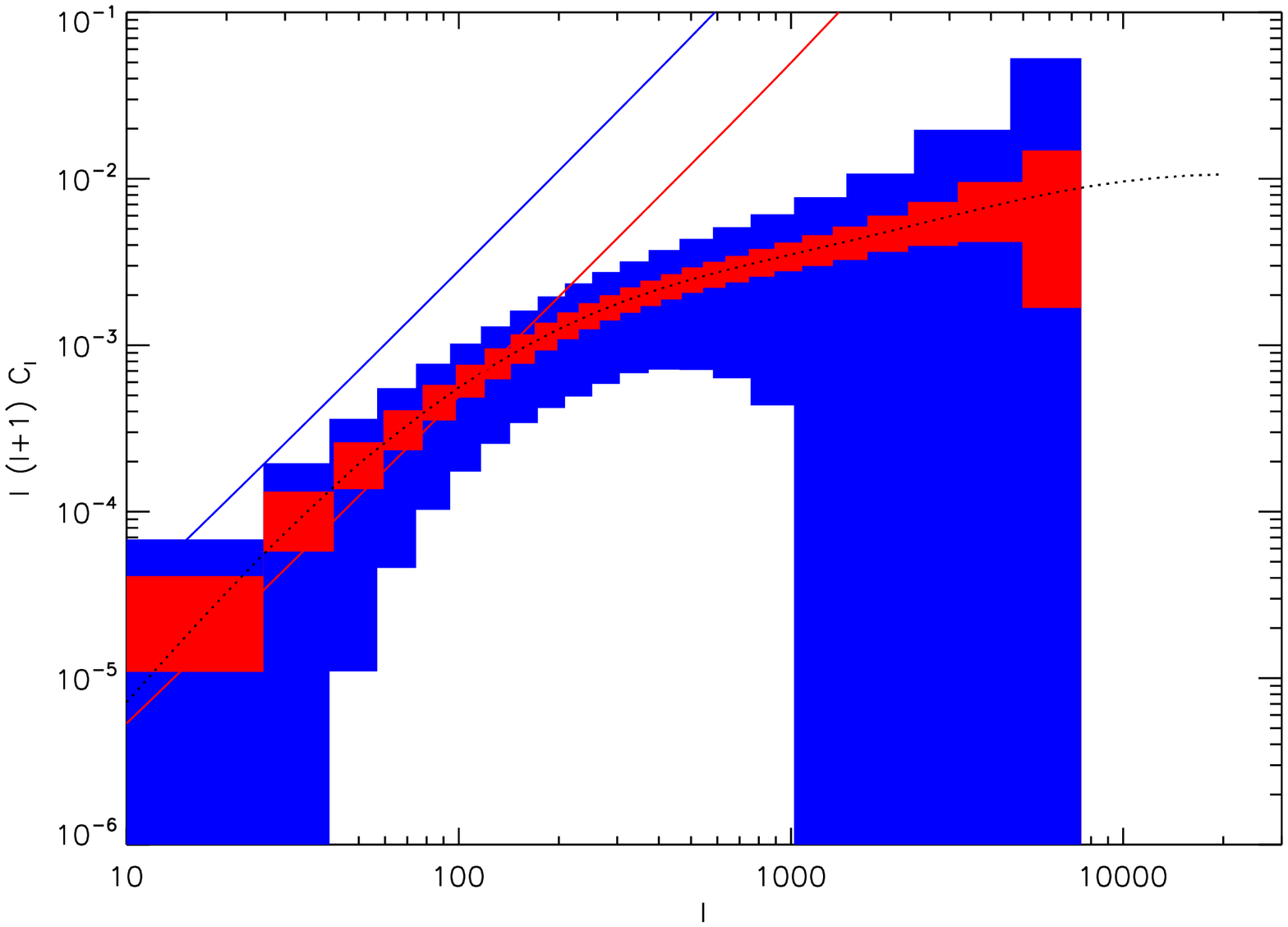}
\includegraphics[angle=90, width=14.5pc]{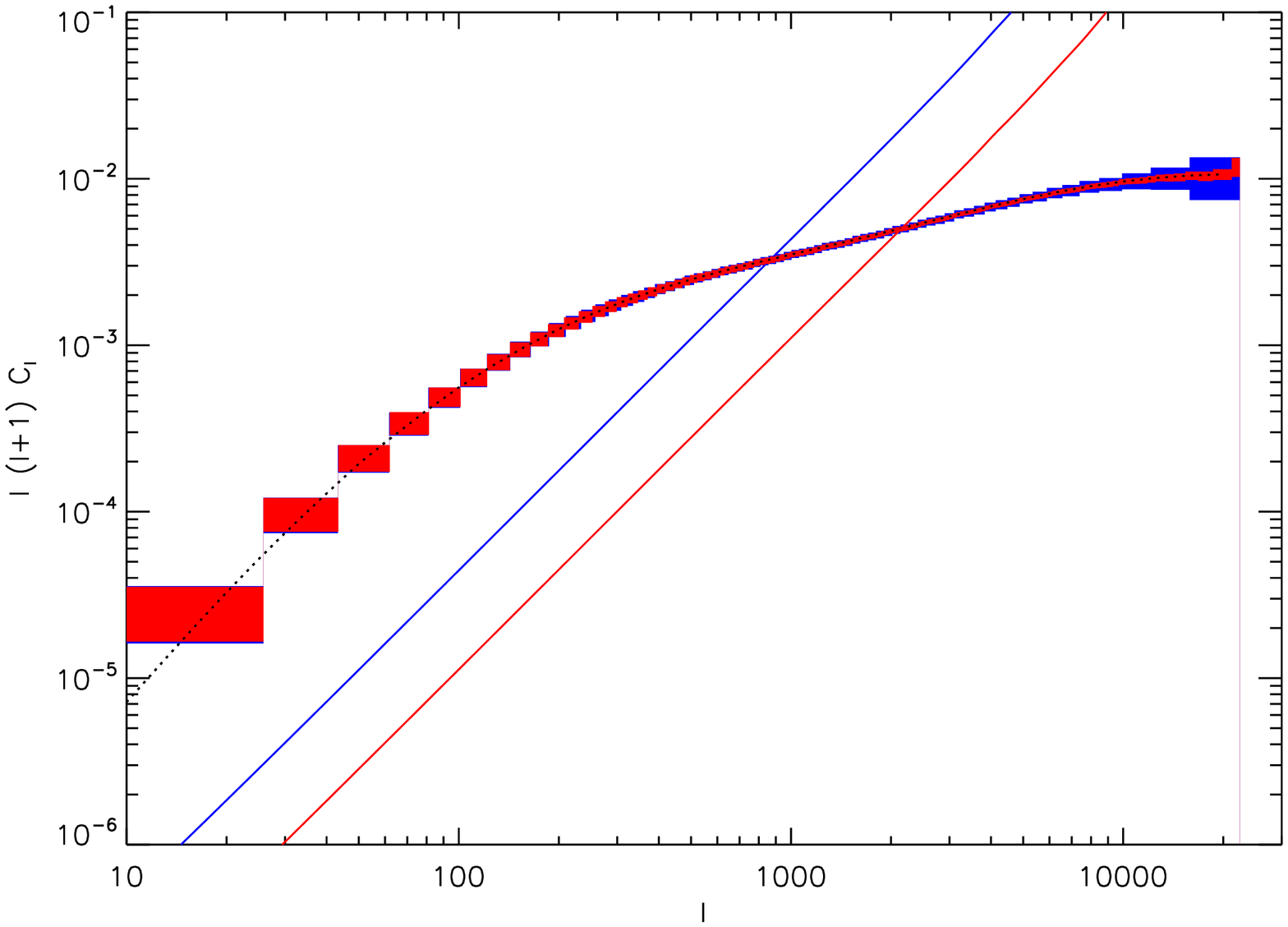}}
\caption{
{\it Left panel:} Forecasts of the 1$\sigma$ uncertainties in estimates of the
convergence power spectrum for a {\tt LOFAR}-1yr (blue) and {\tt LOFAR}-3yr (red)
type experiment of 21~cm weak lensing. The solid straight lines give the noise per
mode, while the dotted curve is the underlying model power spectrum.
{\it Right panel}: As the left panel but for a {\tt SKA} type experiment.
For details see Metcalf \& White (2008).} 
\label{21cmlens}
\end{figure*}

Another 21~cm related experiment that could be used to reconstruct matter density
fluctuations and to obtain constraints on cosmological parameter is weak lensing,
using 21~cm emission as background source (see Fig.~\ref{21cmlens}). 
Zahn \& Zaldarriaga (2006) show that
the lensing effect is lower compared to the case in which CMB fluctuations are
used, but it has the advantage of retrieving information from a range of redshifts.
In any case, 21~cm lensing experiments could provide constraints on cosmological
parameters better than those obtained from comparably sized surveys of galaxy 
lensing (Metcalf \& White 2007; Hilbert, Metcalf \& White 2007; Lu \& Pen 2008;
Metcalf \& White 2008).

\subsubsection{Dark Matter Halo Density Profile}

To prove or disprove the existence of a universal density profile for dark
matter halos, accurate and extensive observations are needed. A method largely
applied in the past to the study of the density profiles relies on
HI rotation curves, although the spatial resolution of these observations is
rarely good enough to set meaningful constraints. This is true both for the rotation
curves of Low Surface Brightness Galaxies (LSBGs) and late-type dwarf galaxies
(e.g. van den Bosch et al. 2000; Borriello \& Salucci 2001; de Blok et al. 2001;
van den Bosch \& Swaters 2001; de Blok, Bosma \& McGaugh 2003; Kleyna et al.
2003), 
which are sometimes
consistent with both a constant density core and an inner cusp, indicating
that HI rotation curves do not have enough resolution to discriminate between
the two models. Only in a few cases it has been possible to derive a meaningful
value of $\beta$ ($\beta$ is the power-index of the density profile in the central regions), 
giving $0.55<\beta<1.26$ at a 99.73\% confidence level
for an LSBG, which would be consistent with the NFW slope, and a $\beta<0.5$
for two dwarf galaxies at the same confidence level (van den Bosch et al. 2000),
more consistent with a flat core, as the nearby Ursa Minor dwarf spheroidal
measured by Kleyna et al. (2003) or the sample of LBGs analysed by de Blok, Bosma
\& McGaugh (2003), who find an average $\beta=0.2$. 
H$_\alpha$ rotation curves have a higher spatial resolution
and rise more steeply in the inner parts than the HI rotation curves
(Swaters, Madore \& Trewhella 2000). Also optical rotation curves have been
used, indicating that the NFW profile provides a good fit to 66\% of the 400
galaxies in the sample analyzed by Jimenez, Verde \& Oh (2003), while 68\% of the
galaxies are well fitted by an isothermal profile with a core. Based on optical
and radio rotation curves, Persic, Salucci \& Stel (1996) have confirmed that
spiral galaxies have a universal rotation curve, characterized by one single
free parameter, the (I-band) luminosity: low-luminosity spiral galaxies show rising
rotation curves out to the optical radius, while high-luminosity ones are
flat or even decreasing. These findings are reiterated in Salucci et al. (2007).
On the other hand, recent {\tt HST} observations
have revealed that elliptical galaxies have cusps which continue toward
the center until the resolution limit. Bright elliptical galaxies have a
shallow cuspy core with $0.5<\beta<1$, while faint ones have $\beta \approx 2$
(Merritt \& Fridman 1996).

Alternatively, the density profile of dark matter halos can be investigated
through gravitational lensing experiments.
It has been argued that radially distorted, gravitationally
lensed images of background sources in galaxy clusters, the so-called radial
arcs, require a flat core in the cluster density profile. This would be
consistent with the high resolution map of a cluster in which a very
smooth, symmetric and non-singular core is observed (Tyson, Kochanski \&
Dell'Antonio 1998). Nevertheless, Bartelmann (1996) shows that the NFW
profile can produce radial arcs despite its central singularity, as indicated
by more recent observational data (Oguri, Taruya \& Suto 2001).
Observations of radial arcs though are so scarce that larger samples are needed
for a more significant study.

Finally, as integral measures of weak gravitational lensing by dark matter halos,
like the aperture mass, are sensitive to the density profile, these can be used to
discriminate between an isothermal and a NFW profile. In particular, as
the halo mass range probed by the aperture mass is much wider for a NFW profile,
counts of halos with
significant weak lensing signal are powerful discriminators (Bartelmann,
King \& Schneider 2001).

In summary, {\it available observations are not sufficient
to significantly constrain the profile of dark matter halos}.

\subsubsection{Formation of Protogalaxies}

As primordial gas condenses within dark matter potential wells, forming 
luminous galaxies, it emits copious radiation. 
The energy radiated in this process is comparable to the gravitational binding energy of the baryons.
Most of this cooling radiation is emitted by gas with $T<20,000$~K. 
As a consequence, roughly 50\% 
of it emerges in the Ly$\alpha$ line. As such emission is predominantly emitted
from the outskirts of the collapsing system, it is less likely to 
be extinguished by dust (if present) than the more deeply embedded stellar light. 
This radiation has been advocated (Haiman, Spaans \& Quataert 1999; Fardal et al. 2001) 
to explain the large ($\approx 100$~kpc), luminous ($L\approx 10^{44}$~erg~s$^{-1}$)
``blobs'' of Ly$\alpha$ emission found in narrow-band 
surveys of $z\approx 3$ protoclusters (Steidel et al. 2000) and similar arguments
could be applied to higher redshift objects.  

In addition to Ly$\alpha$ radiation, molecules could also produce a detectable signal. 
Flower \& Pineau des
For\^{e}ts (2001) calculated the spectrum of H$_2$ that is produced by
collisional excitation in the shock waves generated when the speed of
collapse becomes locally supersonic, finding that the rotational transitions
within the vibrational ground state might be observable as inhomogeneities
in the CMB. The H$_2$ cooling emission of forming galaxies has been
studied by Omukai \& Kitayama (2003), who discuss its observability
(in particular the rotational lines 0-0S(3) at 9.7~$\mu$m and 0-0S(1) at
17~$\mu$m)  
through {\tt SAFIR}\footnote{http://safir.jpl.nasa.gov/index.asp}
(Single Aperture Far-Infrared observatory) and by Mizusawa, Omukai \& Nishi
(2005), who find that both rovibrational and pure rotational H$_2$ lines
can be detected by {\tt SPICA}\footnote{http://www.ir.isas.ac.jp/SPICA/}
(SPace Infrared telescope for Cosmology and Astrophysics), if emitted by
metal-free clouds at $z<10$.

\section{First Stars}
\label{firststars}

As described in the previous Section, hierarchical models of structure
formation predict that the first collapsed objects have small masses
and are of primordial composition and thus the formation and cooling
of these objects is mainly governed by the molecular hydrogen chemistry. As
the collapse proceeds, the gas density increases and the first stars
are likely to form. However, the primordial star formation process
and its final products are presently quite unknown. This 
largely depends on our persisting ignorance of the fragmentation
process and on its relationship with the thermodynamical conditions
of the gas. Despite these uncertainties, it is presently accepted that
{\it the first stars, being formed out of a gas of primordial composition,
are metal-free} ({\it Pop~III stars}). 
In addition to this theoretical
justification, the existence of Pop~III stars have been historically invoked for
many different reasons. Among other issues (see e.g. Larson 1998 and Chiosi 2000
for more complete reviews), Pop~III stars could help explaining:
\begin{itemize}
\item the gap between the Big Bang nucleoshynthesis (BBN)
 metal abundances ($Z\approx 10^{-10}-
10^{-12}$) and those observed in the lowest metallicity Pop~II stars
($Z\approx 10^{-4}-10^{-3}$);
\item the G-dwarf problem, i.e. the paucity of metal-poor stars in the solar 
neighborhood relative to predictions from simple models of chemical evolution
(van den Bergh 1962; Norris, Peterson \& Beers 1993; Primas, Molaro \& Castelli 1994; Sneden
et al. 1994);
\item the oxygen anomaly or enhancement of $\alpha$-elements 
in galactic metal-poor stars (Sneden, Lambert \& Whitaker 1979) and 
the existence of extremely metal-poor stars showing s-process elements
in their envelopes (Truran 1980); 
\item the missing mass in clusters of galaxies and galactic halos
(White \& Rees 1978) formed by their dark remnants;
\item the reionization of the universe and the starting engine for the
formation of the first galaxies;
\item the constant lithium abundance of metal-poor halo stars (the Spite
Plateau) which is lower that the predictions from the BBN (Piau et al. 2006; Asplund et al. 2006);
\item the contaminants of the intergalactic medium as inferred from metallic
absorption lines in the Ly$\alpha$ forest seen in quasar light;
\item the cosmological helium abundance (Talbot \& Arnett 1971; Marigo et al. 2003; 
Salvaterra \& Ferrara 2003b);
\item the formation of massive black holes.
\end{itemize}
In the following Section we will discuss the physical processes governing
the formation and evolution of the first stars.

\subsection{Formation Process}

To study the formation of the first stars it is crucial to understand the
physics of the cooling in the early universe, primarily
the number of channels available for the gas to cool, their
efficiency (Rees \& Ostriker 1977) and the physical processes
that set the scales of fragment masses and hence the stellar mass spectrum.
For a recent review see McKee \& Tan (2008).

In general (e.g. Schneider et al. 2002), cooling is efficient 
when $t_{\rm cool} \ll t_{\rm ff}$,
where $t_{\rm cool} = 3 n k T/2 \Lambda(n,T)$ is the cooling time, 
$t_{\rm ff} = (3 \pi/32 G \rho)^{1/2}$ is the free-fall time, 
$n$ ($\rho$) is the gas number (mass)
density and $\Lambda(n,T)$ is the net radiative cooling rate
(in units of erg $\mbox{cm}^{-3} \mbox{s}^{-1}$).  This efficiency criterium
implies that the energy deposited by gravitational contraction cannot
balance the radiative losses; as a consequence, temperature decreases with
increasing density. When the above condition is satisfied, the cloud cools and 
possibly fragments. At any given time, fragments form on a scale, $R_F$, that is small
enough to ensure pressure equilibrium at the corresponding
temperature, i.e. $ R_F \approx \lambda_J
\propto c_{\rm s} t_{\rm ff} \propto n^{\gamma/2 -1}$
where the sound speed $c_{\rm s} = ({\cal R}T/\mu)^{1/2}$, $T \propto
n^{\gamma-1}$, ${\cal R}$ is the universal gas constant
and  $\gamma$ is the polytropic index.
Since $c_{\rm s}$ varies on the cooling timescale,
the corresponding $R_F$ becomes smaller as $T$
decreases. Similarly, the corresponding fragment mass is the Jeans
mass,
\begin{equation}
M_F \propto n R_F^{\eta} \propto n^{\eta\gamma/2
+ (1-\eta)},
\label{eq:Jm}
\end{equation}
with $\eta = 2$ for filaments and $\eta = 3$ for spherical
fragments (Spitzer 1978). This hierarchical fragmentation process
comes to an end when  cooling becomes
inefficient because {\it (i)} the critical density for Local Thermodynamical
Equilibrium (LTE) is reached
or {\it (ii)} the gas becomes optically thick to
cooling radiation. In both cases, 
$t_{\rm cool}$ becomes larger than $t_{\rm ff}$. At this stage, the
temperature cannot decrease any further and it either remains constant
(if energy deposition by gravitational contraction is exactly balanced
by radiative losses) or increases. The necessary condition to stop
fragmentation and start gravitational contraction within each fragment
is that the Jeans mass does not decrease any further, thus favoring
fragmentation into sub-clumps. From eq.~\ref{eq:Jm}, this implies
the condition $\gamma \gsim 2 ({\eta-1})/{\eta}$,
which translates into $\gamma \gsim 4/3$ for a spherical fragment and
$\gamma \gsim 1$ for a filament. Thus, a filament is marginally stable
and contracts quasi-statically when $t_{\rm cool} \approx
t_{\rm ff}$, and the gas becomes isothermal. Finally,
when $t_{\rm cool}\gg t_{\rm ff}$ or the fragments become optically
thick to cooling radiation, the temperature increases as the
contraction proceeds adiabatically.

The very first generation of stars must have formed out of probably
unmagnetized, pure H/He gas, since heavy elements can only be produced
in the interior of stars. These characteristics render the primordial
star formation problem very different from the present-day case, and lead
to a significant simplification of the relevant physics.
Nevertheless,
the complexity and interactions of the hydrodynamical, chemical and radiative
processes, have forced many early studies
to use the steady state shock assumption, a spherical or highly
idealized collapse model. Only recently have more realistic models
been proposed. The first self-consistent three-dimensional
cosmological hydrodynamical simulations were presented by Abel et al. (1998).
Because of the limited spatial resolution, it was not
possible to study the collapse to stellar densities and address the nature of
the first objects.
A higher resolution simulation has been performed by Bromm, Coppi \& Larson (1999).
They investigate, by means of a Smooth Particle Hydrodynamics (SPH) simulation,
the evolution of an isolated 3-$\sigma$ peak of mass $2 \times 10^6$~M$_\odot$
collapsing at $z\approx 30$. They find that the gas dissipatively settles
into a rotationally supported disk, which eventually fragments in high density
clumps ($n \approx 10^8$~cm$^{-3}$) of mass $M_{clump}\approx 10^2-10^3$ ~M$_\odot$.
Mass accretion and merging could subsequently raise the clump masses up to
$\approx 10^4$~M$_\odot$. Changing initial conditions does not affect the disk
formation and fragmentation, although the accretion and merging history
can change (Bromm, Coppi \& Larson 2002).
Bromm \& Loeb (2004) extended the results of the simulations by Bromm and
collaborators down to a scale of $\approx$100~AU, following the
collapse of the cloud to a fully-molecular protostellar core and the subsequent
accretion phase. Although improving on previous simulations, radiative and
mechanical feedback from the protostar on the accretion flow are not treated
self-consistently.

Similar results have been found by Abel, Bryan \& Norman (2000). Their simulation 
is based on a 3-D Adaptive Mesh Refinement (AMR), covering comoving scales from 128~kpc
down to 1~pc and follows
the collapse of primordial molecular clouds and their subsequent fragmentation
within a cosmologically representative volume. The authors find that these
molecular clouds, of mass $M_{cloud}\approx 10^5$~M$_\odot$, will  eventually form a cold
($\approx 200$~K) region at the center of the halo. Within this cold region
a quasi-hydrostatic contracting clump with mass $\approx 200$~M$_\odot$ and
density $n>10^5$~cm$^{-3}$ (and as high as $\approx 10^8$~cm$^{-3}$) is found.
\begin{figure}
\centerline{\includegraphics[width=26pc]{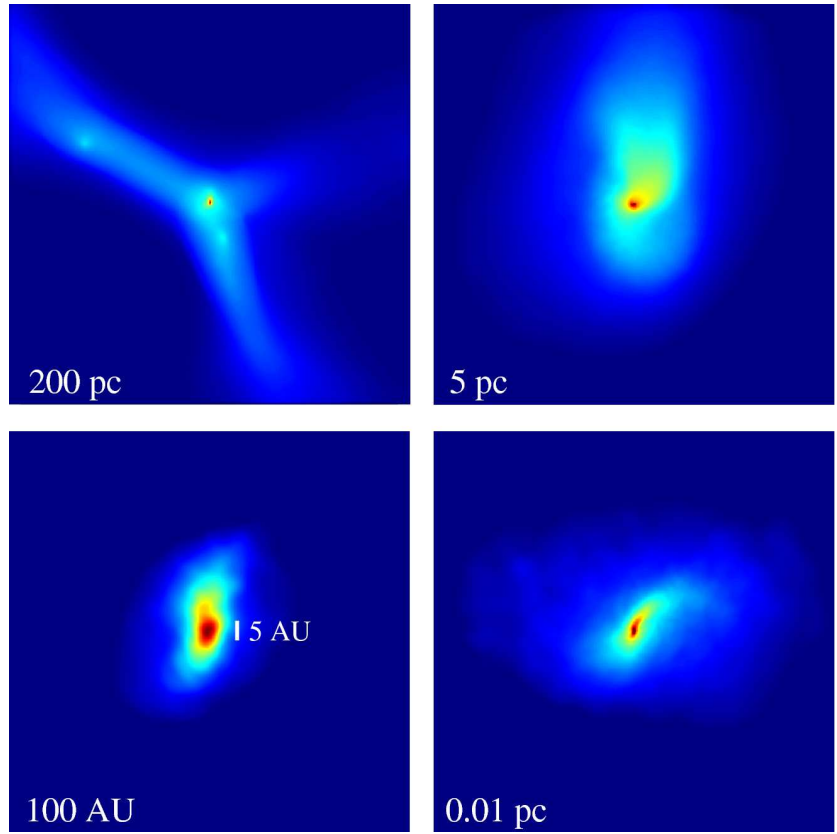}}
\caption{
Projected distribution for a first star cosmological simulation at $z \approx 19$.
The physical side length is indicated in each panel. See Yoshida et al. (2006)
for details.
}
\label{sf_sim}
\end{figure}
The authors speculate that more than
one clump could be found in the same object and that their number is likely
to be proportional to the total amount of cooled gas. Clearly, the ability of these
clumps to form stars is closely related to the strength of feedback processes (see 
Sec.~\ref{feedback}). 
In addition, they suggest that the formation of a disk and its fragmentation,
found by Bromm, Coppi \& Larson (1999, 2002),
is related to their highly idealized initial  conditions
(top-hat spheres that initially rotate as solid bodies
on which smaller density fluctuations are imposed). But, although the process
that leads to the formation of dense clumps in the two simulations is different,
the final result is quite similar, with very dense, massive protostellar
clumps out of which the first star will eventually form. See Fig.~\ref{sf_sim}
for a typical simulation of Pop~III star formation.

It should be noted that the fragmentation and the thermal evolution 
depend strongly on the geometry of the clouds. For filamentary clouds
the gravitational contraction is slower and the compressional heating rate is
lower than for spherical ones. Hence, for the filamentary cloud the 
isothermality breaks down at higher density, leading to a smaller mass of
the final clumps.
Nakamura \& Umemura (2001) have studied the collapse and fragmentation
of a filamentary cloud using a
two-dimensional hydrodynamic simulation, finding that the mass of
the resulting clumps depends on the initial density of the collapsing cloud:
filaments with low initial density ($n<10^5$~cm$^{-3}$)
tend to fragment into dense clumps before
the cloud becomes optically thick to the H$_2$ lines and the H$_2$ cooling with
the three-body reaction is effective. This would result in clumps
much more massive than some tens M$_\odot$. However, relatively dense
filaments result in clumps of a few M$_\odot$ (the result is in practice somewhat 
ambiguous as proper ``initial conditions'' are rarely known). 
The more effective
H$_2$ cooling with the three-body reaction allows the filament to contract
up to $n \approx 10^{12}$~cm$^{-3}$, when it becomes optically thick to H$_2$ lines
and the radial contraction stops.

As the first stars form in regions of the universe corresponding to peaks of the
large-density field, a significant contribution to the local overdensity comes
from large wavelength fluctuations. Thus, simulations that take into account this
effect are needed to properly study the formation of the first stars. Gao et
al. (2007) have run a variety of SPH simulations (from cosmological scales down
to densities at which the gas is no more optically thin to molecular hydrogen lines,
i.e. $\approx 10^{10}$~cm$^{-3}$) to investigate whether there
is a significant scatter in the properties of primordial star-forming clouds
and whether they depend on redshift. They find that the formation path of primordial
stars is similar in all simulations, independently on redshift and environment.
The only difference is that the timescale of the formation is much shorter at
high redshift. Moreover, the star-forming clouds exhibit a variety of morphologies
and different accretion rates. In general though, at the end of the simulation,
none of the clouds have fragmented and stars of more than a few tens of solar
masses can be expected, although nothing more precise can be said on the mass
of the first stars.

It should be noted that SPH simulations that do not resolve the Jeans mass at any
time might suffer from the problem called artificial fragmentation, which, if
not properly addressed, can induce errors in the determination of the mass of
the stars and their Initial Mass Function (IMF). 
A relatively new technique, the particle splitting
(Kitsionas \& Whitworth 2002), has been used by Martel, Evans \& Shapiro (2006)
to investigate the artificial fragmentation of a molecular cloud.

O'Shea \& Norman (2007) have run AMR simulations with the code {\tt ENZO} to study
systematic effects in the formation of primordial protostellar cores. As
previous studies, they find no evidence of fragmentation in any of their
simulations, suggesting that Pop~III stars forming in halos of $\sim 10^5-10^6$
M$_\odot$ form in isolation. The minimum mass of halos hosting Pop~III stars
($\sim 10^5$~M$_\odot$), as well as other bulk halo properties, is independent
on redshift. 

Small mass clumps can be obtained when the collapse of a cloud is driven
by HD or atomic cooling rather than molecular cooling. For example,
Uehara \& Inutsuka (2000) find that, including HD cooling, fragmentation
of primordial gas can produce clumps as small as 0.1~M$_\odot$, but more
typically of few $\times$~M$_\odot$ (Johnson \& Bromm 2006; Yoshida, Omukai 
\& Hernquist 2007). Thus, any
physical process promoting HD formation might induce the formation of
metal-free, small mass stars. Possible locations for such stars are
relic \HII regions (Johnson \& Bromm 2006; Yoshida, Omukai \& Hernquist 2007),
shocked gas during merger of halos (Vasiliev \& Shchekinov 2007),
regions shocked by protostellar jets (Machida et al. 2006), SN explosions
(Johnson \& Bromm 2006) or during structure formation (Vasiliev \& Shchekinov 2005; 
Johnson \& Bromm 2006), in presence of cosmic rays (Jasche, Ciardi \& En{\ss}lin 2007; 
Stacy \& Bromm 2007). Omukai (2001) argues that, in
the absence of molecular hydrogen (which can be easily dissociated by
feedback effects, Sec.~\ref{feedback}), sufficiently massive clouds can start
dynamical collapse by Ly$\alpha$ emission, two-photon emission and H$^-$ 
free-bound emission and continue to collapse
almost isothermally at several thousand K along the ``atomic cooling track'', until
the clump mass is possibly reduced to about 0.03~M$_\odot$.
In conclusion, depending on the cooling mechanism, {\it a clump of $M_{clump}
\approx 10^2-10^3$~M$_\odot$ (H$_2$ cooling) or $M_{clump}\approx 0.1- {\it few} \times 10$~M$_\odot$ (H or HD 
cooling)} forms out of the collapsing molecular cloud. 

\begin{figure}
\centerline{\includegraphics[width=28pc]{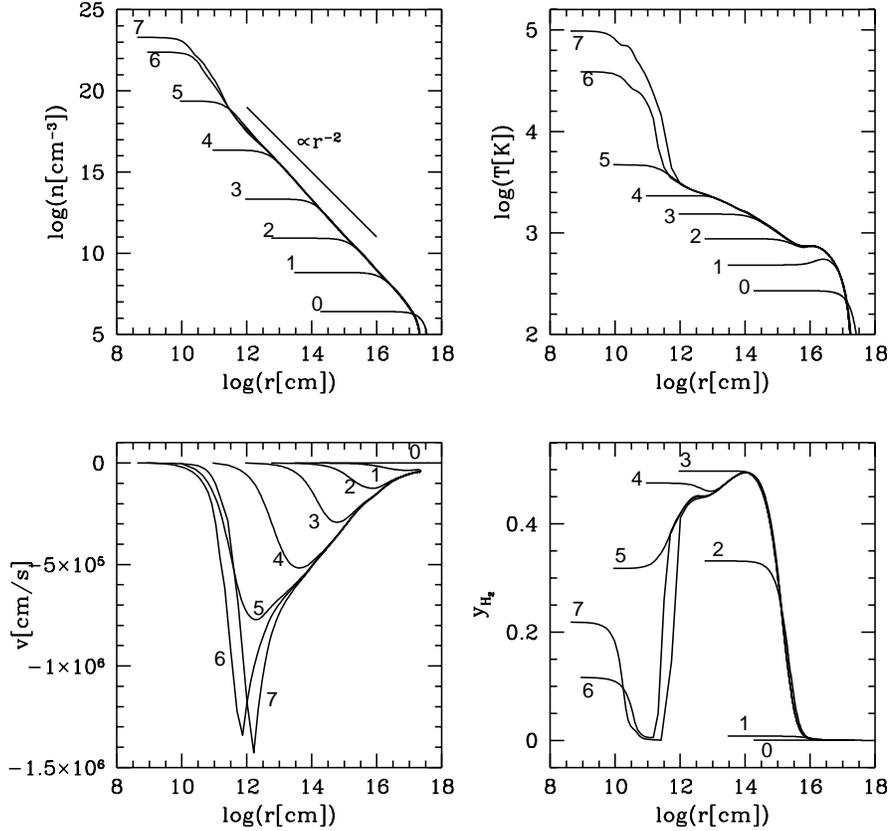}}
\caption{Evolutionary sequence for a contracting protostellar clump. Gas number
density (upper left panel), temperature (upper right), velocity (lower left)
and H$_2$ concentration (lower right) as a function of radial distance.
(0) Initial conditions. (1) After $5.7 \times 10^5$~yr three-body processes are
active in the central region and temperature inversion occurs: (2) $8.5 \times
10^3$~yr after (1). The cloud becomes optically thick to some lines: (3) 
$2.8 \times 10^2$~yr after (2). The central region becomes fully molecular: (4)
12~yr after (3). The central region becomes optically thick to H$_2$ CIA continuum:
(5) 0.32~yr after (4). In the midst of dissociation: (6) $2.4 \times 10^{-2}$~yr
after (5). Shortly after the core formation: (7) $4.1 \times 10^{-2}$~yr after (6).
Final state. See Omukai \& Nishi (1998) for details.}
\label{protos}
\end{figure}
Once a clump has reached densities of the order of $n \approx 10^8$~cm$^{-3}$, the
three-body formation of molecular hydrogen becomes
dominant, the assumption of optically thin cooling begins to break down
and radiative transfer effects become important. Tracking the subsequent
evolution of these clumps is a very challenging problem, as it requires the
simultaneous solution of the hydrodynamic equations and of line radiative
transfer. 
In their pioneering work, Omukai \& Nishi (1998) studied
the contraction of protostellar clumps into stars, for various masses and
initial conditions, by means of 1-D hydrodynamical calculations (see 
Fig.~\ref{protos} for a summary of their results).
Coincidentally, some of their initial conditions are very close to the
final state of Abel, Bryan \& Norman (2000). The authors find that
the evolution of a gravitationally unstable clump
proceeds in a highly non-homologous fashion, with the central parts
collapsing first. This runaway phase is induced by
H$_2$ line radiation cooling up to densities of $n \approx 10^{14}$~cm$^{-3}$, 
and by H$_2$
collision-induced emission at higher densities. The resulting gas
temperature is nearly constant at several $10^2$~K and the innermost
region of $\approx 1$~M$_\odot$ becomes fully molecular due to the three-body
reaction. At densities $n \approx 10^{16}$~cm$^{-3}$ the clump becomes
optically thick to collision-induced absorption and H$_2$ dissociation
works as an effective cooling agent. Finally, lacking further cooling
mechanisms, at $n \approx 10^{22}$~cm$^{-3}$, {\it a small hydrostatic core of
mass $\approx 10^{-3}$~M$_\odot$ is formed. The core mass is highly independent
of initial conditions}. 

\begin{figure}
\centerline{\includegraphics[width=28pc]{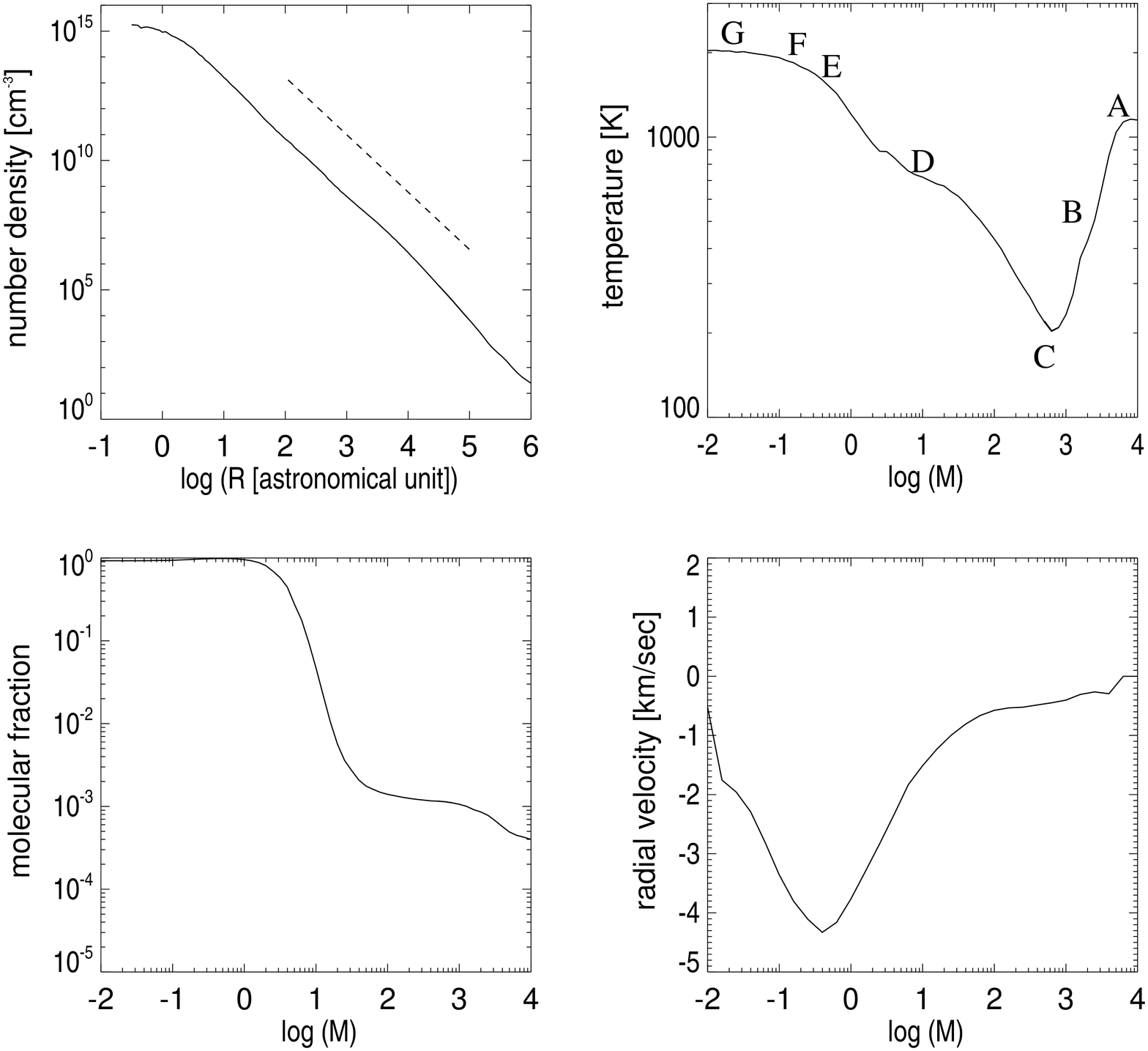}}
\caption{
Radial profiles for density, temperature, molecular fraction, and infall velocity of a protostellar clump
of primordial composition collapsing at redshift $z\approx 19$.
The density is plotted as a function of distance from the center, whereas the other three quantities are plotted
as a function of enclosed gas mass. The density profile is close to the power law $\propto R^{-2.2}$.
The characteristic features in the temperature profile (marked AG) are described in Yoshida et al. (2006).
}
\label{yoshida06}
\end{figure}
The same results have been confirmed by other
authors (e.g. Ripamonti et al. 2002; Machida et al. 2006). 
Among them, Yoshida et al. (2006) and Yoshida, Omukai \& Hernquist (2007)
pushed 3-D simulations up to number densities of $\approx 10^{18}~$cm$^{-3}$,
by improving in both physics implementation (a treatment of the regime in which
the gas is optically thick to H$_2$ lines which allows to reach such densities)
and resolution (which allows to reach a particle mass of
60 times the Earth mass and scale of about 1~AU).
The run-away collapse of the internal core is followed
down to the scales of the central 1~M$_\odot$ fully molecular core and the 0.01~M$_\odot$
which is completely opaque to H$_2$ lines. No indication of fragmentation is found.
Their results are quantitatively summarized in Fig.~\ref{yoshida06} in terms of the
gas density, H$_2$ fraction, temperature and velocity profiles, which can be readily
compared with earlier semi-analytical results obtained by Omukai \& Nishi (1998)
and discussed in Fig.~\ref{protos}.
In spite of some minor discrepancies, the degree of agreement between the two widely different
methods is outstanding.

As the core is surrounded by a large amount of reservoir gas 
(corresponding to the mass of the initial clump), in
the absence of any effect quenching accretion, the protostar
can grow by several orders of magnitude in mass by accreting the envelope
matter. It has yet to
be firmly established whether standard mechanisms proposed to halt the
infall continue to work under primordial conditions: the radiation force
could be opacity limited; bipolar flows need some magnetohydrodynamic
acceleration process, and therefore seem to be excluded by the weak
primordial magnetic field (e.g. Gnedin, Ferrara \& Zweibel 2000; Langer,
Puget \& Aghanim 2003; Banerjee \& Jedamzik 2005; Langer, Aghanim \& Puget 2005).

The mass accretion rate onto the protostar can be written as $\dot{M}_{acc} 
\sim c_s^3/G$ (Stahler, Shu \& Taam 1980), where $c_s$ is the isothermal
sound speed of the protostellar clump. Thus, the mass accretion rate is    
higher for protostars formed by atomic cooling because of the higher 
temperature of the protostellar clump, although the mass available for
accretion is smaller. The evolution of zero-metal protostars
in the main accretion phase has been recently studied by Omukai \& Palla (2001)
under a constant mass accretion rate of $4.4 \times 10^{-3}$~M$_\odot$~yr$^{-1}$.
The high accretion rate used, combined with a low opacity of the 
infalling gas due to the lack of dust grains, implies a milder 
effect of radiation pressure by protostellar photons and a
higher inflow momentum. 
In fact, fast accreting protostars enter a phase of rapid expansion at
a mass of $\approx 300$~M$_\odot$ when the luminosity becomes close to the
Eddington limit. This event may determine the onset of a powerful stellar
wind driven by radiation pressure that can effectively quench further accretion.
The effect of a time-dependent accretion rate, initially as high as 
$\approx 10^{-2}$~M$_\odot$~yr$^{-1}$ and rapidly decreasing once the stellar mass exceeds
$M_\star \approx 90$~M$_\odot$, has been studied by Omukai \& Palla (2003). 
They find that if $\dot{M}_{acc}<\dot{M}_{crit} \approx 4 \times 10^{-3}$
M$_\odot$~yr$^{-1}$, stars with mass $\gg 100$~M$_\odot$ can form, provided there is
sufficient matter in the parent clumps. For $\dot{M}_{acc}>\dot{M}_{crit}$ instead,
the maximum mass limit decreases with $\dot{M}_{acc}$. 
In the context of cosmological simulations, a large scatter in the mass 
accretion rates ($10^{-4}- 10^{-2}$~M$_\odot$/yr) is found, with a general trend of
lower accretion rates at higher redshift (O'Shea \& Norman 2007). 
Although this points toward {\it a minimum
mass of Pop~III stars $\gg 1$~M$_\odot$, the upper limit remains uncertain.}
Similar numbers ($10^{-4}- 10^{-1}$~M$_\odot$/yr) are obtained by Yoshida et al.
(2006), who also estimate a final stellar mass of 60-100~M$_\odot$.

An ingredient which has so far received little attention but which could be quite
important to determine the final mass of the first stars, is rotation of the infalling
envelope. If protostellar clumps are rotating above a critical value of the rotational
energy, which is found to be about $10^{-5}$ times the gravitational energy (Machida et al. 2007), fragmentation
of the clump might occur. Under the conditions prevailing at the time of first stars formation,
such value is not implausible and as a results a non-neglible fraction of this ancient stellar
population might be constituted by binary or multiple systems. In addition,   
as pointed out by Tan \& McKee (2004), rotation might also lead to dramatic differences 
in the way the \HII region breaks out of the infalling envelope and in radiation pressure effects.  
Radiation pressure, in particular,  appears to be the dominant mechanism for 
suppressing infall, becoming dynamically important around 20~M$_\odot$. 

\begin{figure}
\centerline{\includegraphics[width=30pc]{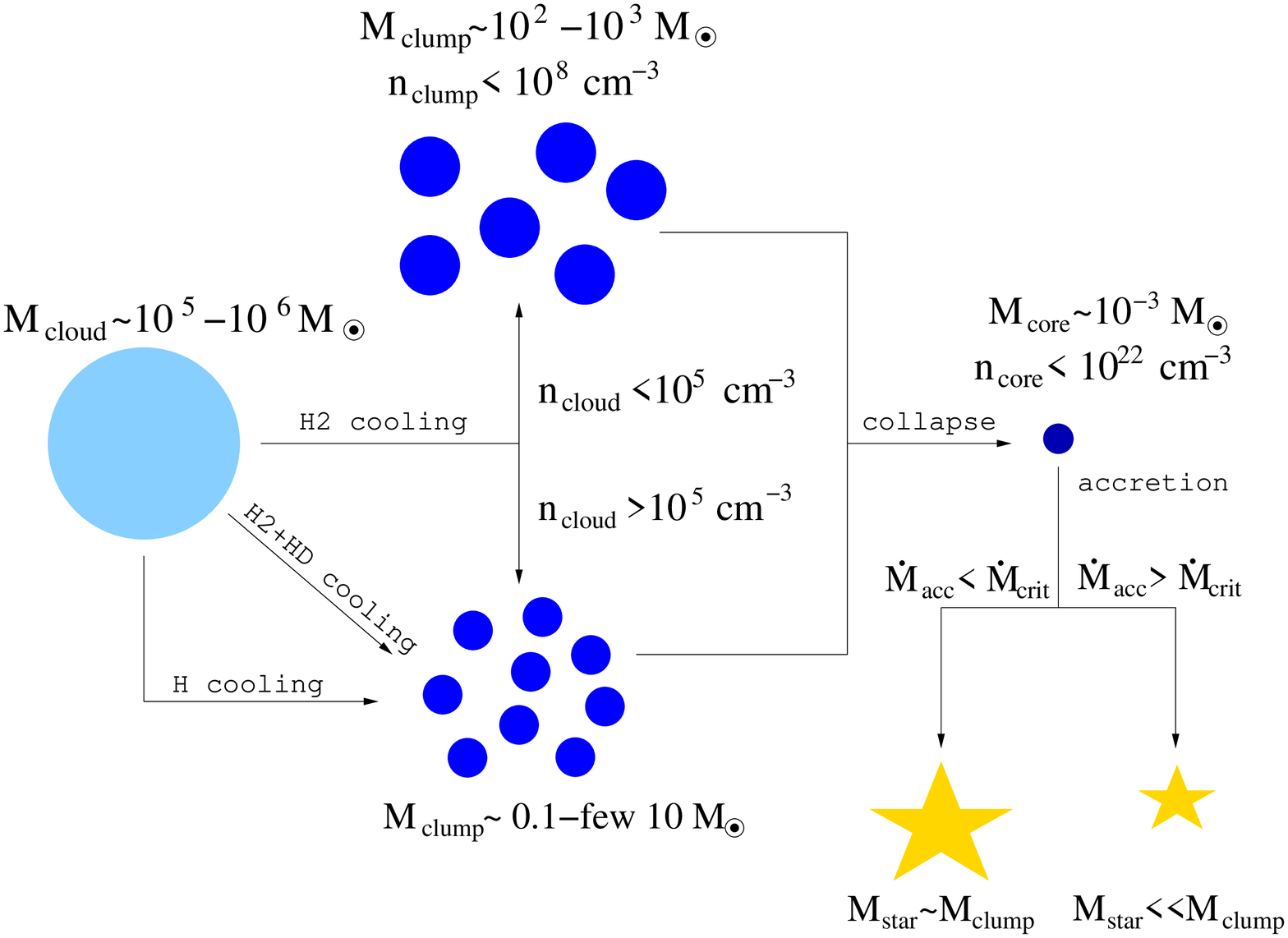}}
\caption{Summary of the relevant physical processes and possible evolution paths for Pop~III
star formation.}
\label{sf_scheme}
\end{figure}
As previously mentioned, primordial magnetic fields seem to be extremely weak. But
during the collapse magnetic pressure increases more rapidly than gas pressure. 
For this reason, although the initial magnetic field is small, it can be
amplified by 10 orders of magnitude and thus affect the formation of a star.
Machida et al. (2006) use 3-D magneto-hydrodynamical (MHD) simulations to
follow the collapse of a slowly rotating spherical cloud of $5 \times 10^4$~M$_\odot$
with central number density $n_c$,
permeated by a uniform magnetic field. They find that if the initial field is
$> 10^{-9} (n_c/10^3 {\rm cm}^{-3})^{2/3}$~G, a protostellar jet forms blowing
away 3-10\% of the accreting matter and reducing the final mass of the star. 
Studies in this area are still 
in their infancy and definite conclusions have to await for more detailed calculations.
To summarize, {\it if $\dot{M}_{acc}<\dot{M}_{crit}$ protostellar cores can accrete
all the matter of the parent clumps and possibly become very massive stars of
$M_\star\sim M_{clump}$; otherwise accretion is less efficient and $M_\star \ll
M_{clump}$}.
In Fig.~\ref{sf_scheme} a summary of the relevant physical processes and possible
paths for Pop~III star formation is sketched.
A possible classification scheme of Pop III and Pop II stars has been recently 
agreed upon at the First Stars III Conference held in Santa Fe
(O'Shea et al. 2008). A slightly modified version of such classification is reported
in Tab. \ref{star_class}. Here III, II.5 and II indicate a metallicity sequence,
while (differently from the original classification), {\it a} and {\it b} indicate
a temporal sequence, i.e. first and second generation of Pop~III stars.

\begin{table*}
\caption[]{Classification of stars. The columns indicate: the name 
of the stellar type, their typical metallicity $Z$ and mass M$_\star$,
their formation mechanism or site.}
\label{star_class}
\begin{tabular}{llll}
\hline
Name & $Z$ & M$_\star$ [M$_\odot$] & Formation \\
\hline
Pop IIIa             &  0                & $\simgt 100$  & minihalos, H$_2$ cooling \\
Pop IIIb             &  0                & $\simgt 10 $  & HD, H cooling            \\
Pop II.5             & $\simlt Z_{crit}$ & $\simgt 10 $  & pre-enriched SF sites     \\ 
Pop~II               & $> Z_{crit}$      & local IMF     & metal enriched SF sites  \\
\hline
\end{tabular}
\end{table*}

\subsubsection{Initial Mass Function}

As we have seen above, recent studies seem to indicate that
the first stars were massive. Nevertheless, as the primordial fragmentation
process is very poorly understood, the distribution of masses with which stars are
formed, the so-called IMF, is still very uncertain.
The determination of the IMF is also of basic importance because, although the
luminosities of galaxies depend
primarily on stars of masses $\approx 1$~M$_\odot$, metal enrichment 
and feedback effects on galactic scales   
depend on the number of stars with masses above $\approx 10$~M$_\odot$.

For many years,
beginning with Schwarzschild \& Spitzer (1953), there have been speculations
in the literature that the IMF was dominated by massive stars at early times.
More recently, Larson (1998) has proposed a modified Salpeter IMF for the first stars.
If we define $N$ as the number of stars formed per logarithmic mass interval,
the standard parameterization of the Larson IMF reads:
\begin{equation}
dN/d{\rm log}M_\star \propto (1+M_\star/M_c)^{-1.35},
\end{equation}
where $M_c$ is a characteristic mass scale, related to the Jeans mass,
and more in general to the scale at which there is a transition from  a 
chaotic regime dominated by non-thermal pressure on larger scales to  a 
regular regime dominated by thermal pressure on smaller scales.
As the Jeans mass depends on temperature and pressure
of a star-forming cloud as $T^2 p^{-1/2}$, a variable   
mass scale $M_c$ naturally arises with a higher value at earlier times.
The essential effect of this type of variability would be to
alter the relative number of low-mass to large-mass stars formed, with an
{\it IMF biased toward massive stars at high redshift}.

If indeed the mass of the clumps originating the protostars depends on
the density of the collapsing cloud, as discussed earlier, this could
produce a bimodal IMF with peaks of $\approx 10^2$ and 1 M$_\odot$ (Nakamura
\& Umemura 2001).
A similar IMF would arise    in a scenario in which the first very massive
stars create intense radiation fields which deplete the H$_2$ content
inside their parent clouds, decreasing, as mentioned earlier, the 
subsequent fragmentation mass scale. The resulting IMF would be bimodal,
with a high-mass peak at $\approx 40$~M$_\odot$ and a low-mass peak at 
$\approx 0.3$~M$_\odot$ (Omukai \& Yoshii 2003). The
high-mass portion of the IMF is a very steep function of mass, with a
power-law index of -4; thus, although some of the  first stars are 
very massive, their typical mass scale is smaller.
A Larson IMF with an upper cut off of $\sim 100$~M$_\odot$ seems to
be favored also by theoretical models of the Pop~III to Pop~II transition
(see below) which include constraints from the {\tt NICMOS UDFs} field (Bouwens
et al. 2005) and the {\tt WMAP} satellite (Spergel et al. 2003).

On the contrary, it has been known since the work of Salpeter (1955)
that the present-day IMF of
stars in the solar neighborhood can be approximated by a declining 
power-law for masses above 1~M$_\odot$,   
flattening below $0.5$~M$_\odot$, and possibly even
declining below $0.25$~M$_\odot$ (Scalo 1986, 1998; Kroupa 2001). While the
behavior of the IMF at the lowest masses remains uncertain because of the poorly known
mass-luminosity relation for the faintest stars, above $\approx 1$~M$_\odot$
there is a consensus on the universality of the power-law part of the IMF,
with $dN/d{\rm log}M_\star \propto M_\star^{-1.35}$ 
(von Hippel et al. 1996; Hunter et al.  1997; Massey 1998).

\begin{figure*}
\centerline{\includegraphics[width=28pc]{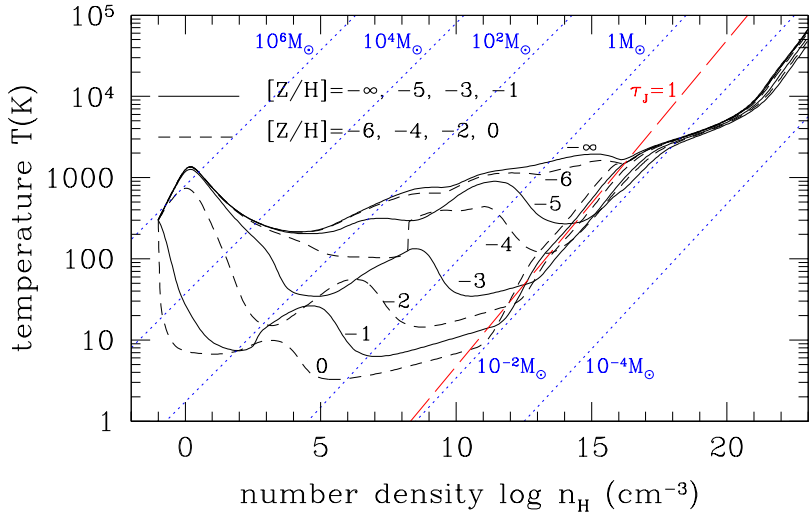}}
\caption{Temperature evolution of prestellar clouds
with different metallicities. Those with metallicities [Z/H]=-$\infty$ (Z=0),
-5, -3, and -1 (-6, -4, -2, and 0) are shown by solid (dashed) lines.
Only the present-day CMB is considered as an external radiation field.
The lines for constant Jeans mass are indicated by thin dotted lines.
The positions where the central part of the clouds becomes optically thick
to continuum self-absorption is indicated by the thin solid line.
The intersection of the thin solid line with each evolutionary trajectory
corresponds to the epoch when the cloud becomes optically thick
to the continuum.
To the right of this line, the clouds are optically thick and there is little
radiative cooling. See Omukai et al. (2005) for details.}
\label{omukai05}
\end{figure*}
Following the early study by Yoshii \& Sabano (1980), it has been proposed 
(Bromm et al. 2001; Schneider et al. 2002; Bromm \& Loeb 3003; Santoro \& Shull 2006;
Smith \& Sigurdsson 2007) that {\it the mechanism
inducing a transition from a top-heavy to a more conventional IMF is
metal enrichment}. In fact, it is only when metals change the
composition of the gas that further fragmentation occurs producing
stars with significantly lower masses. 
Bromm et al. (2001) have numerically studied the effect of metals on the
evolution of the gas in a collapsing dark matter minihalo, using an
SPH simulation, for two different values of metallicity.
They have shown that fragmentation of the gas cloud proceeds very
differently in the two cases and concluded that exists a critical
metallicity, $Z_{\rm cr}$, which marks a transition from a high-mass to a low-mass
fragmentation mode. The value of $Z_{\rm cr}$ is estimated to be
$10^{-4}-10^{-3}\;Z_\odot$. These values should be taken as a first order
approximation since the model does not include some key cooling agents, such as H$_2$,
other molecules and dust grains.
Santoro \& Shull (2006), though, found similar values including cooling from CII,
OI, SiII and FeII lines and from H$_2$. The same values are found by 
Smith \& Sigurdsson (2007) using the code {\tt ENZO} to follow the formation and fragmentation
of clouds starting from cosmological initial conditions.
Schneider et al. (2002) though have investigated the problem
including a much larger number  ($> 500$) of chemical reactions. Using the
1-D code described in Omukai (2000), they have studied the fragmentation properties
of a gravitationally unstable gas cloud for different initial metallicities in the 
range $0-1$ $Z_\odot$.
Including the effect of H$_2$ and molecular cooling as well as the presence of dust grains
in the thermal evolution, they concluded that the transition between the two fragmentation modes
takes place for metallicities in the range $10^{-6}-10^{-4}\;Z_\odot$. 
This critical threshold is almost two orders of magnitude lower than that 
found by Bromm and collaborators
because the fragmentation properties of a gas cloud with $Z=Z_{\rm cr}$
depend on the fraction of metals depleted onto dust grains rather than 
on the metals present in the gas phase (Schneider et al. 2003). A more
recent study by Omukai et al. (2005), which includes also the chemistry of D
previously neglected, confirms the above results (see Fig.~\ref{omukai05}). The equation of state provided by
the authors at different metallicities has been used in SPH simulations of primordial
star formation by Tsuribe \& Omukai (2006) and Clark, Glover \& Klessen (2008). Both
simulations confirm the results of the 1-D calculations (see also Tsuribe \& Omukai 2008)
but show also that, if substantial rotation or elongation of the collapsing
core is present, fragmentation can happen at metallicities $Z>10^{-6} \; Z_\odot$
(Tsuribe \& Omukai 2006) or even lower.

Finally, Omukai \& Palla (2003) have considered the effect of heavy elements with abundances in
the range $Z=5 \times 10^{-5}-5 \times 10^{-3}$~$Z_\odot$ on the accretion
process. The main evolutionary features of protostars are similar to those
of metal-free objects, except that the value of $\dot{M}_{crit}$ increases
for metal-enriched protostars. Since the accretion rate is lower in a slightly
polluted environment, the condition $\dot{M}_{acc}<\dot{M}_{crit}$ is expected
to be more easily met and the formation of massive stars is favored, provided
the clump mass is large enough.

\subsection{Emission Spectrum}

\begin{figure*}
\centerline{\includegraphics[width=30pc]{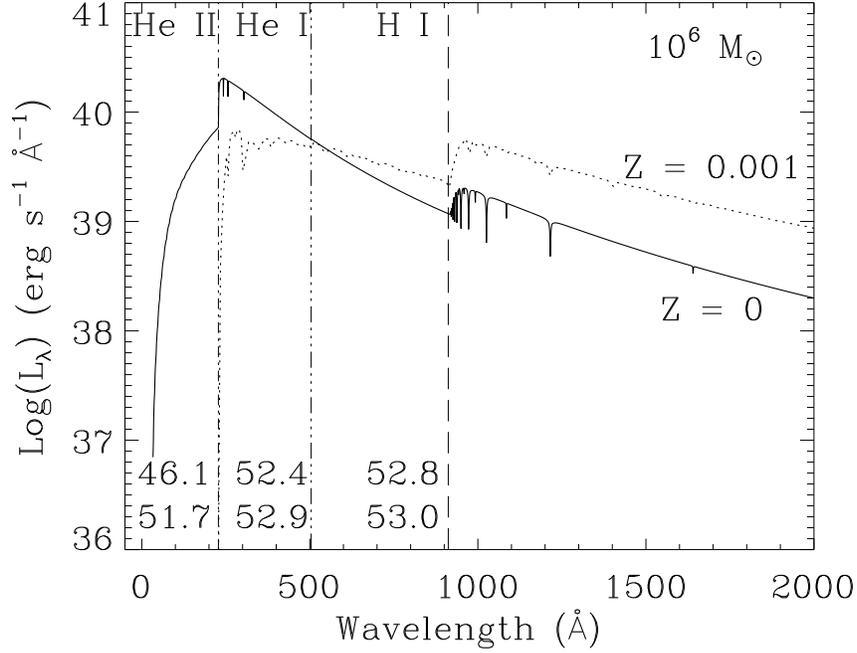}}
\caption{Synthetic spectra of Pop~II and Pop~III clusters of 10$^6$~M$_\odot$
with a Salpeter IMF. The numbers in the lower left corner, near each continuum
mark, represent the rate (in units of photons s$^{-1}$)
of ionizing photons production (log) for that continuum
(Tumlinson \& Shull 2000).}
\label{popstars}
\end{figure*}

The metal-free composition of {\it the first stars}
 restricts the stellar energy source
to proton-proton burning rather than to the more efficient CNO cycle. Consequently,
they {\it are hotter and have harder spectra then their present-day counterparts
of finite metallicity}. These unique physical characteristics enhance the ionizing
photon production of Pop~III stars, particularly in the He~$\scriptstyle\rm II\ $ 
continuum, in which they produce up to $\approx 10^5$ times more photons than
Pop~II (see Fig.~\ref{popstars} for a comparison of  Pop~II and 
Pop~III emission spectra). If, in addition, the first stars were also massive,
they would be even hotter and have harder spectra. Bromm, Kudritzki \& Loeb (2001) 
find that metal-free stars with mass above 300~M$_\odot$ resemble a blackbody
with an effective temperature of $\approx 10^5$~K, with a production rate of ionizing
radiation per stellar mass larger by $\approx 1$ order of magnitude for H and 
He~$\scriptstyle\rm I\ $ and by $\approx 2$ orders of magnitude for He~$\scriptstyle\rm
II\ $ than the emission from Pop~II stars. 
In the less extreme case of metal-free stars with masses $<100$~M$_\odot$,
Tumlinson, Shull \& Venkatesan (2003) predict that the H-ionizing photon production
takes twice as long as that of Pop~II to decline to 1/10 of its peak value.
Nevertheless, due to the redward stellar evolution and short lifetimes of the
most massive stars, the hardness of the ionizing spectrum decreases rapidly,
leading to the disappearance of the characteristic He~$\scriptstyle\rm II\ $
recombination lines after $\approx 3$~Myr in instantaneous bursts (Schaerer 2002, 2003).

Models of emission spectra from primordial stars have been calculated by
different authors for stars with masses up to $\approx 50$~M$_\odot$ (Brocato et al. 2000
and Cassisi, Castellani \& Tornamb\`e 1996 in the implementation of Ciardi et al.
2001), $\approx 100$~M$_\odot$ (Tumlinson \& Shull 2000; Tumlinson, Shull \&
Venkatesan 2003) and $\approx 1000$~M$_\odot$
(Bromm, Kudritzki \& Loeb 2001; Marigo et al. 2003). 
In particular,
Marigo et al. (2003) have carried out extensive evolutionary calculations over
a large range of stellar masses ($0.7$~M$_\odot \le M_\star \le 1000$~M$_\odot$),
covering the H- and He-burning phases and allowing for a moderate overshooting
from convective cores, but neglecting rotation. Additionally, for
very massive stars ($M_\star>120$~M$_\odot$), they apply recent mass-loss rate
prescriptions for the radiation-driven winds at very low metallicities
and the amplification of the loss rate caused by stellar rotation.
Nevertheless, the above studies are based on some strong simplifying assumptions;
e.g. all the stars are assumed to be on the Zero Age Main Sequence
(ZAMS) (i.e. stellar evolution is
neglected) and nebular continuum emission is not included. In particular,
this last process cannot be neglected for metal-poor stars with 
strong ionizing fluxes, as it increases significantly the total continuum 
flux at wavelengths redward
of Ly$\alpha$ and leads in turn to reduced emission line equivalent widths
(Schaerer 2002, 2003).  Nebular emission has been included in a more 
complete and extended study by Schaerer (2002, 2003), who presents realistic 
models for massive Pop~III stars
and stellar populations based on non-LTE model atmospheres, recent stellar
evolution tracks and up-to-date evolutionary synthesis models, including
also different IMFs. For more recent tabulated emission rates refer also to
Heger \& Woosley (2008).
In Table~\ref{zams_new} we have summarized the emission properties of Pop~III
stars. The numbers have been derived integrating the ionizing photon rate 
in the absence of stellar winds given in Schaerer (2002) over
different IMFs, i.e. Salpeter, Larson and Gaussian. 

\begin{table*}
\caption[]{ZAMS Properties of Pop~III stars. The columns show: 
the IMF type; the IMF mass interval [M$_\odot$]; the characteristic (central) 
mass scale for the Larson  (Gaussian) IMF [M$_\odot$]; rms deviation for 
the Gaussian IMF [M$_\odot$]; logarithm of H~$\scriptstyle\rm I\ $--, 
He~$\scriptstyle\rm I\ $--, He~$\scriptstyle\rm II\ $--ionizing and H$_2$--dissociating
photons per baryon in Pop~III stars.}
\label{zams_new}
\begin{tabular}{llllllll}
\hline
IMF & $\Delta$M & M$_c$ & $\sigma_c$ & H~$\scriptstyle\rm I\ $ & 
He~$\scriptstyle\rm I\ $ & He~$\scriptstyle\rm II\ $ & H$_2$ \\
\hline
Salpeter &  1-100 &     &    & 4.235 & 3.919 & 2.010 & 4.374 \\
Salpeter &  1-500 &     &    & 4.355 & 4.070 & 2.750 & 4.469 \\
Salpeter & 10-100 &     &    & 4.705 & 4.410 & 2.518 & 4.799 \\
Salpeter & 10-500 &     &    & 4.753 & 4.483 & 3.175 & 4.833 \\
Salpeter & 50-500 &     &    & 4.869 & 4.632 & 3.540 & 4.925 \\
Larson   &  1-100 &   5 &    & 4.505 & 4.201 & 2.342 & 4.622 \\
Larson   &  1-100 &  10 &    & 4.578 & 4.280 & 2.454 & 4.686 \\
Larson   &  1-100 &  50 &    & 4.703 & 4.420 & 2.698 & 4.791 \\
Larson   &  1-500 &   5 &    & 4.604 & 4.331 & 3.056 & 4.699 \\
Larson   &  1-500 &  10 &    & 4.670 & 4.404 & 3.160 & 4.757 \\
Larson   &  1-500 &  50 &    & 4.782 & 4.533 & 3.402 & 4.851 \\
Larson   &  1-500 & 100 &    & 4.808 & 4.567 & 3.490 & 4.872 \\
Larson   & 10-100 &  50 &    & 4.778 & 4.500 & 2.781 & 4.856 \\
Larson   & 10-500 &  50 &    & 4.822 & 4.576 & 3.447 & 4.887 \\
Larson   & 10-500 & 100 &    & 4.834 & 4.594 & 3.518 & 4.895 \\
Larson   & 50-500 & 100 &    & 4.865 & 4.636 & 3.629 & 4.920 \\
Gaussian &  1-500 &  50 &  5 & 4.835 & 4.568 & 2.659 & 4.902 \\
Gaussian &  1-500 &  50 & 10 & 4.832 & 4.564 & 2.697 & 4.900 \\
Gaussian &  1-500 & 100 & 10 & 4.887 & 4.634 & 3.370 & 4.949 \\
Gaussian &  1-500 & 100 & 20 & 4.885 & 4.632 & 3.354 & 4.947 \\
Gaussian & 10-500 &  50 &  5 & 4.835 & 4.568 & 2.659 & 4.902 \\
Gaussian & 10-500 &  50 & 10 & 4.832 & 4.564 & 2.697 & 4.900 \\
Gaussian & 10-500 & 100 & 10 & 4.887 & 4.634 & 3.370 & 4.949 \\
Gaussian & 10-500 & 100 & 20 & 4.885 & 4.632 & 3.354 & 4.947 \\
Gaussian & 50-500 & 100 & 10 & 4.887 & 4.634 & 3.370 & 4.949 \\
Gaussian & 50-500 & 100 & 20 & 4.885 & 4.632 & 3.357 & 4.947 \\
\hline
\end{tabular}
\end{table*}

\subsection{Final Fate}

As the pp-chain is never sufficiently efficient to power
massive stars, stars of initial zero metallicity contract until central
temperatures $\ge 10^8$~K are reached and CNO seed isotopes are produced by
the triple-$\alpha$ process. As a consequence of their
peculiar behavior in hydrogen burning, the post-main sequence entropy
structure of stars of zero initial metallicity is different from that of stars
having a metallicity above $10^{-5} Z_\odot$. A major uncertainty in the evolution
of these stars is mass loss. Although
radiative mass loss is probably negligible for stars of such low metallicity,
they still might lose an appreciable fraction of their mass because of
nuclear-driven pulsations.
Smith \& Owocki (2006) started from the observational evidence that nebulae
around luminous blue variables (LBV) and LBV candidates (e.g. $\eta$ Car)
show high ejecta masses, to discuss the possibility that the mass loss during
the evolution of very massive stars may be dominated by optically thick,
continuum-driven outbursts or explosions, rather than by steady line-driven
winds. Unlike the latter, the former may be independent of metallicity and
play an important role in the evolution of Pop~III stars. In any case,
recent theoretical analysis on the evolution of metal-free stars predict that
their fate can be classified as follows (see also the comprehensive paper
by Heger \& Woosley 2008):

\begin{itemize}
\item Stars  with masses 10~M$_\odot \simlt M_\star \simlt $40~M$_\odot$
proceed through the entire  series of nuclear burnings accompanied by strong
neutrino cooling: hydrogen to helium, helium to carbon and oxygen, then carbon,
neon, oxygen and silicon burning, until finally iron is produced. When the star
has built up a large enough iron core, exceeding its Chandrasekhar mass, it
collapses, followed by a supernova explosion (Woosley \& Weaver 1995). In 
particular, stars with $M_\star \simgt 30$~M$_\odot$ would eventually collapse 
into a Black Hole (BH) (Woosley \& Weaver 1995; Fryer 1999).
\item For stars of 40~M$_\odot \simlt M_\star \simlt $100~M$_\odot$ the neutrino-driven
explosion is probably too weak to form an outgoing shock. A BH forms and
either swallows the whole star or, if there is adequate angular momentum,
produces a jet which could result in a Gamma Ray Burst (GRB; Fryer 1999).
\item Stars with $M_\star \simgt 100$~M$_\odot$ (see e.g. Portinari, Chiosi \& Bressan 1998)
form large He cores that reach
carbon ignition with masses in excess of about 45~M$_\odot$. It is known that
after helium burning, cores of this mass will encounter the electron-positron
pair instability, collapse and ignite oxygen and silicon burning explosively.
If explosive oxygen burning provides enough energy, it can reverse the collapse in
a giant nuclear-powered explosion (the so-called Pair Instability Supernova, PISN)
by which the star would be partly or
completely (if $M_\star \simgt 140$~M$_\odot$) disrupted (Fryer, Woosley \& Heger 2001).
The onset of a PISN though depends whether the star can retain enough of its initial
mass to have a helium core bigger than about 64~M$_\odot$ at the end of the core
He-burning phase, or rather lose the mass during the previous stages of its evolution
(Meynet, Ekstr\"om \& Maeder 2006).
For even more massive stars
($M_\star \simgt 260$~M$_\odot$) a new phenomenon occurs as a sufficiently large fraction
of the center of the star becomes so hot that the photodisintegration instability
is encountered before explosive burning reverses the implosion. This uses up all
the energy released by previous burning stages and accelerates the collapse leading 
to the prompt formation of
a massive BH, and, again, either a complete collapse or a jet-powered explosion
(Bond, Arnett and Carr 1984; Fryer, Woosley \& Heger 2001; Ohkubo et al. 2006;
Suwa et al. 2007a).
\item At even higher masses ($M_\star \simgt 10^5$~M$_\odot$) the evolution depends on
the metallicity (Fuller, Woosley \& Weaver 1986): if $Z < 0.005$ the star
collapses to a BH as a result of post-Newtonian instabilities without
ignition of the hydrogen burning; for higher metallicities it explodes, as it could
generate nuclear energy more rapidly from $\beta$-limited cycle.
\end{itemize}

\begin{figure}
\centerline{\includegraphics[angle=270,width=26pc]{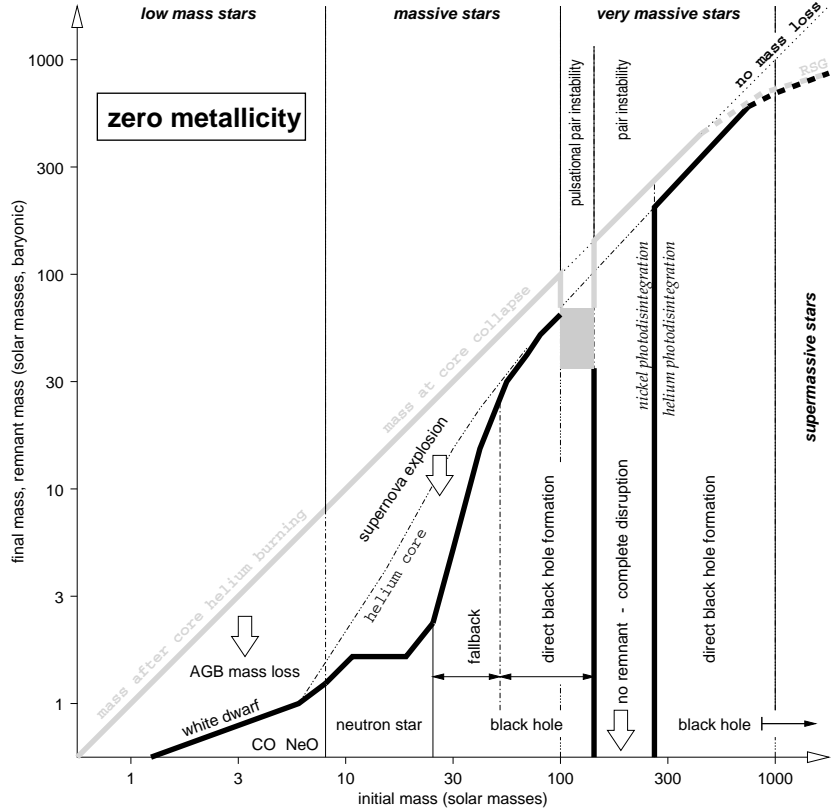}}
\caption{Initial-final mass function of Pop~III stars (Heger \& Woosley 2002). The
horizontal axis gives the initial stellar mass. The $y$-axis gives the final mass of the
collapsed remnant (thick black curve) and the mass of the star at the beginning of the event
producing that remnant (thick gray curve). Since no mass loss is
expected from Pop~III stars before the final stage, the gray curve is approximately
the same as the line of no mass loss (dotted). Exceptions are $\approx 100-140$~M$_\odot$
where the pulsational pair-instability ejects the outer layers of the star before
it collapses, and above $\approx 500$~M$_\odot$ where pulsational instabilities in
red supergiants may lead to significant mass loss (dashed). The hydrogen-rich
envelope and parts of the helium core (dash-double-dotted curve) are ejected in
a supernova explosion.}
\label{finalfate}
\end{figure}

A summary of the final fate of Pop~III stars is sketched in Fig.~\ref{finalfate}.

\subsection{Black Holes and Gamma Ray Bursts from the First Stars}

As discussed in the previous Section, metal-free (and also moderately metal-poor) 
stars more massive than approximately 260~M$_\odot$ collapse completely to BHs. 
Similar arguments apply to stars in a lower mass window (30~M$_\odot - 
140$~M$_\odot$), which are also expected to end their evolution as BHs.
If this is the case, a numerous population of Intermediate Mass Black Holes (IMBHs) 
-- with masses $M_\bullet \approx 10^2-10^3$~M$_\odot$,
between those of stellar and SuperMassive Black Holes (SMBHs) -- 
may be the end-product of those episodes of early star formation and offer a possible
observable of primordial stars. 
As these pre-galactic BHs become incorporated through a series of mergers into larger and 
larger halos, they sink to the center because of dynamical friction, accrete a fraction of 
the gas in the merger remnant to become supermassive, form a binary system, and eventually coalesce. 

Volonteri, Haardt \& Madau (2003) thoroughly studied this scenario
including a number of physical ingredients such as gas accretion, hardening of the 
IMBH binary and
triple interactions. Their results show that this hierarchical growth can reproduce the
observed luminosity function of optically selected quasars in the redshift range $1 < z < 5$. 
A prediction of the model is that a population of ``wandering'' black holes
in galactic halos and in the IGM should exist (and possibly observed), contributing around 10\% of
the present day total black hole mass density, $4\times 10^5$~M$_\odot$~Mpc$^{-3}$ 
(a similar study has
been performed by Islam, Taylor \& Silk 2003).
IMBHs that have not yet ended up in SMBHs could also be either {\it (i}) en route toward 
galactic nuclei, thereby accounting for the X-ray-bright off-center sources 
detected locally by {\tt ROSAT}\footnote{http://wave.xray.mpe.mpg.de/rosat}, 
or {\it (ii}) constituting the dark matter candidates 
composing the entire baryonic halos of galaxies (Schneider et al. 2002).
A possible method to constrain the existence and nature of BHs at $z>6$
would be with the next generation of X-ray and NIR space telescopes, that could
detect BHs down to masses of $10^5-10^6$~M$_\odot$ (Salvaterra, Haardt \& Volonteri
2007).

\begin{figure*}
\centerline{\includegraphics[width=28pc]{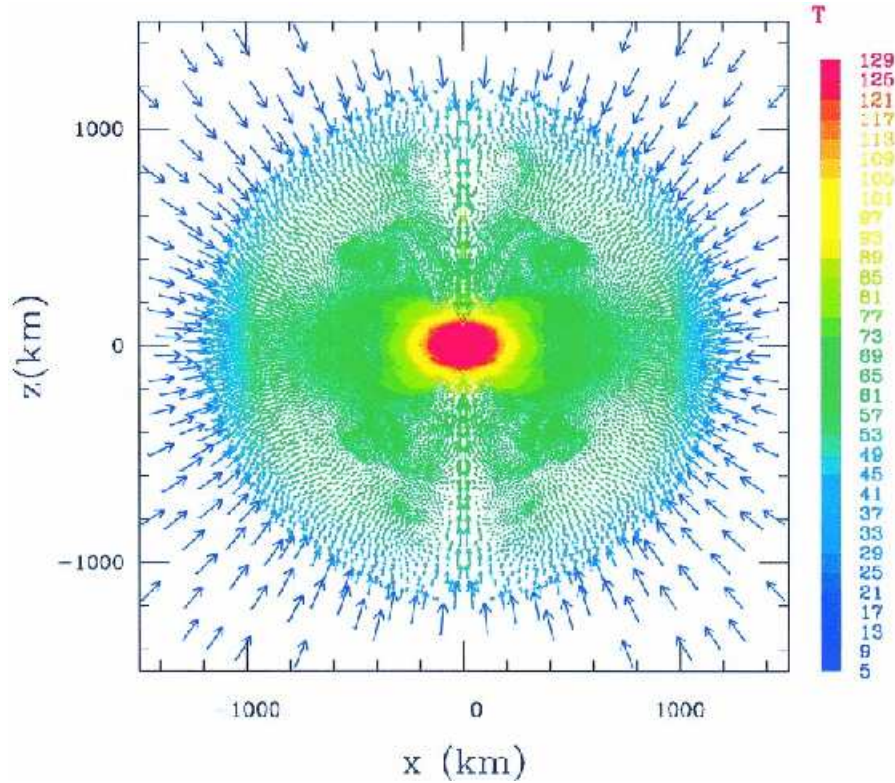}}
\caption{
Collapse of a 300 M$_\odot$ Pop~III star (Fryer, Woosley \& Heger 2001). 
The proto-BH         0.5~s before BH         formation is shown. 
Color denotes temperature in $10^9$~K, and the vectors represent the direction 
and magnitude of the particle velocity. At this time, the proto-BH           
has a mass of roughly 78 M$_\odot$ and size of 1100~km. The inner core first 
forms a black hole, but as soon as it collapses, the entire proto-BH
(which is 90 M$_\odot$ at collapse) quickly accretes onto the BH.}
\label{jet}
\end{figure*}

Once a proto-BH has formed into the stellar core, accretion continues 
through a disk. It is widely accepted, though not confirmed, that magnetic fields 
drive an energetic jet (Fig.~\ref{jet}) which produces a burst of TeV neutrinos 
(Linke et al. 2001;  M{\'e}sz{\'a}ros, \& Waxman 2001; Schneider, Guetta \& Ferrara 
2002) by photon-meson interaction, and eventually breaks out of the stellar envelope
appearing as a GRB. Based on recent numerical simulations and 
neutrino emission models, the expected neutrino diffuse flux from these Pop~III GRBs 
could be within the capabilities of present and planned detectors as 
{\tt AMANDA}\footnote{\tt http://amanda.berkeley.edu/www/amanda.html} (Antartic Muon And 
Neutrino Detector Array) and {\tt IceCube}\footnote{\tt http://icecube.wisc.edu/}. 
High-energy neutrinos from Pop~III GRBs could dominate the overall flux in two energy
bands, $10^4-10^5$ GeV and $10^5-10^6$ GeV, of neutrino telescopes. The enhanced 
sensitivities of forthcoming detectors in the high-energy band ({\tt AMANDA-II}, 
{\tt IceCube}) 
will provide a fundamental insight into the characteristic explosion energies of 
Pop~III GRBs, and will constitute a unique probe of the IMF of the first stars. 
Based on such results, Pop~III GRBs could be associated (e.g. Schneider, Guetta \&
Ferrara 2002; Arefiev, Priedhorsky \& Borozdin 2003) with a new class of events 
detected by {\tt BeppoSax}\footnote{\tt http:www.asdc.asi.it/bepposax/}, 
the Fast X-ray Transients (FXTs), bright X-ray 
sources with peak energies in the $2-10$~keV band and durations between 10~s 
and 200~s (e.g. Heise et al. 2001). Iocco et al. (2007) though, using updated models
of Pop~III star formation history and neutrino emission yields show that the planned
generation of neutrino telescope will not be able to detect the contribution from
Pop~III stars and that, in any case, this would be highly contaminated by the
contribution from Pop~II stars.

\subsection{Key Observations}

Despite the efforts made in the past decades, the search for zero-metallicity
stars has been proven unfruitful as no Pop~III star has been detected yet. The
first very low metallicity star, with a logarithmic abundance relative
to solar of [Fe/H]=$-4.5$\footnote{In the usual notation,
[X/H] = log (X/H) $-$ log (X/H)$_{\odot}$.}, 
was identified more than 15 years ago (Bessel \&
Norris 1984). Since then, more stars with $-4.5<$ [Fe/H] $< -2$ have been
detected (McWilliam et al. 1995; Beers et al. 1998; Norris, Ryan \& Beers 2001;
Carretta et al. 2002). The record was established by
Christlieb et al. (2002), who measured an abundance of [Fe/H]=$-5.3$ in 
HE0107-5240, a star with a mass of $\approx 0.8$~M$_\odot$. 
Does the existence of this star suggest that
Pop~III also contained low-mass and long-lived objects?
To answer this question, we should be able to infer the metallicity
of the gas cloud out of which HE0107-5240 formed from the abundance
of the elements observed on its surface. In particular, the observed abundance
pattern in elements heavier than Mg (which cannot be synthesized in the
interior of a 0.8~M$_\odot$ star) is a distinct signature of the previous
generation of stars.

HE0107-5240 is very iron-deficient but overabundant in C, N
and O (Bessel et al. 2004; Christlieb et al. 2004). The origin of these light
elements can be twofold: ({\it i}) either they were already present
in the parent gas cloud out of which the star was born, or ({\it ii}) their origin is due
to a post-formation mechanism, such as mass transfer from a companion star,
self-enrichment from the star itself or accretion due to repeated passage through
the Galactic disk (for a thorough discussion of these issues we refer to
Christlieb et al. 2004).
In the first scenario, the gas cloud out of which
the star formed would have had to be pre-enriched to a metallicity as high as
$Z=10^{-2}Z_{\odot}$. This metallicity is significantly higher than
$Z_{\rm cr}$ and low-mass stars can form. Moreover, Umeda \& Nomoto (2005) have
shown that if the polluter is a zero-metallicity star with mass
in the range  $\approx 25-130$~M$_\odot$ exploding as a sub-luminous Type-II SN,
a good fit of the observed abundance pattern can be obtained (although some
problems remain with the updated oxygen abundance; Bessel et al. 2004;
Christlieb et al. 2004).
In the second scenario, only the elements heavier than Mg, which cannot
be synthesized in the interior of such a low-mass star, were already present
before its formation. The composition of these elements can be equally well
reproduced by the predicted elemental yields of either an intermediate mass star
that exploded as SN (Christlieb et al. 2002; Limongi, Chieffi \&
Bonifacio 2003; Weiss et al. 2004; Iwamoto et al. 2005) or of a  $200-220$~M$_\odot$
star that exploded as a PISN (Schneider et al. 2003). 
Thus, HE0107-5240 may have
formed from a gas cloud which had been pre-enriched by such supernovae up to
an estimated metallicity of $10^{-5.5} \le Z/Z_{\odot} \le 10^{-5.1}$. These
values fall within the proposed range for $Z_{\rm cr}$.

A second very iron deficient star (HE1327-2326, with [Fe/H]$=-5.4$) with an
overabundance of C and N has been detected by Frebel et al. (2005). 
Differently from HE0107-5240, which is a giant and has evolved off the 
main-sequence, HE1327-2326 is relatively unevolved.

\begin{figure}
\centerline{\includegraphics[width=30pc]{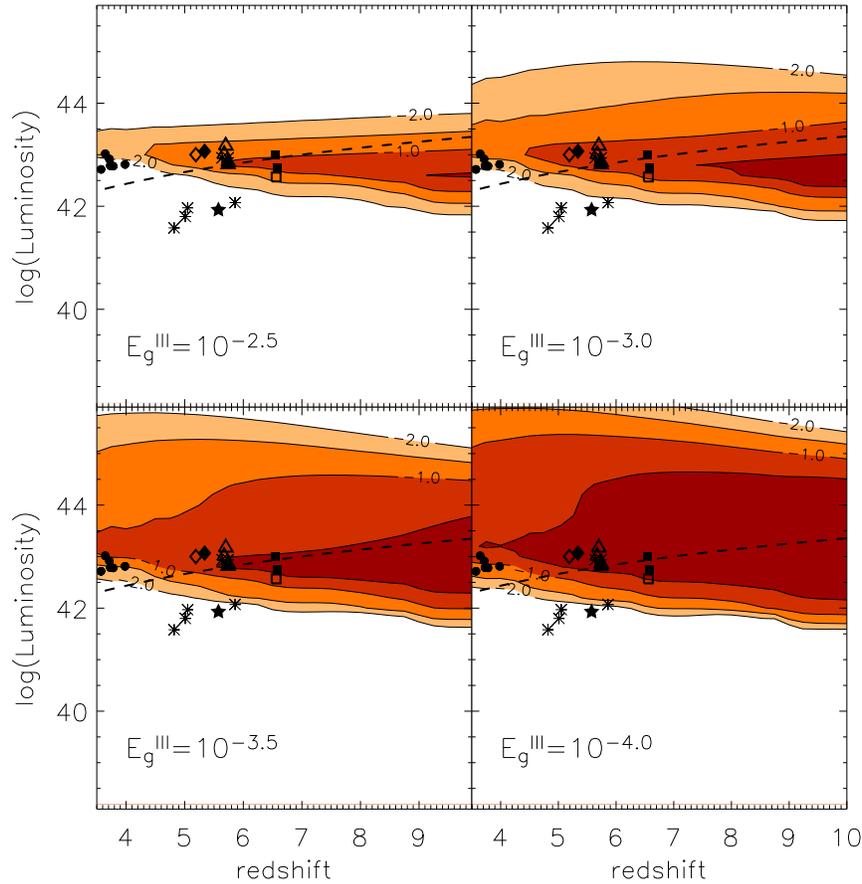}}
\caption{Fraction of Pop~III star-hosting galaxies as a function of Ly$\alpha$
luminosity and redshift.  Isocontours of fractions $\ge 10^{-2},
10^{-1.5}, 10^{-1}$ and $10^{0.5}$ are shown for a
burst-mode star formation.
Each panel is labeled by the assumed feedback parameter $E_g$ value.
For
reference, the dashed line gives the luminosity corresponding to an
observed flux of $1.5 \times 10^{-17}$ ergs cm$^{-2}$ s$^{-1}$,
and the various points correspond to observed galaxies (see details
in Scannapieco, Schneider \& Ferrara 2003).
}
\label{fig:short}
\end{figure}

The most massive star is the Pistol star,
with an estimated initial mass of $200-250$~M$_\odot$ (Figer et al. 1998).
Although there are no indications that the star is a binary, it might
have recently experienced a rejuvenation through a merger with another
star. Figer (2005) has investigated the presence of very massive stars
in the Arches cluster, which has an estimated age of 2-2.5~Myr. Thus,
stars with masses above $\sim 150$~M$_\odot$ should still be visible if
they were formed. The typical mass function would predict 18 stars with
masses greater than 130~M$_\odot$, but none were found, suggesting a
firm upper limit for the star masses of 150~M$_\odot$.
Also Oey \& Clarke (2005), based on a statistical analysis of OB associations,
find that, if the IMF has a slope close to a Salpeter, it would be extremely
unlikely to have stars with masses greater than 120-200~M$_\odot$.

As a novel strategy to search for Pop~III stars, Scannapieco, Schneider \& Ferrara (2003)
have suggested that at least some
intermediate redshift Ly$\alpha$-emitting galaxies might have their flux powered by Pop~III
stellar clusters. In Fig.~\ref{fig:short} the isocontours in the Ly$\alpha$
luminosity-redshift plane indicate the probability to find Pop~III star-hosting galaxies
in a given sample of Ly$\alpha$ emitters for various feedback
efficiencies, $E_g$. Such objects populate a well defined
region of the Ly$\alpha$-redshift plane, whose extent is governed by
the feedback strength.  Furthermore, their fraction increases with redshift,
independently of the assumed feedback strength,
suggesting that Ly$\alpha$ emission from already observed high-$z$ sources
can be due to Pop~III stars.
Hence, collecting large data samples to increase the
statistical leverage may be crucial for detecting the elusive first
stars.

\begin{figure}
\centerline{\includegraphics[width=30pc]{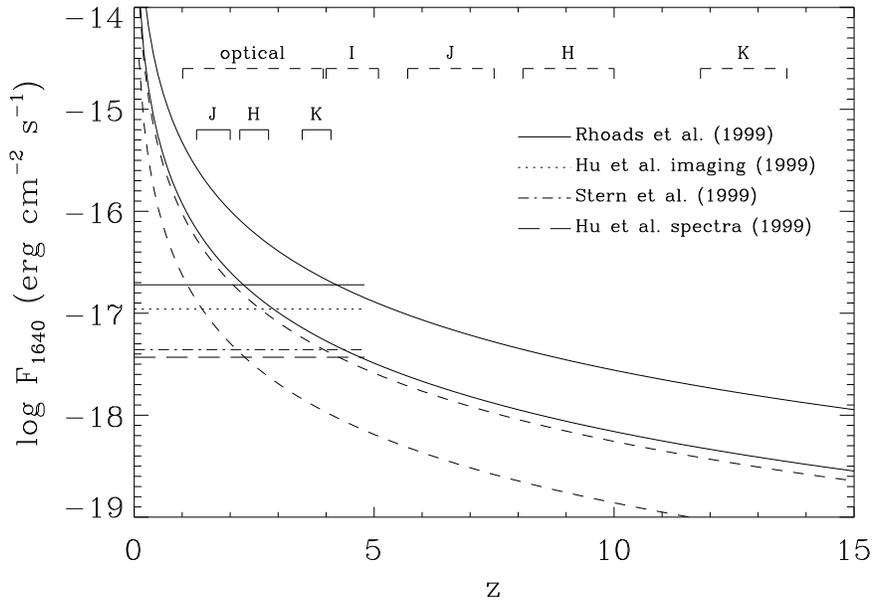}}
\caption{
Flux of \HeII $\lambda$1640 line as a function of the source redshift.
The different lines refer to different models for the time evolution of the
stellar ionizing radiation. The horizontal lines show the limits of the
current Ly$\alpha$ emission-line surveys. At the top, the ranges of redshift
probed by the \HeII $\lambda$1640 (above) and $\lambda$4686 (below) lines
for optical and infrared searches are marked. See Tumlinson, Giroux \& Shull
(2001) for details.}
\label{helines}
\end{figure}

Confirmation of the presence of Pop~III stars in the Ly$\alpha$ emitters
can come from follow-up spectroscopy.
Tumlinson and collaborators (Tumlinson, Giroux \& Shull 2001; Tumlinson,
Shull \& Venkatesan 2003) have suggested that the harder ionizing spectrum
expected from metal-free stars would result in stronger \HeII recombination lines.
Their calculations indicate that line fluxes of the \HeII $\lambda$1640 and
$\lambda$4686 are sufficiently large to be detected by narrow-band and
spectroscopic searches for high-redshift emission-line sources (see Fig.~\ref{helines}).
Together with measurements of the H$\alpha$ emission line, the \HeII
recombination lines can be used also to infer the ratio of the \HeII to \HI
ionizing photons (Oh, Haiman \& Rees 2001).

Nagao et al. (2008) performed a photometric search of Pop~III stars by looking at
galaxies with strong emission both in Ly$\alpha$ and \HeII $\lambda$1640
in the redshift intervals $3.93 < z < 4.01$ and $4.57 < z < 4.65$. As none of the
observed dual emitters could be identified with a galaxy hosting a Pop~III star,
an upper limit of $5 \times 10^{-6}$~M$_\odot$~yr$^{-1}$~Mpc$^{-3}$ to the Pop~III 
SFR has been set.

In addition to He lines, also H$_2$ lines could be used to observe the first
stars. For example, H$_2$ lines
(principally $J=5-3 (v=0)$, $J=2-0 (v=0)$ and $\Delta v=1$ lines) emitted by
collapsing primordial clouds, although not observable from an individual object,
could produce cumulative fluxes in the
sub-mJy level, marginally observable with {\tt ASTRO-F}\footnote
{http://www.ir.isas.ac.jp/ASTRO-F/index-e.html} and {\tt ALMA}\footnote
{http://www.alma.nrao.edu} (Atacama Large Millimeter Array). If HD is an
important coolant as discussed in the previous Section, then additional
line emission might be observed. In this case, the lines emitted during
the collapse would show a double-peak feature, with the higher peak associated
with the H$_2$ $J=2-0 (v=0)$ transition and the lower peak with the HD
$J=4-3 (v=0)$ transition (Kamaya \& Silk 2002; Kamaya \& Silk 2003).
A similar strategy has been proposed by Ripamonti et al. (2002), who
suggest that the sum of the H$_2$ infrared line and continuum emission
radiation produced during the collapse and accretion on
a small hydrostatic protostellar core could be detected by {\tt JWST}.
In contrast with the above studies, Mizusawa, Nishi \& Omukai (2004), who
include in their calculation the emission of H$_2$ lines both in the
collapse and in the accretion phase, find that observation of such lines
with forthcoming facilities will be highly improbable.

While direct observations of the early stars are very difficult (if not
impossible) to obtain, one can search for evidence of their past existence
with other observable. Such evidence can in particular be seen in
binary systems, where one of the components is a sub-solar-mass star and the
other was a larger mass star. The presently observed spectrum of the
lower-mass companion would exhibit the chemical signature of the larger-mass
companion. The high frequency of carbon-enhanced stars found among very
metal-poor stars in the halo of our Galaxy provides a constrain on the past
IMF and indicates that it was different from the present and shifted toward
higher masses (e.g. Lucatello et al. 2005). These findings are consistent with
an early prediction by Hernandez \& Ferrara (2001) of an upward shift of the characteristic 
stellar mass with redshift based on an analysis of the observational data of Galactic 
metal poor stars. Such mass increases above the Jeans mass fixed by the 
increase of the CMB temperature with redshift.  

Ballero, Matteucci \& Chiappini (2006) though reach a different conclusion.
They have studied the effect of Pop~III star metal pollution on the chemical
evolution of the Milky Way, focusing their analysis on $\alpha$-elements,
carbon, nitrogen and iron, finding that the effect of Pop~III stars is
negligible and it is not possible to prove or disprove their existence based
on chemical abundance in the Milky Way. Moreover, their analysis favors
a constant or slightly varying IMF to a strongly variable one.
Along the same lines, Damped Lyman Alpha (DLA) systems have been searched for 
the nucleosynthetic ashes of Pop~III stars. Erni et al. (2006) 
compare the chemical enrichment pattern of the most metal-poor DLAs 
with predictions from different explosive nucleosynthesis models and find 
that the observed abundance pattern favors massive (10-50~M$_\odot$) 
metal-free supernovae/hypernovae. Pair-instability SNe seem to be excluded; 
however, given the metallicity level of DLAs ($\approx 10^{-2} Z_\odot$), this
is not surprising as it is likely that the gas in these systems has already been 
enriched by a large number of stellar populations. 
 
Signatures of Pop~III stars might be seen in 
puzzling observations of the abundance of lithium isotopes (e.g. Asplund et al. 2006).
While $^7$Li is produced both in BBN and through cosmic ray (CR) spallation, $^6$Li is produced only
via the second process. On this basis a correlation between $^6$Li abundance
and metallicity is expected. Two problems arise: {\it (i)} $^7$Li/H abundance observed in metal-poor
halo stars is $\sim 10^{-10}$, i.e. much lower than  expected from BBN alone ($\sim 4.5 \times
10^{-10}$); {\it (ii)} $^6$Li abundance does not correlate with metallicity and rather shows a plateau
that extends to low metallicity values. Different scenarios have been proposed to explain the
plateau (CR in shocks during the formation of the Galaxy, decay of relic particles during BBN,
production in connection with gamma rays, initial burst of cosmological CRs) without
overproducing $^7$Li. A tentative solution
has been found by Rollinde, Vangioni \& Olive (2006) who have calculated the production
of $^6$Li from CR produced by Pop~III and Pop~II stars using a prescription for the
cosmic SN rate.  
However, the models makes extreme and somewhat unphysical assumptions
concerning the cosmic star formation history and it is not clear to what extent it
can reproduced the metallicity distribution function of halo stars.
Useful reviews for readers interested in the subject can be found in Coc et al. (2004) and
Steigman (2005).

Jimenez \& Haiman (2006) explain with the presence of 10\%-30\% of metal-free
stars at $z\sim 3-4$, four puzzling observations that cannot be accounted
for with a more normal stellar population: {\it (i)} the significant UV emission from
LBGs at $\lambda < 912$~$\AA$; {\it (ii)} the strong Lyman-$\alpha$ emission from extended
blobs at $z=3.1$ with little or no associated apparent ionizing continuum;
{\it (iii)} a population of $z\sim 4.5$ galaxies with a very wide range of Lyman-$\alpha$
equivalent width emission lines; {\it (iv)} strong HeII emission line in a composite of
LBGs.

GRBs have been proposed as possible tools to study the high-$z$ universe and, in
principle, a fraction of them could be triggered by Pop~III progenitors.
To produce a GRB though, a star needs to lose its outer H envelope.
As the mass loss from Pop~III stars is not expected to be significant, 
mass transfer to a companion star in binaries could help.
If Pop~III binaries are common, Pop~III star formation at high-$z$ could be
probed by {\tt SWIFT}\footnote{http://www.swift.psu.edu/xrt/} (Bromm \& Loeb 2006).
Belczynski et al. (2007) find that, although, about 10\% of Pop~III binaries
lose the H envelope, only 1\% acquire sufficient angular momentum to produce
a GRB. Thus, it will be extremely difficult for {\tt SWIFT} to detect GRBs from
Pop~III binaries.

\begin{figure}
\centerline{\includegraphics[width=26pc]{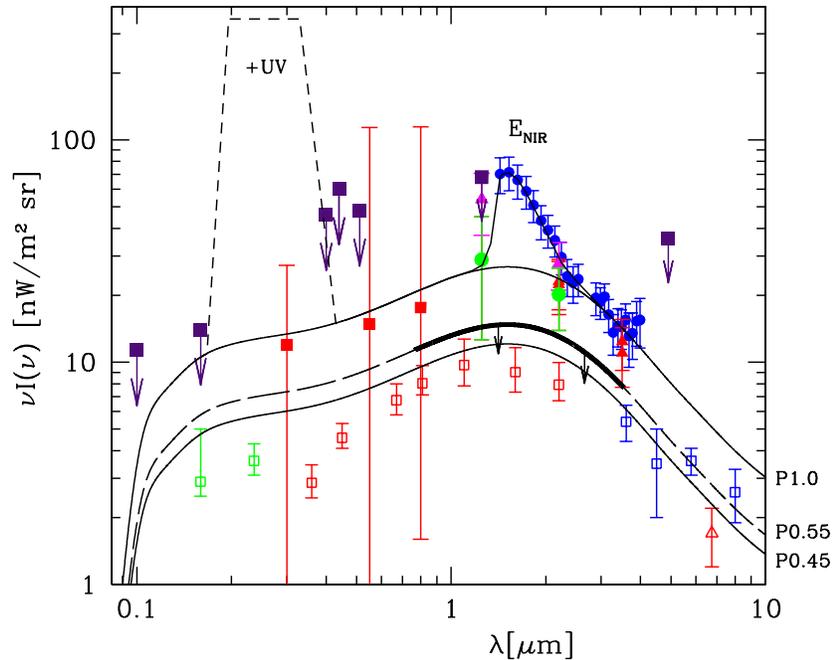}}
\caption{
Extragalactic background light data compilation from various IR and UV experiments. Also shown are 
(lower) limits coming from galaxy counts (open symbols). Lines denote different 
theoretical models (for details see Aharonian et al. 2006).
}
\label{irback}
\end{figure}
In the last years the alleged discovery of an excess in the near IR background (an
updated compilation on the extragalactic background light measurements is shown in Fig. \ref{irback}) and
of its fluctuations above the ``normal'' galaxy contribution claimed by 
Matsumoto et al. (2005; but see, e.g., Aharonian et al. 2006 for a different view), 
has produced considerable excitement given the fact that the excess could be
interpreted as an imprint of redshifted light of the first (massive) stars, 
(Santos, Bromm \& Kamionkowski 2002; Salvaterra \& Ferrara 2003a; Madau \& Silk 2005;
Kashlinsky 2005).
In addition, the emission from metal-free stars is needed to explain the level of
fluctuations in the same background, at least at the shortest wavelengths. In fact,
clustering of unresolved Pop~III stars can account for the entire signal at almost
all the $\approx 1-30$~arcsec scales probed by observations in the $J$ band, while
their contribution fades away at shorter frequencies and becomes negligible in
the $K$ band, where ``normal'' galaxies are dominant (Magliocchetti, Salvaterra
\& Ferrara 2003; Cooray et al. 2004; Kashlinsky et al. 2004; Kashlinsky et al. 2005;
Cooray et al. 2007).
The new measurements of fluctuations with the {\tt IRAC} instrument on board of {\tt Spitzer} 
and the associated analysis confirm that the anisotropies on scales $\sim 0.5'-10'$ must
be of extragalactic origin (Kashlinsky et al. 2007a).
Salvaterra et al. (2006) find that Pop~III stars can contribute only less than
40\% to the NIRB and that the fluctuations can be reproduced by the clustering
of Pop~II hosting galaxies at $z>5$. That the fluctuations must arise from a population
of sources at those redshifts is confirmed also by Kashlinsky et al. (2007b), although they
find that such sources should be much brighter per unit mass than the present stellar population.

Dwek, Arendt \& Krennrich (2005) also noticed that explaining the
excess with Pop~III contribution only, would require uncomfortably high
formation efficiencies. In fact, more than 5\% of all baryons must be processed into
Pop~III stars and, to avoid excessive metal pollution, they should end up in IMBHs
with steep UV/X-ray spectra (not to overproduce the present-day soft X-ray background).
In addition, if the excess in the NIRB were of extragalactic
origin (rather than due to zodiacal light), its $\gamma-\gamma$ interactions with
$\gamma$-ray photons emitted by blazars would produce a distinct absorption feature
in the spectra of these sources. Dwek, Krennrich \& Arendt (2005) show that, if
the NIRB excess were due to Pop~III stars it would be difficult to reproduce
the absorption-corrected spectra of some observed distant TeV blazars.          

A substantial IR background could arise from Pop~III stars not only due to the
direct emission associated with these stars, but also due to indirect processes
that lead to free-free and Ly$\alpha$ emission from the ionized nebulae or \HII
regions surrounding the stars. In view of more recent observations though, a 
substantial contribution to the near IR background for Pop~III stars seems to be excluded.
Salvaterra \& Ferrara (2006) find that if this is the case, a large fraction
(in some case all) of {\tt Spitzer} counts (at 3.5~$\mu$m and 4.5~$\mu$m; Fazio at al.
2004) at the faintest fluxes is due to galaxies at $z>8$, but worse, the number of 
J-dropout and Ly$\alpha$ emitters identified in {\tt NICMOS} (Bouwens et al. 2005) and 
{\tt ZEN} (Willis \& Courbin 2005) deep surveys would be overpredicted by a factor of thousands.

An additional NIRB contribution could come from
supernovae (although this is usually minor) and high redshift quasars. The latter could 
give a substantial contribution if Pop~III were indeed very massive and ended up
as IMBH (Cooray \& Yoshida 2004). In this case though they would probably overestimate the measured
soft X-ray background (Salvaterra, Haardt \& Ferrara 2005). 

Regardless of the precise contribution from primordial stars to the IR  background,
they would leave enough photons to provide a large optical depth for high-energy
photons from high-$z$ GRBs. Observations of their spectra could thus provide
information on the emission from the first stars (Kashlinsky 2005).

\begin{figure}
\centerline{\includegraphics[width=30pc]{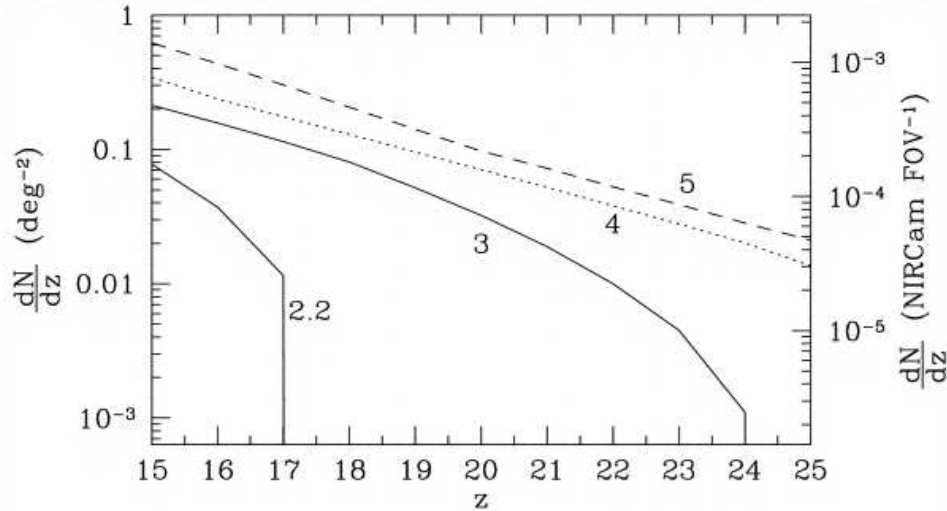}}
\caption{
Predicted surface density of PISNe on the sky per unit redshift as a function
of redshift for four different observing wavelengths
(labeled in $\mu$m), assuming that survey images reach 1 mag below
maximum light, after Weinmann \& Lilly (2005).
}
\label{lilly}
\end{figure}
{\tt JWST} could observe the upper mass range of pair instability 
supernovae from the first stars, which would provide, in addition to
an observational proof of the existence of very massive metal-free
stars, stringent constraints on the understanding of primordial star
formation and IMF (Wise \& Abel 2005). Although PISN are bright enough to
be detected by {\tt JWST}, Weinmann \& Lilly (2005) claim that 
their rate is so small (0.2-4~deg$^{-2}$~yr$^{-1}$)
that either a dedicated wide-angle search or a serendipitous search is
needed. The numbers quoted by the latter authors are less optimistic than 
previous estimates (0-2200~deg$^{-2}$~yr$^{-1}$; e.g. Cen 2003a; Mackey,
Bromm \& Hernquist 2003; Wise \& Abel 2005) because very approximate light 
curves were adopted. Although, as
noted by Scannapieco et al. (2005), the light curves used by Weinmann \& Lilly
(2005) contain a normalization error, the numbers derived are in agreement with
those derived adopting the correct light curves (see Fig.~\ref{lilly}). 
In any case, Smith at al. (2007) reported the detection of the peculiar Type II SN
2006gy which could be the first SN powered by a pair-instability explosion.

If indeed the first stars were very massive and left BH remnants,
then gravitational waves emitted during each collapse (Schneider et al. 2000;
Fryer, Holz \& Hughes 2002; Buonanno et al. 2005; Sandick et al. 2006; Suwa et al. 2007b) 
would be seen as a
stochastic background, whose predicted spectral strain  peaks in the frequency 
range $5\times 10^{-4}-5\times 10^{-3}$~Hz (where it has a value 
$10^{-20}-10^{-19}$~Hz$^{-1/2}$) and might be detected by
{\tt LISA}\footnote{http://lisa.jpl.nasa.gov}(Laser Interferometer Space
Antenna).
Similarly, the coalescence of massive black hole binaries due to galaxy
mergers has also been considered as a source of low-frequency gravitational 
radiation (Wyithe \& Loeb 2003c; Sesana et al. 2004; Belczynski, Bulik \&
Rudak 2004). Several discrete mHz massive 
black hole sources detectable by {\tt LISA} are likely to originate at 
redshifts $z > 7$, 
while the advanced {\tt LIGO}\footnote{http://www.ligo.caltech.edu} 
should detect at least several events per year with high signal to noise ratio
($\simgt 10$).
Core collapse SN from primordial stars would produce a gravitational
wave background that peaks at 30-100~Hz (compared to 300-500~Hz from a normal
mode of star formation) and is within the sensitivity of planned interferometers
as {\tt LIGO}, {\tt BBO} (Big Bang Observatory) and 
{\tt DECIGO} (DECi-hertz Interferometer Gravitational wave Observatory) 
(Sandick et al. 2006). Using an improved spectrum, Suwa et al. (2007b) find that
the range of observability is 0.1-1~Hz.

\section{Feedback Effects}              
\label{feedback}

Once the first sources have formed, their mass deposition, energy injection
and emitted radiation can deeply affect the subsequent
galaxy formation process and influence the evolution of the IGM via a number
of so-called ``feedback'' effects. 
The word ``feedback'' is by far one of the most used in modern cosmology, where
it is applied to a vast range of situations and astrophysical objects. However,
for the same reason, its meaning in the context is often unclear or fuzzy.
Generally speaking, the concept of feedback
invokes a {\it back reaction of a process on itself or on the causes that have 
produced it.} Secondly, the character of feedback can be either {\it negative or 
positive.} Finally, and most importantly, the idea of feedback is intimately 
linked to the possibility that {\it a system
can become self-regulated}. Although some types of feedback processes are disruptive,
the most important ones in astrophysics are probably those that are able to drive the systems towards
a steady state of some sort. To exemplify, think of a galaxy which is witnessing a 
burst of star formation.
The occurrence of the first supernovae will evacuate/heat the gas thus suppressing the 
subsequent SF process.   
This feedback is then acting back on the energy source (star formation); it is of a
negative type, and it could drive  the SF activity in such a way that only
a sustainable amount of stars is formed. 
However, feedback can fail to produce
such regulation either in small galaxies, where the gas can be ejected by a handful of 
SNe, or in cases when the star formation timescale is too short compared to 
the feedback timescale.
In the spirit of the present review we will only discuss feedback
occurring at high redshifts and hence shaping the first structures.

\begin{table*}
\caption[]{Classification of different feedback effects.}
\label{tablefeedback}
\begin{tabular}{lll}
\hline
       & {\bf NEGATIVE}& \\
{\it Radiative}                 &  {\it Mechanical} & {\it Chemical} \\
\hline
1. Photoionization/evaporation & 1. Blowout/blowaway & 1. Fragmentation   \\
2. H$_2$ Photodissociation     & 2. Impinging shocks &                   \\
3. Photoheating filtering      & 3. Preheating       &                     \\
\\
\hline
        & {\bf POSITIVE} & \\
{\it Radiative}                 &  {\it Mechanical} & {\it Chemical}      \\
\hline
1. In front of \HII regions     & 1. Behind shocks  &                      \\
2. Inside relic \HII regions    & 2. Shell fragmentation &                 \\
3. X-ray background             &                   &                      \\
\hline
\end{tabular}
\end{table*}
Although a rigorous classification of the various effects is not feasible,
they can be divided into three broad classes: {\it radiative}, 
{\it mechanical} and {\it chemical} feedback. In the first class fall all those effects associated,
in particular, with ionization/dissociation of hydrogen atoms/molecules; the second
class is produced by the mechanical energy injection of massive stars in form of winds or SN explosions;
and chemical feedback is instead related to the postulated existence of a critical metallicity 
governing the cosmic transition from very massive stars to ``normal'' stars. 

Attempting a classification of all proposed feedback effects is almost desperate, due to the
large number of applications and definitions, often discordant, present in the literature.  
Nevertheless, we offer in Table~\ref{tablefeedback} a working classification aimed 
essentially at organizing the material discussed in this Section.   
Whereas the radiative feedback is intimately connected with cosmic reionization,
chemical feedback is instead strongly dependent on the history of metal enrichment of the universe. 
Although reionization and metal enrichment could then be analyzed as aspects of feedback,
given their importance and the large amount of literature
on these subjects, we have reserved a separate discussion for them
in Sec.~\ref{reionmet}.

\subsection{Radiative Feedback}

Radiative feedback is related to the ionizing/dissociating radiation
produced by massive stars or quasars. This radiation can have local effects
(i.e. on the same galaxy that produces it) or long-range effects, either affecting
the formation and evolution of nearby objects or joining the radiation produced
by other galaxies to form a background.
In spite of the different scenarios implied, the physical processes are very similar.

\vspace{0.3cm}
\centerline {\it Photoionization/evaporation}
\vspace{0.3cm}

The collapse and formation of primordial objects exposed to a UV radiation field
can be inhibited or halted for two main reasons: {\it (i)} cooling is considerably 
suppressed by the decreased fraction of neutral hydrogen, and  {\it (ii)} gas can be
photoevaporated out of the host halo.
In fact, the gas incorporated into small mass objects
that were unable to cool efficiently, can be boiled out of
the gravitational potential well of the host halo if it is heated by UV
radiation above the virial temperature. 
Such effects are produced by the same radiation field and act simultaneously.
For this reason, it is hard to separate their 
individual impact on the final outcome. The problem has been  
extensively studied by several authors (e.g.
Thoul \& Weinberg 1996; Nagashima, Gouda \& Sugiura 1999; Barkana \& Loeb 1999; Susa \&
Kitayama 2000; Chiu, Gnedin \& Ostriker 2001; 
Kitayama et al. 2000, 2001; Machacek, Bryan \& Abel 2001; Haiman, Abel \& Madau 2001;
Susa \& Umemura 2004b; Ahn \& Shapiro 2007; Whalen et al. 2007).
In particular, Susa \& Kitayama (2000) and Kitayama et al. (2000, 2001),
assuming spherical symmetry, solve self-consistently
radiative transfer of photons, non-equilibrium H$_2$ chemistry and
gas hydrodynamics of a collapsing halo. They find that at weak UV
intensities ($J< 10^{-23}$~erg~s$^{-1}$~cm$^{-2}$~sr$^{-1}$~Hz$^{-1}$), 
objects as small as $v_c \sim 15$~km~s$^{-1}$
are able to collapse, owing to both self-shielding of the gas and H$_2$
cooling. At stronger intensities though, objects as large as $v_c \sim
40$~km~s$^{-1}$ can be photoevaporated and prohibited from collapsing, in
agreement with previous investigations based on the optically thin
approximation (Thoul \& Weinberg 1996). Similar results have been found by
Susa \& Umemura (2004b) by means of a 3-D hydrodynamic simulation which
incorporates the radiative transfer of ionizing photons. Also the
3-D hydrodynamic simulation of Machacek, Bryan \& Abel (2001) confirms 
that the presence of UV radiation delays or suppresses the
formation of low mass objects. 
In contrast with these studies, Dijkstra et al. (2004), applying the 1-D
code by Thoul \& Weinberg (1996) to $z>10$ objects, find that
objects as small as $v_c \sim 10$~km~s$^{-1}$ can self-shield and collapse
because the collisional cooling processes at high redshift are more
efficient and the amplitude of the ionizing background is lower. The second
condition might not always apply, as 
the ionizing flux at high redshift is dominated by the direct radiation 
from neighboring halos rather than the background (Ciardi et al. 2000).

\begin{figure}
\centerline{\includegraphics[width=3.2in]{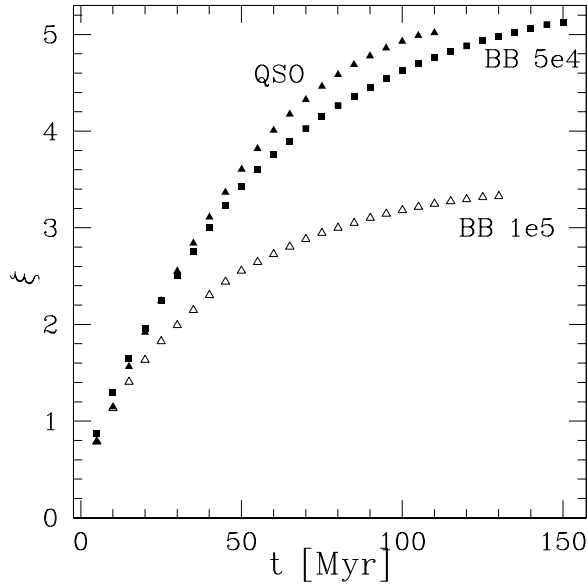}}
\caption{Evolution of the cumulative number of ionizing photons absorbed per 
initial minihalo atom for a QSO and two black-body spectra with $T=5\times 10^4$~K and 
$T=10^5$~K, as labelled (see Shapiro, Iliev \& Raga 2004 for detailed
definitions).}
\label{consump_time}
\end{figure}

The photoevaporation effect might be particularly important for 
Pop~III objects, as their virial temperatures are below the typical
temperatures achieved by a primordial photoionized gas ($T\approx 10^4$~K).
For such an object, the ionization front gradually burns 
its way through the collapsed gas, producing a wind that blows backwards into the
IGM and that eventually evaporates all the gas content. 
According to recent studies (Haiman, Abel \& Madau 2001; Shapiro, Iliev \& Raga 2004;
Whalen et al. 2007) 
these sub-kpc galactic units were so common 
to dominate the absorption of ionizing photons.
This means that estimates of the 
number of ionizing photons per H atom required to complete reionization
should not neglect their contribution to absorption.
As Fig.~\ref{consump_time} (Shapiro, Iliev \& Raga 2004) shows, the 
number of ionizing photons absorbed per initial minihalo atom,
$\xi$, increases gradually with time; in addition,
it depends on the ionizing spectrum assumed.  For the hard QSO spectrum,
the ionization front is thicker and penetrates deeper into the denser
and colder parts
of the halo, increasing the rate of recombinations per atom, compared to stellar
type sources. However, this same
pre-heating effect shortens the evaporation time, ultimately
leading to a rough cancellation of the two effects and the same
total $\xi$ as for a black-body spectrum with $T=5\times 10^4$~K (mimicking a 
low metallicity Pop~II stellar emission). An even lower $\xi$ is needed for
a black-body spectrum with $T=10^5$~K (more typical of a Pop~III stellar emission), because of an increased evaporation vs. penetration ability.
Thus, overall, Pop~III stellar sources appear significantly more efficient than
Pop~II or QSO sources in terms of the total
number of ionizing photons needed to complete the photoevaporation process.
It should be noted though that the typical timescales for photoevaporation are larger than
the lifetime of massive Pop~III stars, and thus minihalos generally survive photoevaporation
(Alvarez, Bromm \& Shapiro 2006).
The degree of photoevaporation also depends sensively on the evolution stage of halos,
with diffuse halos with central densities below few cm$^{-3}$ being completely photoevaporated,
while more evolved halos generally retaining their core (Whalen et al. 2007).

\vspace{0.3cm}
\centerline {\it H$_2$ Photodissociation}  
\vspace{0.3cm}

As intergalactic H$_2$ is easily photodissociated, a soft-UV background
in the Lyman-Werner bands could quickly build up and have a negative
feedback on the gas cooling and star formation inside small halos (Haiman, Rees \& Loeb
1997; Ciardi, Ferrara \& Abel
2000; Ciardi et al. 2000; Haiman, Abel \& Rees 2000; Ricotti, Gnedin \&
Shull 2002; Mackey, Bromm \& Hernquist 2003; Yoshida et al. 2003a;
Wise \& Abel 2008; Johnson, Greif \& Bromm 2008). 
In addition to an external background, the evolution
of structures can also be affected by internal dissociating radiation. In fact,
once the first generation of stars has formed in an object, it can affect the
subsequent star formation process by photodissociating molecular hydrogen
in star forming clouds (Omukai \& Nishi 1999;
Nishi \& Tashiro 2000; Glover \& Brand 2001; Oh \& Haiman 2002).
In particular, Nishi \& Tashiro (2000) show that if the molecular
cloud has a metallicity smaller than about $10^{-2.5} Z_\odot$,
a single O star can seriously deplete the H$_2$ content so that subsequent star
formation is almost quenched. Thus, it seems plausible that stars do not form
efficiently before the metallicity becomes larger than about $10^{-2}$ solar.
On the other hand, Susa \& Umemura (2006) performed hydrodynamic simulations
coupled to radiative transfer of ionizing and dissociating photons finding that,
if a star forming cloud exceeds the threshold density of $\sim 10^2$~cm$^{-3}$
and is at a distance $>30$~pc from a 120~M$_\odot$ Pop~III star, it can survive
negative feedback as the H$_2$ shell formed in front of the ionization front
is able to shield the cloud. This suggests that in typical primordial star forming
conditions secondary star formation can actually take place.

\begin{figure}
\centerline{\includegraphics[width=4.2in]{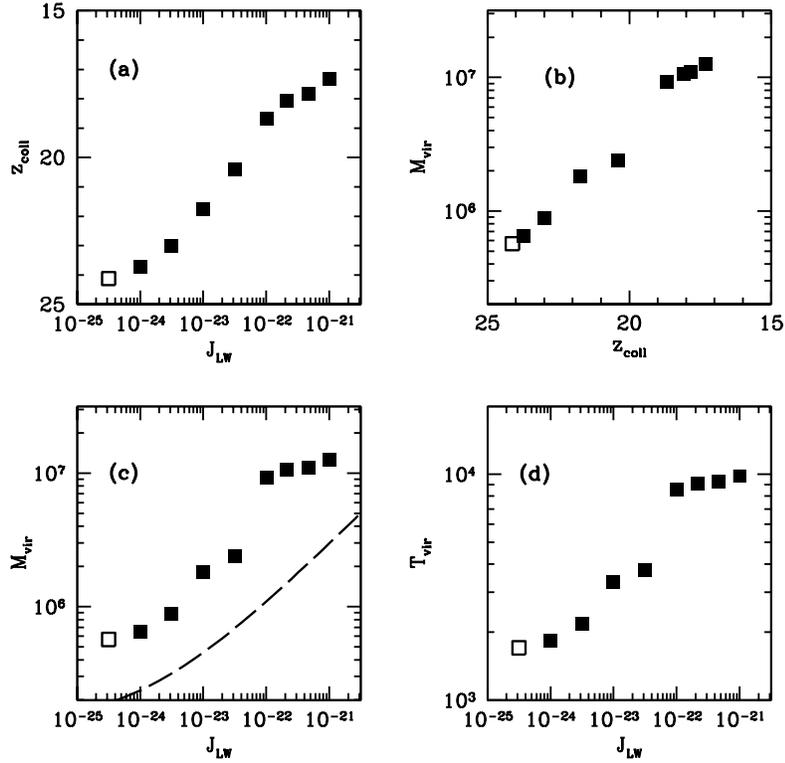}}
\caption{Mean halo quantities for several simulations with the same cosmic
realization but a range of Lyman-Werner molecular hydrogen photodissociating flux
backgrounds (O'Shea \& Norman 2007). Panel (a):  halo collapse redshift vs. J$_{LW}$.
Panel (b): halo virial mass vs. collapse rdshift. Panel (c): halo virial mass vs. J$_{LW}$.
Panel (d): virial temperature vs. J$_{LW}$.
The J$_{21} = 0$ ``control'' results are shown as an open square
(it is at log J$_{LW} = -24.5$ in the panels which are a function of J$_{LW}$).
In the bottom panel, the dashed line corresponds to
the fitting function for threshold mass by Machacek, Bryan \& Abel (2001).}
\label{OShea_Norman}
\end{figure}

Machacek, Bryan \& Abel (2001) find that the fraction
of gas available for star formation in Pop~III objects of mass $M$ exposed to a
flux with intensity $J_{LW}$ in the Lyman-Werner band is $\sim 0.06$~${\rm ln}
(M/M_{th})$, where the mass threshold, $M_{th}$, is given by:
\begin{equation}
(M_{th}/{\rm M}_\odot)=1.25 \times 10^5+8.7 \times 10^5 \left( \frac{J_{LW}}{10^{-21}
{\rm erg}^{-1} {\rm cm}^{-2} {\rm Hz}^{-1}} \right ).
\end{equation}
An AMR simulation by O'Shea \& Norman (2007b) confirms that the formation of
small mass, primordial objects in the presence of a Lyman-Werner flux is delayed
(although not completely suppressed because enough H$_2$ is retained in the core),
and, as a result, the virial mass of the halo at the time of the collapse is
increased compared to a case when no dissociating flux is present (see Fig.~\ref{OShea_Norman}).
The same problem has been analyzed by Susa \& Umemura (2004a) by means of
three-dimensional SPH calculations, where radiative transfer is
solved by a direct method and the non-equilibrium chemistry of primordial
gas is included. 
They find that star formation is suppressed appreciably by UVB, but baryons
at high-density peaks are self-shielded, eventually forming some amount of stars.
Similarly, a 1-D radiation-hydrodynamical simulation by Ahn \& Shapiro (2007) suggests
that minihalos down to total masses of $10^5$~M$_\odot$ are able to self-shield
against dissociating radiation from a nearby massive Pop~III star and form dense cold 
clouds at their center, also due to shock-induced molecule formation. Their
results though suffer from unphysical geometric effects due to the 1-D configuration 
and are not reproduced by the 2-D simulations by Whalen et al. (2007), who find that
shock-induced molecule formation plays no role. Nevertheless, they confirm that a
LW flux does not exert a strong feedback on collapse of minihalos.

The negative feedback described above could be counterbalanced by
the {\it positive feedback} of H$_2$ re-formation, e.g. in front of \HII regions,
inside relic \HII regions, once star formation is suppressed in a halo and ionized gas 
starts to recombine (Ricotti, Gnedin \& Shull 2001, 2002; Mashchenko, Couchman \& Sills
2006; Abel, Wise \& Bryan 2007; Johnson, Greif \& Bromm 2007), in
cooling gas behind shocks produced during the ejection of gas from these
objects (Ferrara 1998). 
Ricotti, Gnedin \& Shull (2002) study the formation/dissociation of
H$_2$ by means of hydrodynamics simulations coupled with a radiative 
transfer scheme following the propagation of ionizing and dissociating radiation. 
Their main result is that the above positive feedback
is usually able to counterbalance the effect of H$_2$ dissociation. As a consequence,
the formation of small mass galaxies is not suppressed, as found by other authors. 
Cen (2005) has also proposed that supersonic collisions of cold atomic gas clouds,
which were originally minihalos and accreted into larger halos, can shock heat
and compress the gas, which then cools and forms H$_2$ molecules. This would 
enhance the star formation process.
Thus, a second burst of star formation might
take place also in the small objects where it has been suppressed
by H$_2$ dissociation.
H$_2$ production could also be promoted by an X-ray background, which would
increase the fractional ionization of protogalactic gas. Such a positive
feedback, though, is not able to balance UV photodissociation
in protogalaxies with $T_{vir}<2000$~K (Haiman, Abel \& Rees 2000;
Glover \& Brand 2003). 
Kuhlen \& Madau (2005) have simulated with the
code {\tt ENZO} the effect of the X-ray radiation produced by the first quasar
on the surrounding medium. They find that the net effect of X-ray is to reduce
the gas clumping with the consequence of suppressing gas infall at overdensities
$\delta < 2000$, but at higher overdensities molecular cooling is increased,
although this is not enough to overcome heating in the proximity of the quasar.
Similar arguments apply to cosmic rays (e.g.
Shchekinov \& Vasiliev 2004; Jasche, Ciardi \& En{\ss}lin 2007;
Stacy \& Bromm 2007). 

\begin{figure*}
\centerline{\includegraphics[width=4.7in]{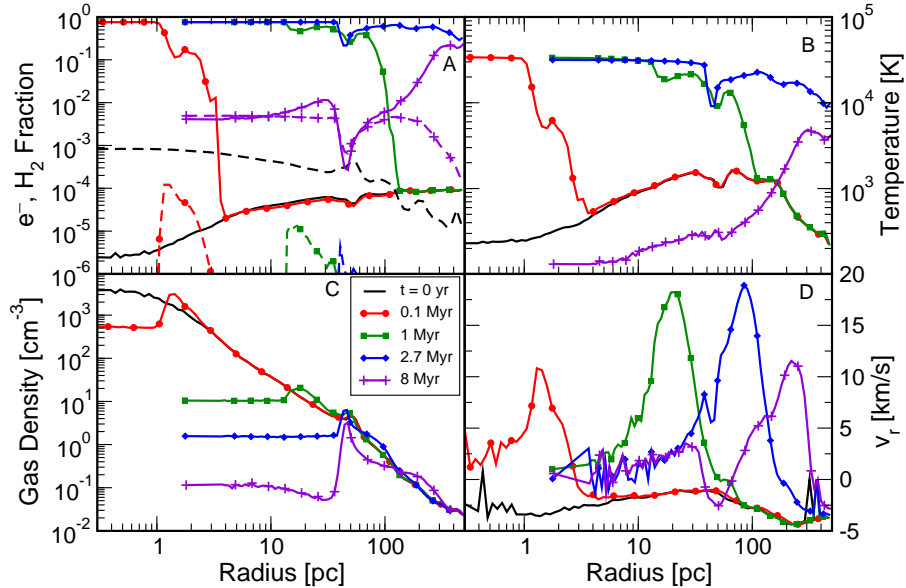}}
\caption{Mass weighted radial profiles around the position of a star with
mass $M_\star=100$~M$_\odot$. Panel A: electron number
fraction (solid) and H$_2$ mass fraction (dashed) for 5 different times,
0, 0.1, 1, 2.7, and 8 Myr after the star is born. The star dies at 2.7~Myr.
Panel B: temperature. Panel C: gas density. Panel D: radial velocity. (Abel, Wise \& Bryan 2007).}
\label{relicHII}
\end{figure*}
An example of positive feedback can be seen in Figure~\ref{relicHII}.

\vspace{0.3cm}
\centerline {\it Photoheating filtering}  
\vspace{0.3cm}

Cosmic reionization might have a strong impact on subsequent galaxy formation, 
particularly affecting low-mass objects. In fact, the heating associated with 
photoionization causes an increase in the temperature of the IGM gas which will 
suppress the formation of galaxies with masses below the Jeans mass. As Gnedin (2000b) 
has pointed out, one can expect that the effect of reionization depends on the 
reionization history, and thus is not universal at a given redshift. More precisely,
one should introduce a ``filtering'' scale, $k_F$, (or, equivalently, filtering mass 
$M_F$) over which the baryonic perturbations are 
smoothed as compared to the dark matter, yielding the approximate relation $\delta_b
 = \delta_{dm} e^{-k^2/k_F^2}$. The filtering mass as a function of time is related 
to the Jeans mass by:
\begin{equation}
M_F^{2/3} = {3\over a} \int_0^a da' M_J^{2/3}(a')\left[1-\left({a'\over a}\right)^
{1/2}\right].
\end{equation}
Note that at a given moment in time the two scales can be very different.   
Also, in contrast to the Jeans mass, the filtering mass depends on the full thermal history 
of the gas instead of the instantaneous value of the sound speed, so it accounts for the finite 
time required for pressure to influence the gas distribution in the expanding universe. 
The filtering mass increases from roughly $10^7$~M$_\odot$ at $z \approx 10$ to 
about $10^9$~M$_\odot$
at redshift $z\approx 6$, thus efficiently suppressing the formation of objects 
below that mass threshold. 
Of course such result is somewhat dependent on the assumed reionization history.

An analogous effect is found inside individual \HII 
regions around the first luminous sources. Once an ionizing source turns off, 
its surrounding \HII 
region Compton cools and recombines. Nonetheless, the ``fossil'' \HII regions 
left behind remain at high adiabats, prohibiting gas accretion and cooling in 
subsequent generations of Pop~III objects (Oh \& Haiman 2003). A similar entropy
floor could be obtained by heating the gas with Ly$\alpha$ photons (Ciardi \& 
Salvaterra 2007) or x-ray photons (Oh \& Haiman 2003), 
although the latter mechanism would also promote H$_2$ formation which in turn would 
increase the cooling (Kuhlen \& Madau 2005). 
The suppression of minihalos formation in photoionized regions is increased if the clustering
of the sources is considered (e.g. Kramer, Haiman \& Oh 2006).
O'Shea et al. (2005), on the other hand, find that it is possible for a
primordial star to form within an \HII region (similar results are found by
Nagakura \& Omukai 2005 by means of 1-D hydrodynamical calculation which include
also the effect of HD). By means of an Eulerian
adaptive mesh refinement simulation they find that the enhanced electron
fraction within the \HII region catalyzes H$_2$ formation that leads to faster
cooling in the subsequent star forming halos, although the accretion rate
is much lower. A similar result is found by Abel, Wise \& Bryan (2007) who, via
a hydrodynamic simulation with radiative transfer, follow the formation of 
a first star and its \HII region, and the consequent collapse inside it of a second
star. They also argue that an entropy floor in relic \HII regions is present,
but not as high as discussed by Oh \& Haiman (2003), because the typically high density 
gas does not get very heated.
This is confirmed by Mesinger, Bryan \& Haiman (2006), who
use a hydrodynamic simulation to study the collapse of minihalos inside \HII
regions. They find that there exist a critical intensity for the UV flux such
that for lower values H$_2$ formation is enhanced, while for higher star
formation is reduced. This critical value is a function of the gas density.

\begin{figure}
\centerline{\includegraphics[width=30pc]{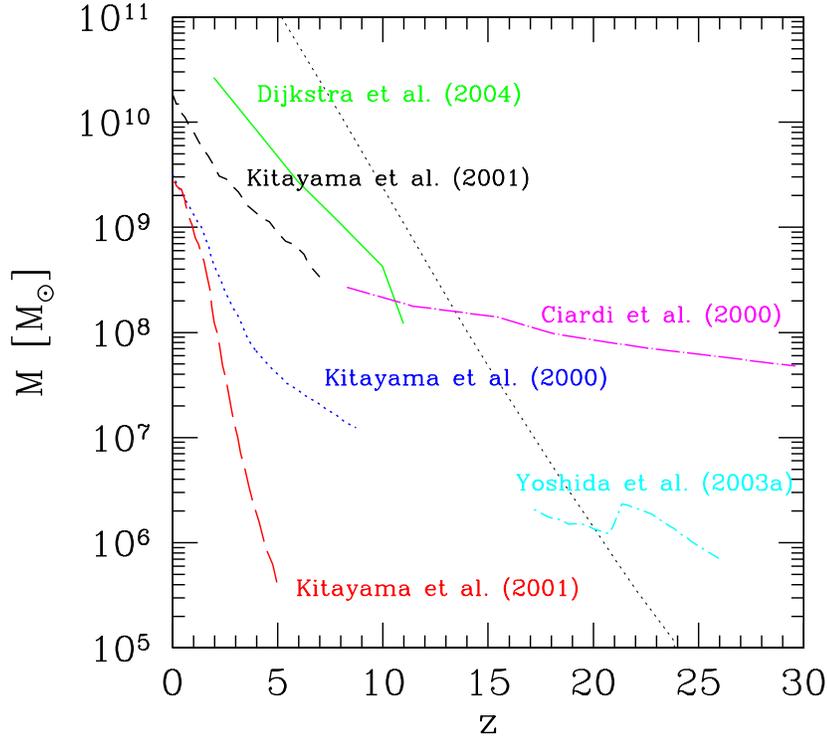}}
\caption{Mass of halos affected by radiative feedback in the formulation of
Ciardi et al. (2000) (long dashed-dotted line), Kitayama et al. (2000) (dotted),
Kitayama et al. (2001) (short dashed and long dashed), Dijkstra et al. (2004)
(solid) and Yoshida et al. (2003a) (short dashed-dotted). The straight dotted line
corresponds to the formation redshift of 3-$\sigma$ peaks in a $\Lambda$CDM model. 
For details refer to the text and the above papers.}
\label{feed_rad}
\end{figure}

\vspace{0.5cm}
Haiman \& Bryan (2006) interpret the low value of the Thomson scattering optical
depth inferred by the 3-yr (confirmed by the 5-yr) {\tt WMAP} release as an evidence for the efficiency of
radiative feedback on small mass object formation. In fact, the ionizing photon
production efficiency in these objects must have been low in order to not
overproduce the optical depth. Nevertheless,
a comprehensive conclusion on the impact of radiative feedback effects is still 
impossible, given the large number of different assumptions, approximations and
parameters used in the (sometimes specific) studies present in the literature. With 
the only aim to give a flavor of the large discrepancies, in
Fig.~\ref{feed_rad} we compare various predictions on  the mass of halos 
affected by radiative feedback (in some of the forms discussed above). 
There we show the results of Ciardi et al. (2000) (long dashed-dotted line), Kitayama et al. (2000)
(dotted), Kitayama et al. (2001) (short dashed and long dashed), Dijkstra et al. (2004)
(solid) and Yoshida et al. (2003) (short dashed-dotted); obviously, the comparison is
far from exhaustive. 
Here we simply summarize the main features addressed in those papers, while
for the details we refer the reader to them.                   
Ciardi et al. (2000) have derived the minimum mass a halo must
have to self-shield against an incident UV flux (in addition
to the background, the contribution from the direct flux from nearby objects is
included). They have studied the non-equilibrium multifrequency radiative transfer
of the incident spectrum in the approximation of a homogeneous gas layer composed
of H, H$^{-}$, H$^{+}$, He, He$^{+}$, He$^{++}$, H$_2$, H$_2^+$ and free electrons.
Kitayama et al. (2000) have studied the collapse of primordial objects in the
presence of an UV background radiation, by means of a 1-D spherical hydrodynamical 
calculation combined with a treatment of the radiative transfer of photons.
An analogous, but more refined calculation, has been performed by Kitayama et al. 
(2001), for different assumptions for the incident flux. The curve of Dijkstra et
al. (2004) is the mass corresponding to the case in which the presence of an
external ionizing flux reduces the gas infall of a factor of 2. The calculations
are performed by means of a 1-D spherically symmetric hydrodynamic code.
Finally, Yoshida et al. (2003a) derive the minimum mass of halos that
host gas clouds in their N-body/SPH simulations, which include the non-equilibrium
evolution of the relevant species. 

In summary, {\it the presence of a UV flux delays the collapse and cooling of small mass halos and the
amount of cold gas available primarily depends on the intensity of the flux
and the mass of the halo. Some controversy though still exists on the efficiency of such feedback.}

\subsection{Mechanical Feedback}

In addition to the radiative feedback, the mechanical feedback associated with
mass and energy deposition from the first stars, can deeply affect the subsequent
SF process in several ways. 

\vspace{0.3cm}
\centerline {\it Blowout and Blowaway}  
\vspace{0.3cm}

\begin{figure}
\centerline{\includegraphics[width=28pc]{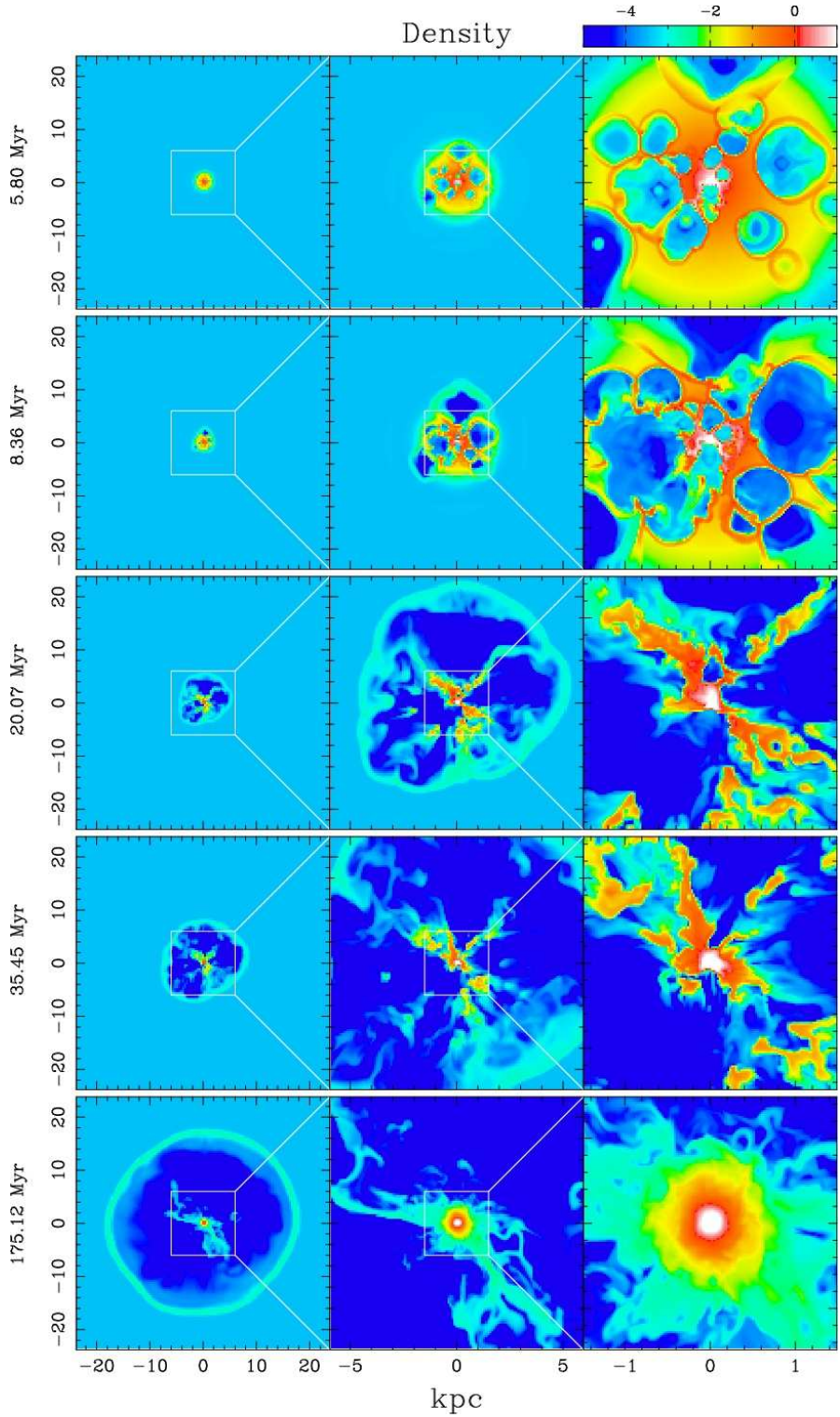}}
\caption{Snapshots of the simulated logarithmic gas number density
due to SN explosions (at five different elapsed times)
in a halo
with mass $M=10^8\,h^{-1} M_\odot$ at redshift $z=9$  (Mori, Madau \& Ferrara 2002).
The three panels in each row show the spatial density distribution in the $X-Y$
plane on the nested grids.
The left, middle, and right panels in each row correspond to the spatial
resolution levels.}
\label{explos}
\end{figure}
\begin{figure}
\centerline{\includegraphics[width=24pc]{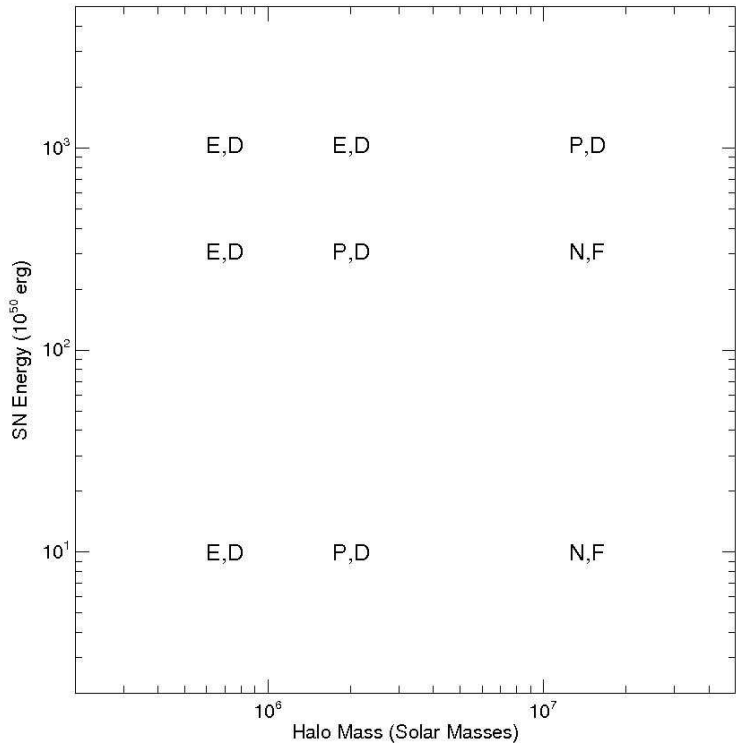}}
\caption{Eventual fate of a halo given the indicated explosion energy.  The first
letter refers to the final state of the halo prior to the explosion (E: photoevaporated;
P: partly ionized, defined as the I-front not reaching the virial radius; N: neutral, or
a failed \HII region).  The second letter indicates outcome of the explosion (D: destroyed,
or F: fallback). See Whalen et al. (2008) for details.}
\label{whalen2008}
\end{figure}

Depending on the mass and the dark matter
content of galaxies, SN and multi-SN events might induce partial ({\it blowout})
or total ({\it blowaway}) gas removal from the galaxy itself, regulating the
SF (e.g. Mac Low \& Ferrara 1999; Nishi \& Susa 1999; Ferrara \& Tolstoy 2000;
Springel \& Hernquist 2003; Bromm, Yoshida \& Hernquist 2003; Silk 2003;
Wada \& Venkatesan 2003; Greif et al. 2007; Whalen et al. 2008). 
It is found that only objects with masses
$M \lsim 5 \times 10^6$~M$_\odot$ can experience a complete blowaway, while
larger objects
have lower mass ejection. Thus, quenching SF in galaxies by ejecting
large fractions of their gas is very difficult.
These results have been substantiated and confirmed by Mori, Ferrara \& Madau (2002) 
who have performed a three-dimensional hydrodynamic simulation, using a nested grid method
to follow the evolution of explosive multi-SN events in an object of total mass
of $10^8$~M$_\odot$ at $z=9$, representing a 2-$\sigma$ fluctuation of the density power spectrum.
They find that, depending on the stellar distribution, less than 30\%
of the available SN energy is converted into kinetic energy of the escaping
material, the remainder being radiated away. It appears that mechanical feedback
is less efficient than expected from simple energetic arguments, as
off-nuclear SN explosions drive inward-propagating shocks that tend to collect
and pile up cold gas in the central regions of the host halo (see Fig. \ref{explos}). 
Thus, also relatively low-mass galaxies at early epochs may be able to retain
a considerable fraction of their gas and continue forming stars. A variant of this 
study, using the SPH method and considering a single supernova explosion in Pop~III 
objects at higher redshift, has been carried out by Bromm, Yoshida \& Hernquist (2003). 
These authors find that the more energetic PISN explosions
($M_\star =250$~M$_\odot$; $E\approx 10^{53}$~erg) lead to a blowaway (this has been
later confirmed by Greif et al. 2007 for a PISN with mass $M_\star =250$~M$_\odot$
and $E\approx 10^{53}$~erg), while the lower explosion energy 
corresponding to a single Type II SN  ($E\approx 10^{51}$~erg) leaves much of the halo intact.
Nevertheless, the extent to which such feedback can operate depends sensitively
on the initial configuration of the medium surrounding the explosion site and the
ionization front expansion around a massive progenitor star can significantly
aid gas evacuation by SN blastwaves (Kitayama \& Yoshida 2005), as can be seen in 
Fig.~\ref{whalen2008}.      

Following the ejection of gas, H$_2$ formation is favored in cooling gas 
behind the shocks, possibly balancing H$_2$ dissociation, as discussed in the
previous Section (Ferrara 1998). This process should be accessible to observations,
as the re-formed H$_2$ emits mid-IR roto-vibrational lines (as for example the 
restframe $2.2$~$\mu$m) detectable by {\tt JSWT} (Ciardi \& Ferrara 2001).

In some cases, SN explosions could promote star formation. In fact,
following an explosion, the interstellar gas is swept up and a dense shell
is formed immediately behind the radiative shock front at radius $R_{sh}$. 
When the dense shell gets dynamically overstable, it breaks into fragments, some of which
will contract and eventually form stars (see e.g. Vishniac 1983;
McCray \& Kafatos 1987; Nakano 1998; Tsujimoto, Shigeyama \& Yoshii 1999;
Reynoso \& Mangum 2001; Mackey, Bromm \& Hernquist 2003; Salvaterra, Ferrara \& Schneider 2003). 
Shell expansion tends to suppress (stabilize) the growth
of  density perturbations owing to stretching, hence counteracting the
self-gravity pull. 
The instantaneous maximum growth rate is:
\begin{equation}\label{eq:w}
\omega=-\frac{3\dot{R}_{sh}}{R_{sh}}+\left[\left(\frac{\dot{R}_{sh}}{R_{sh}}\right)^2+\left(\frac{\pi G\rho_0 R_{sh}}{3c_{s,sh}}\right)^2\right]^{1/2},
\end{equation}
where $c_{s,sh}=(kT_{sh}/\mu m_H)^{1/2}$ is the sound speed in the shell 
and $\rho_0$ the ambient 
gas density. Instability occurs only if $\omega>0$, or
\begin{equation}
\label{eq:w2}
\frac{\dot{R}_{sh}}{R_{sh}}<\frac{1}{8^{1/2}}\frac{\pi G\rho_0 R_{sh}}{3c_{s,sh}}\propto \frac{t_{cross}}{t_{ff}^2},
\end{equation}
where $t_{cross}\sim R_{sh}/c_{s,sh}$ is the crossing time in the shell and $t_{ff}$
is the free-fall time. Furthermore, if the fragment mass is not much larger than 
the Jeans
mass, the $\omega >0$ condition translates into  $\dot{R}_{sh}/R_{sh}<1/t_{ff}$.
Instability occurs only if $\dot{R}_{sh}/R_{sh}<t_{cross}/t_{ff}^2$, where 
$R_{sh}$ is the shell radius, $t_{cross}$ is the crossing time in the shell and $t_{ff}$
is the free-fall time.
Hence, large shell velocity-to-radius ratios inhibit the formation of
gravitationally unstable fragments. Under some conditions (see e.g. Salvaterra, 
Ferrara \& Schneider 2003) the above instability criterion 
is satisfied and (typically low-mass) stars are formed as a result of shell fragmentation.

\vspace{0.3cm}
\centerline {\it Impinging Shocks}  
\vspace{0.3cm}

The formation of a galaxy can be inhibited also by the effects of shocks 
from neighboring objects, causing heating, evaporation and/or
stripping of the baryonic matter (Scannapieco, Ferrara \& Broadhurst 2000).
In the former scenario (mechanical evaporation), the gas in a forming
galaxy is heated above its virial temperature by the shock. 
The thermal pressure of the gas then
overcomes the dark matter potential and the gas expands out of the
halo, preventing galaxy formation.  Only if the cooling
time of the collapsing cloud is shorter than its sound crossing
time, will the gas cool before it expands out of the gravitational
well and continue to collapse.
In the latter scenario, the gas may be stripped from a collapsing 
perturbation by a shock from a nearby source. In this case, the momentum of the
shock is sufficient to carry the gas with it, 
emptying the halo of its baryons and preventing a galaxy from forming.

In practice, the short cooling times for
most dwarf-scale collapsing objects suggest that the baryonic stripping
scenario is almost always dominant. This mechanism has the largest impact in
forming dwarfs in the $\simlt 10^9$~M$_\odot$ range, which is sufficiently
large to resist photoevaporation by UV radiation, but too small to
avoid being swept up by nearby outflows. Sigward, Ferrara \& Scannapieco (2005) studied
this problem by means of numerical simulations, finding that, if the forming galaxy
is virialized, the impinging shock has a negligible effect on its evolution, while,
if it is in the turnaround stage, up to 70\% of its gas content can be stripped away.  

\subsection{Chemical Feedback}
\label{chemfeed}

\begin{figure}
\centerline{\includegraphics[width=26pc]{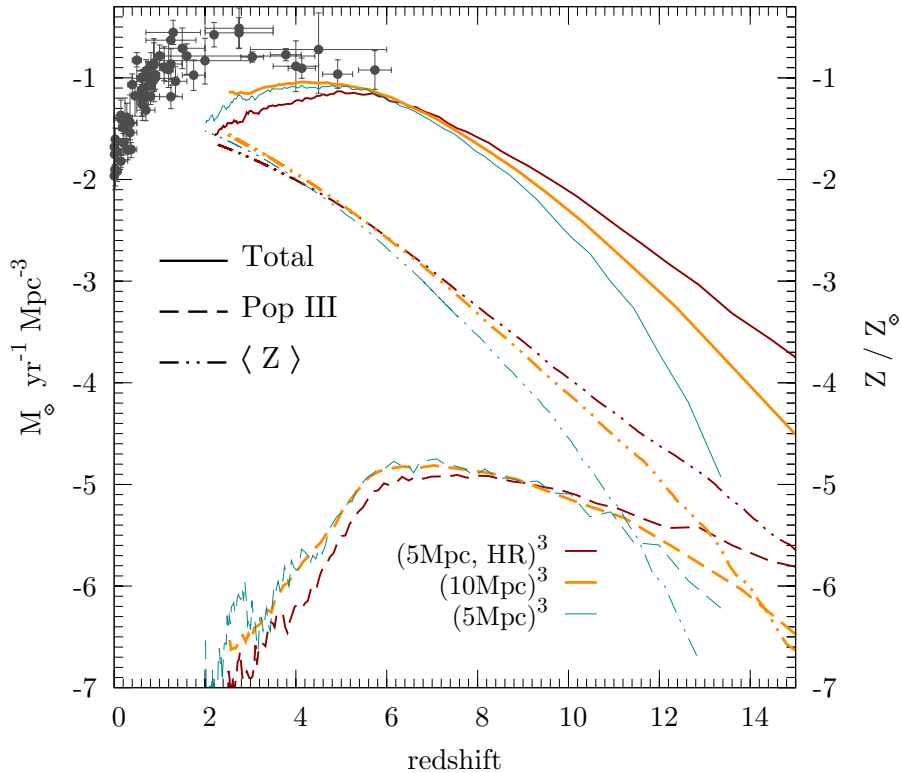}}
\caption{
Predicted evolution of Pop II (solid lines) and Pop III
(dashed) cosmic star formation rates, and mass-averaged metallicity
(dot-dot-dashed) from Tornatore, Ferrara \& Schneider (2007). The results
of three simulation runs at different resolution are shown for each quantity.
As a reference, low-redshift measurements (points) taken from Hopkins (2004) are
  reported.
}
\label{fig:sfr_torna}
\end{figure}
The concept of chemical feedback is relatively recent, having been first explored
by Bromm et al. (2001) (following the early study by Yoshii \& Sabano 1980) and
Schneider et al. (2002), and subsequently discussed by Schneider et al. (2003),
Mackey, Bromm \& Hernquist (2003) and others.
 
According to the scenario outlined in Sec.~\ref{firststars}, the first stars forming out of
gas of primordial composition might be very massive, with masses
$\approx 10^2-10^3$~M$_\odot$. 
The ashes of these first supernova explosions
pollute with metals the gas out of which subsequent generations of low-mass Pop~II/I stars
form, driving a transition from a top-heavy IMF to a ``Salpeter-like'' 
IMF when locally
the metallicity approaches the critical value $Z_{cr}=10^{-5\pm1}Z_\odot$ 
(Schneider et al. 2002, 2003). The uncertainty in the value of the 
critical metallicity is due to the role of dust cooling, which has been
only recently quantitatively assessed by Schneider et al. (2006b). In brief, 
metals depleted onto dust grains enable fragmentation to solar or sub-solar mass scales already
at metallicities $Z=10^{-6} Z_\odot$; on the contrary, in the absence of dust grains, even at 
$Z= 0.01 Z_\odot$ metals diffused in the gas-phase lead to fragment mass scales which are
still of the order of 100 solar masses. 

Thus, the cosmic relevance of Pop~III stars and the transition to a Pop~II/I star
formation epoch depends on the efficiency of
metal enrichment from the first stellar explosions, the so-called chemical feedback,
which is strictly linked to the number of Pop~III stars that explode as PISN,
the metal ejection efficiency, transport and mixing in the IGM.
Depending on the strength of the chemical feedback Pop~III star formation
could last for a very short time and their contribution to the ionizing photon
production could be negligible (see e.g. Ricotti \& Ostriker 2004a for a discussion).
It is very likely though that the transition occurred
rather smoothly because the cosmic metal distribution is observed
to be highly inhomogeneous: even at moderate
redshifts, $z \approx 3$, the clustering properties of 
C~$\scriptstyle\rm IV\ $ and Si~$\scriptstyle\rm IV\ $ QSO absorption systems
are consistent with a metal filling factor $ < 10\%$, showing that metal enrichment is
incomplete and inhomogeneous (see Sec. \ref{reionmet}). 

As a consequence, the use of the critical metallicity as a global criterion
is somewhat misleading because chemical feedback is
a {\it local process}, with regions close to star formation sites rapidly becoming
metal-polluted and overshooting $Z_{\rm cr}$, and others remaining essentially
metal-free.  Thus, Pop~III and Pop~II star formation modes
could have been coeval, and detectable signatures from Pop~III stars
could be found well after the volume-averaged metallicity has become larger
than critical.

In a seminal paper Scannapieco, Schneider \& Ferrara (2003) studied, using an analytical model
of inhomogeneous structure formation, the separate evolution of Pop~III/Pop~II 
stars as a function of star formation and wind efficiencies. Their main finding was
that, essentially independent of the free parameters of the model,  Pop~III stars continue to 
contribute appreciably to the star formation rate density at much lower redshift.  
These findings have been now confirmed by additional numerical work (Tornatore, Ferrara \& Schneider
2007) and are summarized in Fig. \ref{fig:sfr_torna}, where the relative contribution
of Pop~III and Pop~II to the cosmic star formation is shown along with the mean IGM metallicity 
evolution.  Due to inefficient heavy element transport by outflows and slow "genetic" transmission 
during hierarchical growth, large fluctuations around the average metallicity arise; as a result
Pop III star formation continues down to $z=2.5$, but at a low peak rate of $10^{-5}$~M$_\odot 
{\rm yr}^{-1} {\rm Mpc}^{-3}$ occurring at $z\approx 6$ (about $10^{-4}$ of the Pop~II one). 
This finding has important implications for the development of efficient strategies for the
detection of Pop~III stars in primeval galaxies, as discussed previously.  A similar result has 
been found by Wyithe \& Cen (2007).

\section{Cosmic Reionization and IGM Metal Enrichment}
\label{reionmet}

Although cosmic reionization has received great attention in the last
decade due to the availability of high-$z$ absorption line studies and progresses
in the CMB experiments, the nature of the ionizing sources, the reionization
history and its effect on structure formation remain unclear and highly debated.
This is mostly due to uncertainties in the modeling of several physical issues:
properties of first stars and quasars, ionizing photon production and radiative
transfer, just to mention a few. 
Whatever their nature, only a fraction of the emitted ionizing photons
is able to escape from their production site and reach the IGM. Thus, the knowledge
of the escape fraction is crucial for any theoretical modeling of the
reionization process. The same stars that produce the ionizing radiation are
expected to pollute the IGM with heavy elements. In the following we will discuss
the present understanding of the above processes. 

\subsection{Escape Fraction}

The first attempt to derive a value of the escape fraction, $f_{esc}$, 
for a Milky Way-type galaxy
dates to Dove \& Shull (1994), who assumed a smoothly varying \HI 
galactic distribution
and OB associations located in the Galactic plane. By integrating the fraction of
Lyman continuum (Lyc) photons that escapes the disk over the adopted luminosity
function of OB associations, they estimated $f_{esc} \approx 14\%$.
Dove, Shull \& Ferrara (2000) later improved the
calculation by solving the time-dependent transfer problem
of stellar radiation through evolving superbubbles. Their main result
is that the shells of the expanding superbubbles
quickly trap ionizing photons, so that most of the radiation escapes only
after the superbubbles have broken out of the Galaxy, when however the ionizing
power of the central stellar cluster has already largely faded away.
This results in an escape fraction roughly a factor of 2 lower
than the one obtained by Dove \& Shull (1994), although the exact value
depends on the star formation history.
An even more refined approach is the one by Fujita et al. (2003), in which
the effects of repeated supernova explosions, including the formation and
evolution of superbubbles, are modeled using a hydrodynamical simulation.
They confirm that the shells can trap ionizing radiation very effectively until
the bubbles start to accelerate, causing the shells to fragment.
The values found for $f_{esc}$ are roughly consistent with the
Dove, Shull \& Ferrara (2000) estimates.
All the above studies model the idealized case of a smoothly varying \HI
distribution and sources positioned on the Galactic plane. A more realistic
case         would give a higher value of the escape fraction.
The effects of gas density inhomogeneities have been studied by
Ciardi, Bianchi \& Ferrara (2002) for a Milky Way-type galaxy. To this aim,
 a comparison between a smooth Gaussian
distribution and an inhomogeneous fractal one has been made, including realistic
assumptions for the position and emission properties of the ionizing stellar sources
based on the available data, and the 3-D radiative transfer of ionizing photons.
While a fractal interstellar medium results in an escape fraction roughly constant over
a wide range of ionization rates, in a Gaussian distribution $f_{esc}$
decreases with decreasing rates.
Based on a model that provides a good fit to the observed size distribution
of \HI holes in nearby galaxies, Clarke \& Oey (2002) derive a simple
relationship between the star formation rate and the interstellar porosity
of a galaxy. This gives a critical star formation rate for which the porosity
is of the order of 1. The authors expect high escape fractions only in galaxies whose
star formation rates exceed the critical value, as for example Lyman Break Galaxies
(LBGs).

\begin{figure}
\centerline{\includegraphics[width=13pc]{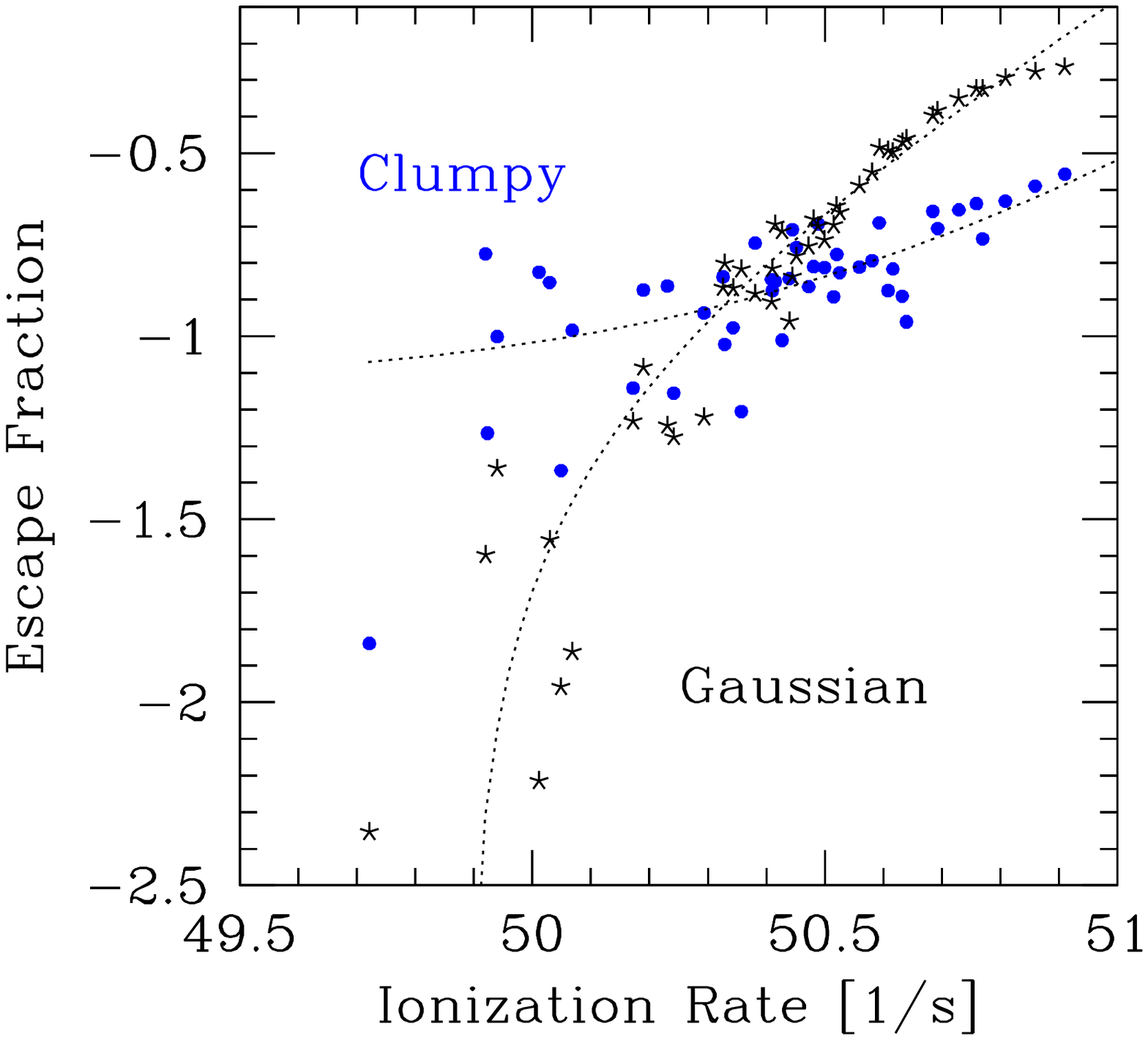}
\hspace{-1.2truecm}  \includegraphics[width=10.5pc]{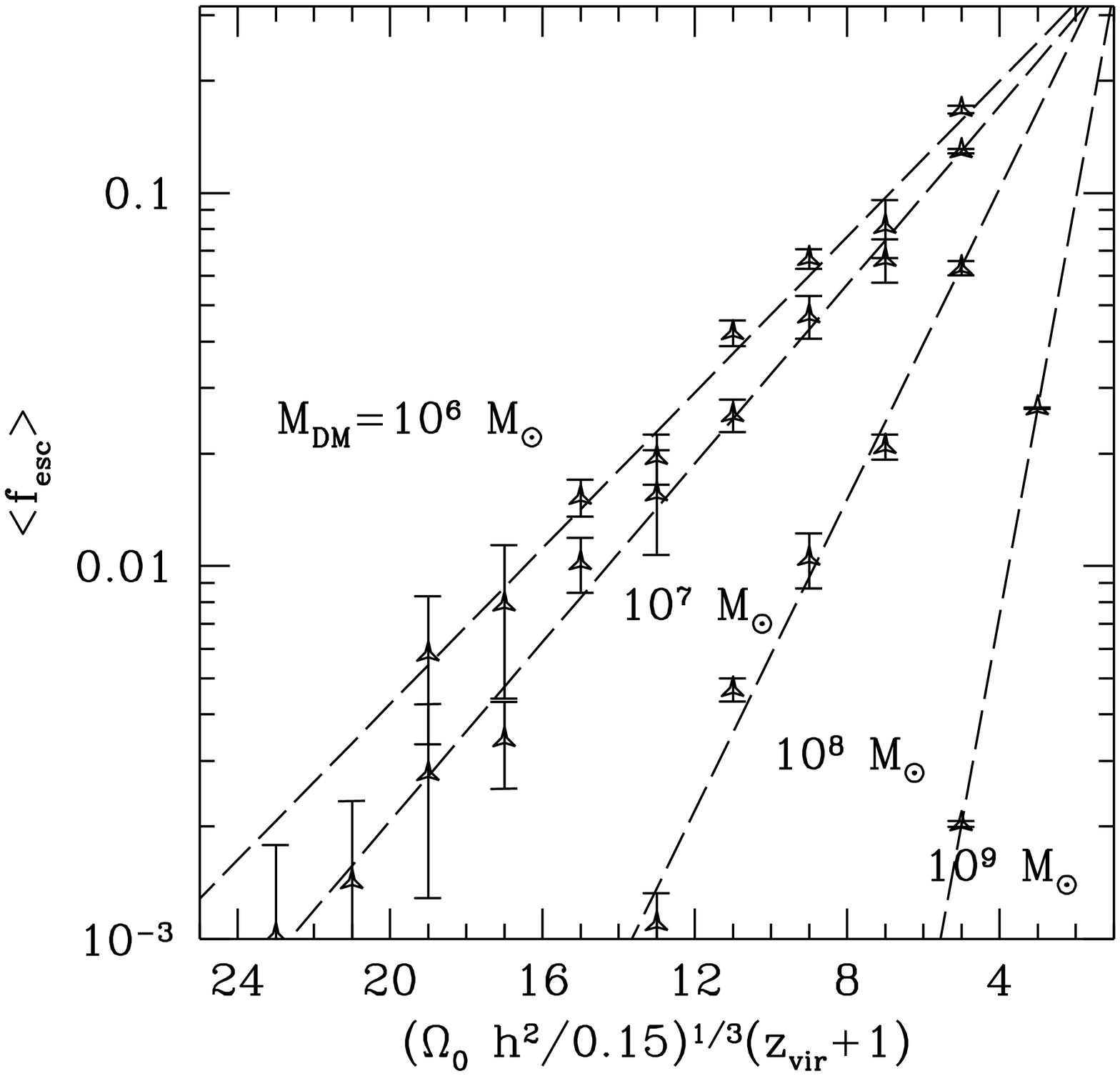}
\hspace{-0.8truecm} \includegraphics[width=13pc]{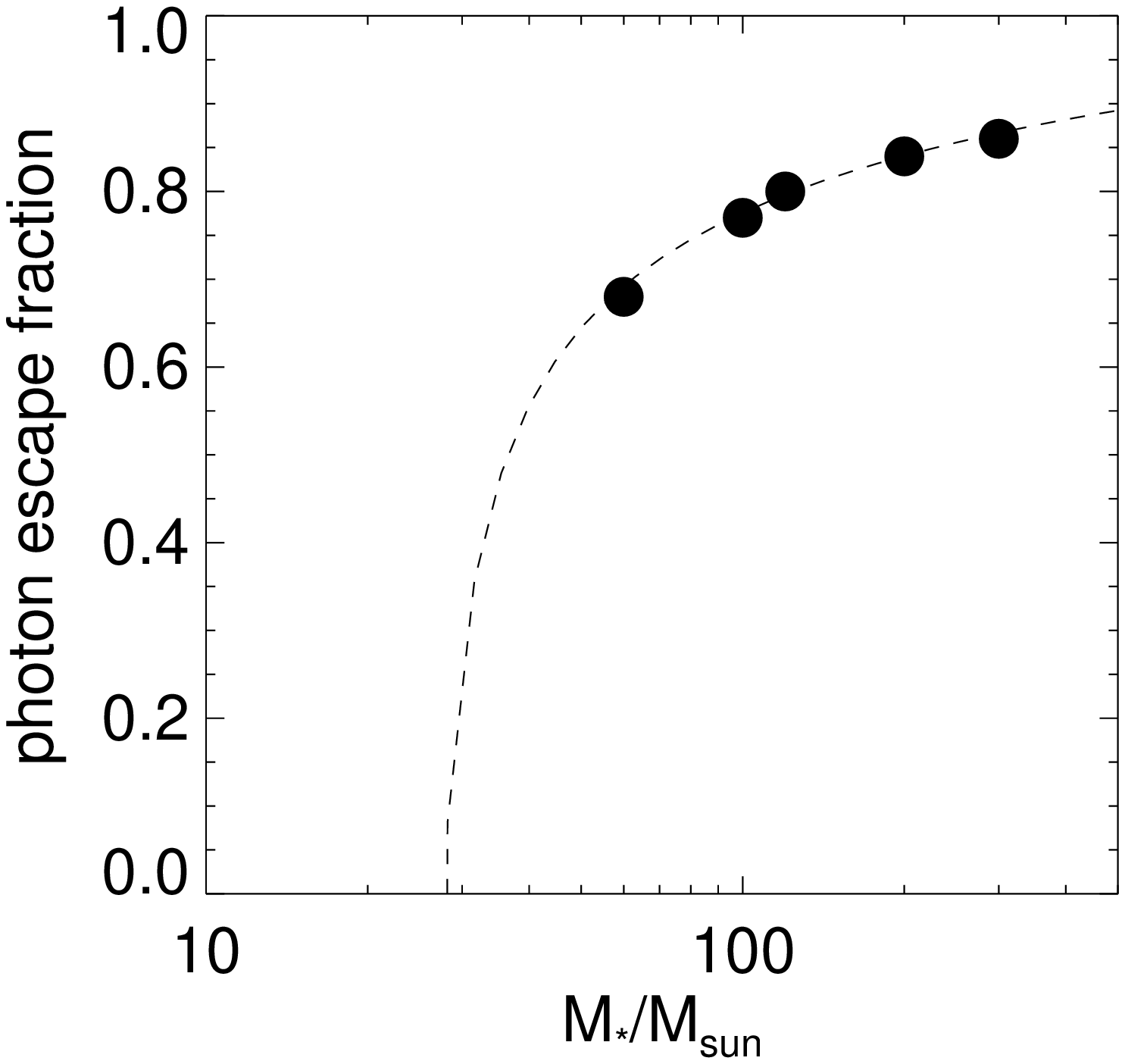}}
\caption{Left panel: Escape fraction as a function of the ionization rate for a clumpy
(circles) and a Gaussian (asterisks) density distribution (Ciardi, Bianchi \& Ferrara 2000).
Central panel: Redshift evolution of the escape fraction for different masses of the
host dark matter halo (Ricotti \& Shull 2000). Right panel: Evolution of the escape fraction
as a function of the mass of the first, massive star (Yoshida et al. 2007).}
\label{fesc}
\end{figure}

Additional theoretical works (Ricotti \& Shull 2000; Wood \& Loeb 2000;
Fujita et al. 2003) have extended the analysis to include high-redshift galaxies,
for which the escape of Lyc photons can be modified by geometrical effects (i.e.
disk vs. spheroidal systems) and by the higher mean galactic Interstellar Medium
(ISM) density.
Ricotti \& Shull (2000) use models of smoothly distributed gas in hydrostatic
equilibrium and estimate $f_{esc}$ by integrating the escaping radiation of each
single OB association over the luminosity function of the OB associations.
They perform a parametric study to understand the dependence of $f_{esc}$ on
redshift, mass, star formation efficiency, stellar density distribution and
OB association luminosity function.
Generally speaking, the escape fraction decreases as the object virialization
redshift or mass become larger or the rate of Lyc photons emitted smaller.
Their analysis is confined to objects with $M\approx 10^{6-9}$~M$_\odot$, but 
a similar result is found by Wood \& Loeb (2000) for objects with 
$M\approx 10^{9-12}$~M$_\odot$.
They improved on the previous calculation by following the propagation
of ionizing photons with a 3-D radiation transfer code and considering
also a case of a two-phase medium. 
They find that the escape fraction is enhanced in a clumpy ISM. Although the derived 
escape fraction is lower than 1\% at $z\approx 10$ for stellar sources, for 
miniquasars, which have higher luminosities and harder spectra, $f_{esc}$ can be 
$\simgt$ 30\% at the same redshift. The higher escape fraction expected for quasars
suggests that, if sufficiently abundant, they might have played a crucial role
in the reionization of the IGM (see next Section).
A similar trend with the mass and the virialization redshift of the galaxies is
found for single metal-free very massive stars (Kitayama et al. 2004), by means
of 1-D hydrodynamic code coupled with radiative transfer.
In contrast with the above works,
Fujita et al. (2003), in the previously mentioned paper, find $f_{esc}
>20$\% at $z>5$. This reversed trend would be due to the fact that ionizing
photons escape more easily
from high-redshift disks because the gas distribution is  highly stratified under
the strong gravitational potentials of the compact, high-redshift halos, and this
allows quick acceleration of the superbubbles. 
The same trend is found by Razoumov \& Sommer-Larsen (2006) who find that $f_{esc}$
increases from 1-2\% at $z=2.39$ to 6-10\% at $z=3.6$ by postprocessing high-resolution
simulations of galaxies with radiative transfer. A value of few \% for galaxies with
$M > 10^{11}$~M$_\odot$ and SFR $\approx 1-5$~M$_\odot$~yr$^{-1}$ over the redshift
range $3-9$ is found by Gnedin, Kravtsov \& Chen (2008) by means of AMR simulations
with on-the-fly radiative transfer.
Escape fractions larger than 70\%
as high as 100\% (depending on the mass of the star and the hosting halo) 
are instead expected from very massive stars
at high redshift (Whalen, Abel \& Norman 2004; Alvarez, Bromm \& Shapiro 2006;
Yoshida et al. 2007).
Thus, {\it the majority of theoretical models converge  on values $f_{esc}
<$15\% (the number though depends e.g. on mass of the host object, redshift, gas distribution),
but a clear consensus has not yet been reached, although
the first, relatively- or very-massive stars seem to have $f_{esc}>$70\% and as high
as 100\%.} For an example of the dependence of the escape fraction see Fig.~\ref{fesc}.

\subsection{Hydrogen Reionization}

It is commonly believed that the reionization process proceeds through different
stages. According to the terminology introduced by Gnedin (2000a), during
the initial, pre-overlap stage, the hydrogen throughout the universe is
neutral except for isolated \HII regions due to individual sources. Following
is the overlap stage, during which individual \HII regions overlap 
and reionize the low-density, diffuse intergalactic gas.
Some neutral hydrogen remains in dense
clumps which are then slowly reionized during the post-overlap stage.
The epoch of complete reionization, $z_{ion}$, is somewhat dependent on the definition
of ``complete''. A possible definition could be the redshift at which the mean 
free path to the ionizing radiation is of the order of the Hubble radius, or the
time derivative of the mean free path has a peak. A more practical definition,
associated with a quantity more easily measurable,
is that $z_{ion}$ corresponds to the epoch when the volume weighted neutral
hydrogen fraction becomes $< 10^{-3}$.
Increasing attention has been dedicated to the theoretical modeling of the
reionization process, adopting semi-analytical (from the pioneering
Shapiro \& Giroux 1987;
Fukugita \& Kawasaki 1994; Miralda-Escud\'e \& Rees 1994 and
Tegmark, Silk \& Blanchard 1994
to the more recent Haiman \& Loeb 1997; Madau, Haardt \& Rees 1999;
Valageas \& Silk 1999; Cojazzi et al. 2000;
Chiu \& Ostriker 2000; Miralda-Escud\'e, Haehnelt \& Rees 2000;
Wyithe \& Loeb 2003a; Cen 2003a; Liu et al. 2004; Furlanetto, Zaldarriaga \&
Hernquist 2004a; Furlanetto \& Oh 2005; Benson et al. 2006; Choudhury \& Ferrara 2006;
Cohn \& Chang 2007) 
perturbative or semi-numerical (e.g. Zhang, Hui \& Haiman 2007; Mesinger \& Furlanetto 2007) 
and numerical approaches 
(e.g. Gnedin \& Ostriker 1997; Gnedin 2000a; Ciardi et al. 2000;
Razoumov et al. 2002; Ciardi, Stoehr \& White 2003; Ciardi, Ferrara \& White 2003;
Ricotti \& Ostriker 2004a; Sokasian et al. 2003; Iliev et al. 2006a; Trac \& Cen 2007;
Croft \& Altay 2008).
Two main ingredients are required for a proper modeling of the reionization
process: {\it (i)} a {\it reliable model of galaxy formation} and {\it (ii)} an 
{\it accurate treatment of the radiative transfer of ionizing photons.}

Despite many applications of hydrodynamical simulations to the structure
formation process, much of our current understanding comes from semi-analytical models,
based on simplified physical assumptions. These models have the advantages
of allowing a larger dynamic range and of being fast enough to
explore a vast range of parameters. Their main disadvantage
is that no information on the spatial distribution of structures
is provided. For this reason a new method has recently been developed
that combines N-body simulations of the
dark matter component, with semi-analytical models predicting galactic properties
``a posteriori'' (e.g. Kauffmann et al. 1999). The limitation of the
above methods is related to the resolution of the N-body simulations,
which turns out to be critical in studies of the
reionization process. In fact, the resolution must be high enough to
follow the formation and evolution of the objects responsible for
producing the bulk of the ionization radiation. At the same time, a
large simulation volume is required to have a region with ``representative''
properties and to avoid biases due to cosmic variance. It should be
noted, however, that although the contribution of small mass objects to
the ionizing photon production might be negligible due to feedback
effects, they are nevertheless sinks of ionizing radiation (e.g. Barkana
\& Loeb 2002 and references therein). Thus, any numerical simulation which is 
not able to resolve these objects is likely to underestimate the gas clumping
factor, and consequently the recombination rate, resulting in an earlier reionization
epoch. These minihalos introduce a substantial cumulative opacity to
ionizing radiation, unless they are photoevaporated by UV radiation quickly
enough (Barkana \& Loeb 2002; Shapiro, Iliev \& Raga 2004; see previous Section).
Ciardi et al. (2006) demonstrated by means of numerical simulations of cosmic
reionization including minihalos, that, depending on the
details of the minihalo formation process, their effect can delay complete 
reionization by $\Delta z\sim 0-2$.

At the present state, while the evolution of dark matter structures is well
understood, the treatment of the physical processes that govern the
formation and evolution of the luminous objects (e.g. heating/cooling of
the gas, star formation, feedback) is more uncertain. In particular,
no study treats the entire range of the feedback effects that can
influence the galaxy formation process (see previous Section).
A first attempt to treat self-consistently a number of feedback effects ranging
from the mechanical energy injection to the H$_2$ photodissociating radiation
produced by massive stars has been done by Ciardi et al. (2000), followed by the
more recent Ricotti, Gnedin \& Shull (2002) and Yoshida et al. (2003a), but the complexity
of the network of processes makes their implementation not trivial.

The next challenge for this kind of study is to develop accurate and fast
radiative transfer schemes that can be then implemented in cosmological
simulations. The full solution of the seven dimension radiative transfer
equation (three spatial coordinates, two angles, frequency and time)
is still well beyond our computational capabilities
and, although in some specific cases it is possible to reduce its
dimensionality, for the reionization process no spatial symmetry
can be invoked. Thus, an increasing effort has been devoted to the
development of radiative transfer codes based on a variety of approaches
and approximations (e.g. Umemura, Nakamoto \& Susa 1999; Razoumov \& Scott 1999;
Abel, Norman \& Madau 1999; Gnedin 2000a; Ciardi et al. 2001; Gnedin \& Abel 2001;
Sokasian, Abel \& Hernquist 2001; Cen 2002; Maselli, Ferrara \& Ciardi 2003;
Whalen \& Norman 2006; Mellema et al. 2006a; Qiu et al. 2006; Rijkhorst et al. 2006; 
Susa 2006; Pawlik \& Schaye 2008;
for a comparison between different codes see Iliev et al. 2006b).
The first attempt to include the radiative transfer in simulations
of the reionization process has been done by Gnedin \& Ostriker (1997).
Ideally, radiative transfer codes should be eventually
coupled with a hydrodynamic simulation. Although such an implementation has
already been attempted by Gnedin (2000a), the Local Optical Depth approximation
adopted for radiative transfer is less accurate for optical depths
$\simgt 1$ and therefore may not be suitable to describe in detail the 
entire reionization history. 
A similar implementation, with the Optically Thin Variable Eddington Tensor 
approximation for radiative transfer, is used by Ricotti \& Ostriker (2004a).
Although in some respect these two works improve on previous calculations
(they couple radiative transfer with hydrodynamic simulations
and include some feedback effects), 
their conclusions might be affected by possible uncertainties introduced by 
the small cosmic volumes (maximum box size of 4~$h^{-1}$ Mpc) considered 
(see e.g. Barkana \& Loeb 2004 for a discussion of box dimension in simulations
of cosmic reionization). Iliev et al. (2006a) have in fact clearly shown that
only boxes larger than 20-30$h^{-1}$~Mpc can properly describe the global
reionization process, while smaller boxes exhibit a large scatter. 
For this reason, boxes with such dimensions are needed (see Trac \& Cen 2007
for a self-consistent calculation on large boxes).
The largest simulations run (applied to the study of high-$z$ QSOs spectra) have 
1280$h^{-1}$~Mpc side (Kohler, Gnedin \& Hamilton
2007), but the resolution is very poor and results from smaller simulations are
used to include the effect of scales below the resolution limit.

\begin{figure}
\centerline{\includegraphics[width=24pc]{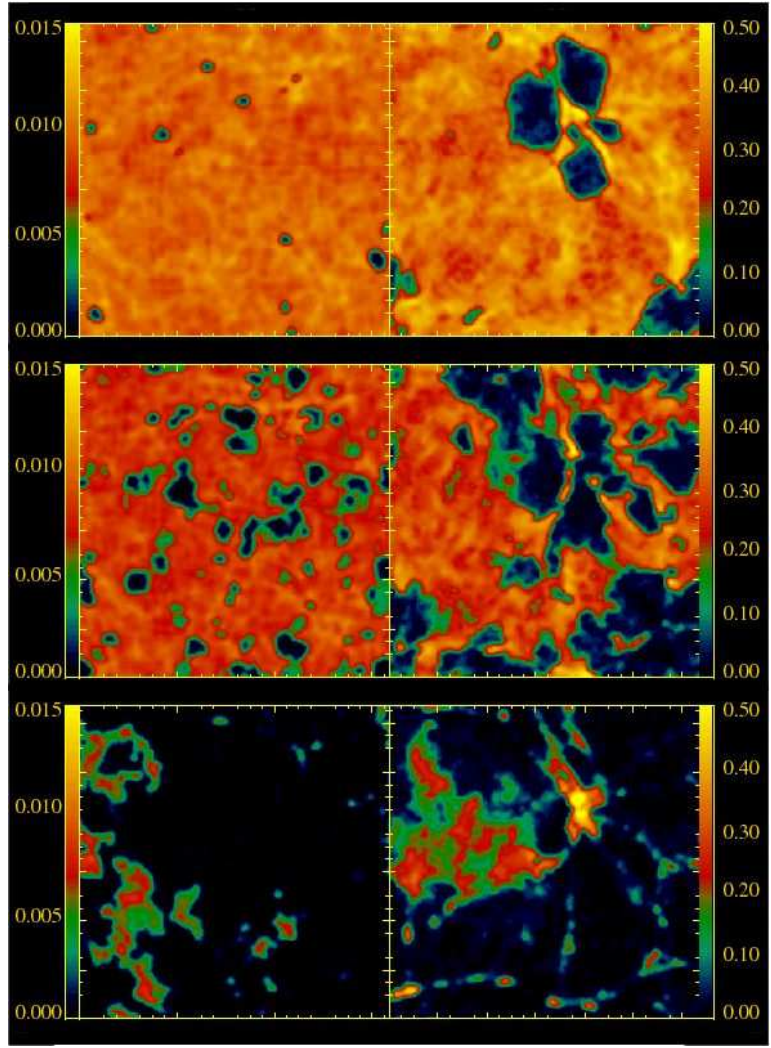}}
\caption{Reionization maps showing the redshift evolution of the number density
of neutral hydrogen in the field ($L=20 h^{-1}$~Mpc comoving; left panels) and
in a protocluster environment ($L=10 h^{-1}$~Mpc comoving; right panels). From
top to bottom the redshift of the simulations is 15.5, 12 and 9 (Ciardi, Stoehr \&
White 2003; Ciardi, Ferrara \& White 2003).} 
\label{reionmaps}
\end{figure}

An illustrative example of models of the reionization process is shown
in Fig.~\ref{reionmaps}. The Figure refers to the simulations by Ciardi,
Stoehr \& White (2003), employing a combination of high resolution N-body
simulations, a semi-analytical model of galaxy formation and the radiative
transfer code {\tt CRASH} (Ciardi et al. 2001; Maselli, Ferrara \& Ciardi
2003). The maps are indicative of how the reionization process is affected
by the environment: although in a protocluster environment 
the ionizing photon production per unit mass is higher at high redshift than
in a field region, as the high-density
gas, more common in a protocluster, is more difficult to ionize and recombines
much faster, filaments of neutral gas are still present
after the field region is almost completely ionized.

The reionization process is affected not only by the environment, but
also by the scale considered. In fact, while on scales of order of few Mpc
reionization might be complete, on larger scales it could still be quite
inhomogeneous due to the large scatter in the abundance of star-forming
galaxies at early epochs (Barkana \& Loeb 2004). 
A useful mathematical tool to study such complex topology is constituted by the 
Minkowski functionals (Gleser et al. 2006)  of isodensity surfaces which can be used to discriminate 
among different reionization histories.
In general, we expect that the properties of the \HII regions depend on many
poorly constrained quantities. McQuinn et al. (2007) have investigated this
issue by means of radiative transfer simulations on large scales, 
including small scale physics via analytic prescriptions. They find that the
morphology of \HII regions is similar for fixed values of the mean ionization
fraction, with other parameters playing a secondary role (the next most
important dependence is on the nature of the ionizing sources, with larger
and more spherical regions for rarer sources).
Clearly topology of the \HII regions, clustering of sources and the clumpiness of 
the IGM are key issues to understand reionization in detail. These have been investigated 
by Furlanetto \& Oh (2005). Their main result is that the radiation background at
any point evolves through a series of discrete jumps until it saturates at the mean 
free path, when a region becomes recombination-limited; afterwards the evolution proceeds
very slowly. Towards the end of reionization fluctuations in the mean free path within 
large contiguous \HII regions dominates over the clustering of the sources. 
However, Liu et al. (2006) noted that if the UVB evolves rapidly the following facts 
can be explained without invoking a fluctuating UVB at $z\sim 6$: rapid increase and large scatter
in the Gunn-Peterson optical depth, long-tail distribution of the transmitted flux, long dark 
gaps in spectra.

{\it The nature of the sources responsible for the IGM reionization is still the
subject of a lively debate, but most theoretical models adopt stellar type
sources, including a variety of spectra and IMF.}
These have to be constrained on the basis of the available observations: {\it (i)} the
value of the IGM temperature from the Ly$\alpha$ forest at $z\approx 2-4$
(e.g. Bryan \& Machacek 2000; Ricotti, Gnedin \& Shull 2000; Schaye et al. 2000b;
McDonald et al. 2001; Meiksin, Bryan \& Machacek 2001; Zaldarriaga, Hui \&
Tegmark 2001); {\it (ii)} the abundance of neutral hydrogen at $z>6$ from the spectra
of high-redshift QSOs or galaxies (e.g. Fan et al. 2000, 2001, 2003, 2006) and {\it (iii)} the 
measurements of the Thomson scattering optical depth (e.g. Kogut et al. 2003;
Spergel et al. 2003; Page et al. 2007; Nolta et al. 2008). A simple scenario in which the IGM 
reionized early ($z_{ion}>10$, as suggested by the last condition) and remains so thereafter, 
is ruled out as the IGM would reach an asymptotic thermal state too cold
compared to observations (Theuns, Schaye \& Haehnelt 2000; 
Hui \& Haiman 2003; Wyithe \& Loeb 2003b).  

\begin{figure}
\centerline{\includegraphics[width=19pc, angle=-90]{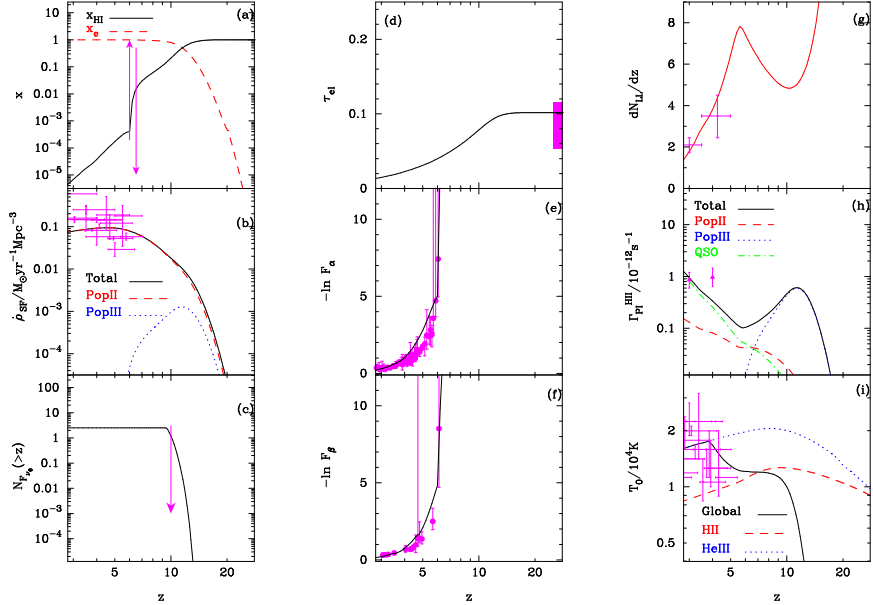}}
\caption{The best-fitting model Choudhury \& Ferrara (2006). The panels show
the redshift evolution of: (a) the volume-averaged electron and HI fraction. The arrows show an observational lower limit from QSO
absorption lines at $z=6$ and upper limit from Ly$\alpha$ emitters at $z=6.5$; (b) the cosmic star formation history, with
the contribution of Pop III and Pop II stars. 
(c) the number of source counts above a given redshift, with the observational upper limit from NICMOS HUDF;
(d) the electron scattering optical depth, with observational constraints from 3-yr {\it WMAP} data; (e) Ly$\alpha$ effective optical
depth; (f) Ly$\beta$ effective optical depth; (g) the evolution of
Lyman-limit systems; (h) photoionization rates for neutral hydrogen, with estimates from numerical simulations; (i) temperature 
evolution of the mean density IGM. For tha data sources, see the original paper.
}
\label{cf06}
\end{figure}

{\it The most attractive scenario requires an enhanced ionizing photon emission at 
high redshift (compared to the low redshift)}, obtained either through a higher
ionizing photon production (metal-free stars and/or top-heavy IMF), star formation
efficiency or escape fraction. A suitable combination of these parameters allows an early 
enough reionization and matches the measured Thomson
scattering optical depth (e.g. Cen 2003b; Ciardi, Ferrara \& White 2003;
Ciardi, Stoehr \& White 2003; Haiman \& Holder 2003; Onken \& Miralda-Escud\'e 2004; 
Sokasian et al. 2003, 2004; Wyithe \& Loeb 2003b; Iliev et al. 2006a;
but see also Ricotti \& Ostriker 2004a). 
Following feedback and metal enrichment,
it is plausible that the ionizing photon production drops, resulting in a
partial/total recombination of the IGM, followed by a second reionization at
$z\approx 6$, produced by more standard sources (Wyithe \& Loeb 2003a; Cen 2003a;
but see also Oh \& Haiman 2003 who propose that the ionizing photon production
is regulated by gas entropy rather than by a transition from Pop~III to Pop~II stars) or
more simply resulting in a temporary slight decrease of the ionizing flux  
(Tumlinson, Venkatesan \& Shull 2004). Such reionization history might also help 
in releasing the tight constraint posed by the IGM temperature at $z\approx 3$
on the reionization redshift.  
The contribution to cosmic reionization from very massive metal-free stars is still
very uncertain. In fact, to produce a number of ionizing photons sufficient to
substantially ionize the IGM, there would be an over production of metals
(e.g. Ricotti \& Ostriker 2004a; Rozas, Miralda-Escud\'e \& Salvador-Sol'e 2006) and 
the transition from Pop~III to Pop~II stars
would occur early on (Fang \& Cen 2004). It has been proposed that while very massive
metal-free stars would form in metal-free halos with $T_{vir}<10^4$~K, larger metal-free
halos would host metal-free stars with masses $<100$~M$_\odot$. In this case, it is
found (Greif \& Bromm 2006) that, due to the impact of radiative feedback, the
contribution of very massive Pop~III stars to reionization (and metal enrichment, see 
Section \ref{metenrich}) is not substantial as their formation terminates at relatively
high redshift. The formation of lower mass Pop~III stars would stop at lower redshift
due to metal enrichment and feedback, and their contribution to reionization would be more important.
This picture is supported by Schneider et al. (2006a). Using a semi-analytic
model to follow the formation and evolution of dark matter halos, coupled with a
self-consistent treatment of chemical and radiative feedback, they find that the
constraints imposed by the {\tt NICMOS} observations (Bouwens et al. 2005) can be fulfilled 
together with those imposed by {\tt WMAP} only if Pop~III stars have 
a slightly top-heavy IMF, but masses $<100$~M$_\odot$. A similar result is found by
Daigne et al. (2004) and Choudhury \& Ferrara (2006). 
The last paper presents a comprehensive model of reionization which satisfies simultaneously 
all the available constraints set by different observations (see caption of Fig. \ref{cf06} for
a complete list). Although many details might be modified by future observations, a solid
conclusion seems to be that reionization has been a gradual and long-lasting event which 
was started by Pop III stars at $z\approx 15$, it was 90\% complete at $z \approx 9$ and completed only
with the help of Pop II stars and quasars at $z\approx 7$.  

Following the first year release of {\tt WMAP} data, that implied a high value for the
Thomson scattering optical depth,
several alternative processes/contributions to enhance the high redshift ionizing photon
emission have been proposed. For example, a component of a non-scale-free
isocurvature power spectrum in addition to the more standard scale-free adiabatic
power spectrum, would induce an earlier structure formation and
yield an efficient source of early star formation, down to
$z\approx 10$, without violating the existing constraints from the Lyman-$\alpha$ forest
observations (Sugiyama, Zaroubi \& Silk 2004). The same effect could be obtained
with non-Gaussian density fluctuations (Chen et al. 2003).
If magnetic fields generated in the early universe had a strength $B\simgt 0.5$~nG
(consistent with BBN and CMB constraints that set an upper limit to the comoving
amplitude of $\sim$~10nG at 1~Mpc scale), they would as well induce additional density
perturbation on small scales (Tashiro \& Sugiyama 2006a).
A contribution could also come from globular clusters, for which a high
escape fraction is expected (Ricotti 2002).
An enhancement of the ionizing photon production caused by an            
increased/earlier formation of low mass galaxies, though, would crucially
depend on the efficiency of feedback effects and their ability to
partially/completely suppress star formation in primordial objects (see previous 
Section). Moreover, e.g. Ricotti, Gnedin \& Shull (2002) have shown that even for the
case of weak feedback, small galaxies can only partially reionize the IGM. 
More specifically, reionization begins earlier with the inclusion of small galaxies
(with the consequence of raising the value of the Thomson scattering optical depth),
but their influence is limited by Jean-mass filtering in the ionized regions and
the time of overlap is dictated by the efficiency of the higher mass halos
(Iliev et al. 2007a).
In more exotic scenarios, loops formed from a cosmic string network 
would induce structure formation at earlier times (Olum \& Vilenkin 2006).

{\it In addition to stellar-type sources, a contribution to the UV photon budget
could also come from other sources. In particular, high-$z$ quasars},
if present, can   
produce a substantial amount of ionizing photons (e.g. Madau, Haardt \& Rees
1999; Valageas \& Silk 1999; Miralda-Escud\'e, Haehnelt \& Rees 2000;
Wyithe \& Loeb 2003a; Ripamonti, Mapelli, Zaroubi 2008), thanks to their 
higher luminosity and escape fraction. 
Madau et al. (2004), for example, have
shown that miniquasars powered by intermediate-mass black holes (the remnants
of the first generation of massive stars) could produce a significant amount
of ionizing photons at $z>15$, under a number of plausible assumptions for 
the amount of gas accreted onto the black holes and their emission spectrum. 
On the other hand, as the recombination rate and hence the 
required emissivity to reionize the IGM, increase both with redshift and
with the luminosity of the
sources (Miralda-Escud\'e, Haehnelt \& Rees 2000), this might reduce the
effectiveness of reionization by luminous quasars. Another crucial
question about QSOs is their relatively short lifetime, which might 
effectively limit the growth of their ionized regions. 
In addition, Salvaterra, Haardt \& Ferrara (2005) find that, unless primordial
quasars are extremely X-ray quite (to avoid exceeding the observed soft X-ray
background), their contribution to reionization is only secondary.

Also more energetic photons, e.g. X-ray photons, can contribute to the IGM
reionization (Oh 2001; Venkatesan, Giroux \& Shull 2001; Madau et al. 2004; Ricotti
\& Ostriker 2004b). This component could come from early quasars, but also from
thermal emission from the hot supernova remnants, or inverse Compton scattering
of soft photons by relativistic electrons accelerated by supernova explosions
(Oh 2001).
As X-rays have much larger mean free paths, compared to photons from stellar
spectra (the mean free path for a 1~keV X-ray photon is $10^5$ times larger
than for a 13.6~eV photon), and their escape fraction is $\approx 1$,
they can permeate the IGM relatively uniformly. But only if redshifted X-rays
are taken into account, their contribution to reionization can be substantial
(e.g. Ricotti \& Ostriker 2004b).

Another possible contribution to reionization at high redshift could come from
decaying particles and neutrinos. 
While massive active neutrinos (Sciama 1990; Scott, Rees \&
Sciama 1991) have been excluded
by current cosmological data (see Spergel et al. 2003 for the latest results),
a decaying sterile neutrino does not violate existing astrophysical limits on the
cosmic microwave and gamma ray backgrounds and could partially ionize the
IGM. While Hansen \& Haiman (2004) find that a decaying sterile neutrino
could account for the observed Thomson 
scattering optical depth, Mapelli \& Ferrara (2005) conclude that it must have
played only a minor role in the reionization process.
But there is no lack of other decaying 
particle-physics candidates; e.g. cryptons, R-parity violating gravitinos, moduli
dark matter, superheavy dark matter particles, axinos and quintessinos. Chen \&
Kamionkowski (2004) have studied these particles, deriving the 
channels in which most of the decay energy ionizes and heats the IGM gas and those
in which most of the energy is instead carried away (e.g. photons with energies 100~keV
$<E<$~1~TeV). They find that decaying particles can indeed produce an optical
depth consistent with the {\tt WMAP} results, but they produce new fluctuations
in the CMB temperature-polarization power spectra. The amplitude of such fluctuations
generally violates current constraints for decay lifetimes that are less than 
the age of the universe, while it is usually consistent with the data
if the lifetime is longer (but see also Kasuya \& Kawasaki 2004, who find that also
particles in a short lifetime are consistent with the {\tt WMAP} data. In a subsequent
version of the paper they favor longer lifetimes to be consistent the {\tt WMAP3}
results).
Avelino \& Barbosa (2004) have shown that a contribution to reionization from the decay
products of a scaling cosmic defect network could help to reconcile a high optical
depth with a low redshift of complete reionization.
Finally, it has been proposed (Sethi \& Subramanian 2005) that the primordial magnetic
field energy could dissipate into the IGM by ambipolar diffusion and by generating
decaying magneto hydrodynamics turbulence. These processes can modify the thermal
and ionization history of the IGM and contribute to the Thomson scattering optical
depth.

Neither X-rays nor decaying particles alone produce a fully ionized IGM, but the
IGM may have been warm and weakly ionized prior to full reionization by stars and
quasars. This scenario would also alleviate constraints on
structure formation models with low small-scale power, such as those with
a running or tilted scalar index, or warm dark matter models, which, alone, would
not provide enough ionizing photons (e.g. Haiman
\& Holder 2003; Somerville, Bullock \& Livio 2003; Yoshida et al. 2003b;
Yoshida et al. 2003c). 

\subsection{Helium Reionization}

The stellar sources that ionize hydrogen can as well produce singly ionized
helium, through photons with energies $h\nu>24.6$~eV. Moreover, as these two
species have comparable recombination rates, their reionization history is
very similar. On the other hand, \HeII reionization requires 
harder photons ($h\nu>54.4$~eV), either from massive Pop~III 
stars (Venkatesan, Tumlinson \& Shull 2003), from quasars
(Madau, Haardt \& Rees 1999; Miralda-Escud\'e, Haehenelt \& Rees 2000; Wyithe 
\& Loeb 2003a; Madau et al. 2004; Ricotti \& Ostriker 2004b) or from thermal
emission from shock-heated gas in collapsed cosmic structures (Miniati et al. 2004).
This last contribution would be comparable to the QSO one at $z\approx 3$ and
dominating at $z>4$, in terms of \HeII ionizing photons.
Since \HI and \HeI do not absorb a significant fraction of photons with energy
higher than 54.4~eV, the problem of \HeII reionization can be decoupled from that
of the other two species.

The \HeII Ly$\alpha$ absorption is generally much stronger than the \HI 
Ly$\alpha$ absorption, by a factor $\eta$=N(\HeII)/N(\HI)$\sim 4 
\tau_{\rm He \scriptstyle\rm II}/\tau_{\rm H \scriptstyle\rm I}$,
where the second equality is valid for optically thin lines. The
larger strength of \HeII arises because it is harder to photoionize than \HI
(although the number of helium atoms is smaller than hydrogen atoms by
a factor of $\approx 10$), owing to lower fluxes and cross sections at its 
ionizing threshold. In addition,
\HeIII recombines $\approx 5.5$ times faster than \HII and \HeII
(this number becomes $\approx 5.9$ when including the increase in the electron
density due to the ionization of helium).
This suggests, supported by semi-analytical (Madau, Haardt \& Rees 1999;
Miralda-Escud\'e, Haehenelt \& Rees 2000; Venkatesan, Tumlinson 
\& Shull 2003; Wyithe \& Loeb 2003a; Furlanetto \& Oh 2008) and numerical (Sokasian, Abel \& Hernquist
2001; Gleser et al. 2005; Paschos et al. 2008) 
calculations, that helium reionization occurred at $z \simlt 5$, i.e.
after completion of hydrogen reionization. 
The known population of quasars is sufficient to completely reionize \HeII before
$z\approx 3$ (e.g. Miralda-Escud\'e, Haehenelt \& Rees 2000), although this is
subject to the assumed gas clumping and a contribution from high-$z$ quasars
(or other sources) could be needed (e.g. Madau, Haardt \& Rees 1999). 
On the other hand, one might worry that the primordial contribution to
\HeII ionizing photons may be too high and induce an almost simultaneous
\HI and \HeII reionization. Assuming a standard power-law spectrum for
miniquasars, it is shown that only a small delay between
the complete overlapping of \HII and \HeIII regions is expected. Thus if
in addition to miniquasars Pop~III stars and thermal
emission from shock-heated gas also contributed to \HeII reionization in the
early universe, one might expect a double reionization, a first
time by the above sources, followed by recombination as Pop~II stars with softer 
spectra took over, and then by a second overlap phase at $z\approx 3$,
driven by the known population of quasars 
(e.g. Venkatesan, Tumlinson \& Shull 2003; Wyithe \& Loeb 2003a).
Ricotti \& Ostriker (2004b) find that reionization of \HeII  
from primordial black holes can be almost complete at $z\approx 15$. In their
model, the full reionization at $z\approx 3$ is due to a hardening of the 
background radiation rather than QSO activity.
The \HeIII ionizing
flux should fluctuate substantially because, in a random region of space, it is 
dominated by only a few active sources. This can produce optical depth fluctuations 
in the \HeII forest on scales comparable to the mean free path, $\approx 7000$~km~s$^{-1}$.
These fluctuations may also be due, however, to the random fluctuations in the 
number of underdense voids in the IGM crossed by the line of sight. A strong
proximity effect is also expected, and it is observed in the spectra of quasars
for which the \HeII Ly$\alpha$ forest has been observed.

Associated with IGM reionization is also an increment of its temperature
(e.g. Miralda-Escud\'e \& Rees 1994; Theuns et al. 2002a; Ricotti \& Ostriker
2004b).
Because its cooling time is long, the low-density IGM retains some memory
of when and how it was reionized. The post-ionization temperature is
generally higher for harder spectra of the ionizing radiation, but it depends 
also on the type of source. 
While the IGM can be heated up to $T\approx 10,000$~K during hydrogen reionization,
it can reach $T\approx 20,000$~K during helium reionization (see also Fig.~\ref{tempIGM}). 

\subsection{IGM Metal Enrichment}
\label{metenrich}

The same stars that ionize the IGM are responsible for its metal enrichment.
Differently from their present counterparts, Pop~III stars do not suffer strong
stellar winds and thus the main contribution to the metal pollution comes
from the later stages of their evolution. 
Wind-driven mass loss is believed to be metallicity-dependent, and a scaling law
$\propto Z^{1/2}$ has been suggested for hot stars (Kudritzki 2000; Nugis \&
Lamers 2000). This scaling breaks down at $Z<10^{-2} Z_\odot$, where the power-law
becomes steeper (Kudritzki 2002). 
Marigo, Chiosi \& Kudritzki (2003) have studied the mass loss by stellar winds
for stars with initial masses in the range $120-1000$~M$_\odot$, 
including the two main mass loss driving processes, i.e. radiative line acceleration 
(e.g. Kudritzki 2002; Krti\u{c}ka \& Kub\u{a}t 2006) and stellar rotation (Maeder \& Meynet 2000). 
According to their calculations, wind mass-loss
for stars with metallicity of $10^{-4}\;Z_\odot$ is significant
only for very massive objects ($750-1000$~M$_\odot$). For
a 1000~M$_\odot$ non-rotating star, 17.33\% (0.03\%) of the initial mass
is ejected via mass loss as helium (oxygen) during the stellar lifetime. A
small amount of carbon and nitrogen is ejected as well. 
Several important effects are introduced if rotation is taken into consideration. 
First, chemical mixing is more efficient for lower
metallicity for a given initial mass and velocity. This favors the evolution into
the red supergiant stage, when mass loss is higher, and it enhances the metallicity
of the surface of the star, boosting the radiatively driven stellar winds. Second,
although metal-poor stars lose less mass to winds compared to their more enriched
counterparts, they also lose less angular momentum, so that they have a larger
chance of reaching the break-up limit (when the outer layers become unbound and
are ejected) during the main sequence phase. Meynet, Ekstr\"om \& Maeder (2006),
including the above effects of rotation, find that a 60~M$_\odot$ star, with
$Z=10^{-8}$ and initial velocity of 800~km/s can lose 22\% of its mass. At $Z=0$
though the mass loss is of the order of few percent (Ekstr\"om, Meynet \& Maeder
2006).
Finally, whereas pulsation-driven mass loss is important, if not dominant,
at high metallicity, Baraffe, Heger \& Woosley (2001)
have shown that it is negligible at very-low metallicity. 
If indeed mass loss in very massive stars is dominated by optically thick, 
continuum-driven outbursts or explosions as suggested by Smith \& Owocki (2006),
nitrogen would be expelled, but no other heavy metal 
because these stars are not in advanced O, Si and subsequent burning stages.
We can conclude then that, {\it as metal pollution from stellar winds is negligible 
for very-low metallicity stars, the main contribution to the IGM metal enrichment comes
from those stars that end up their lives as PISN and core-collapse SNe}
(see Fig.~\ref{finalfate}).

\begin{figure}
\centerline{\includegraphics[width=30pc]{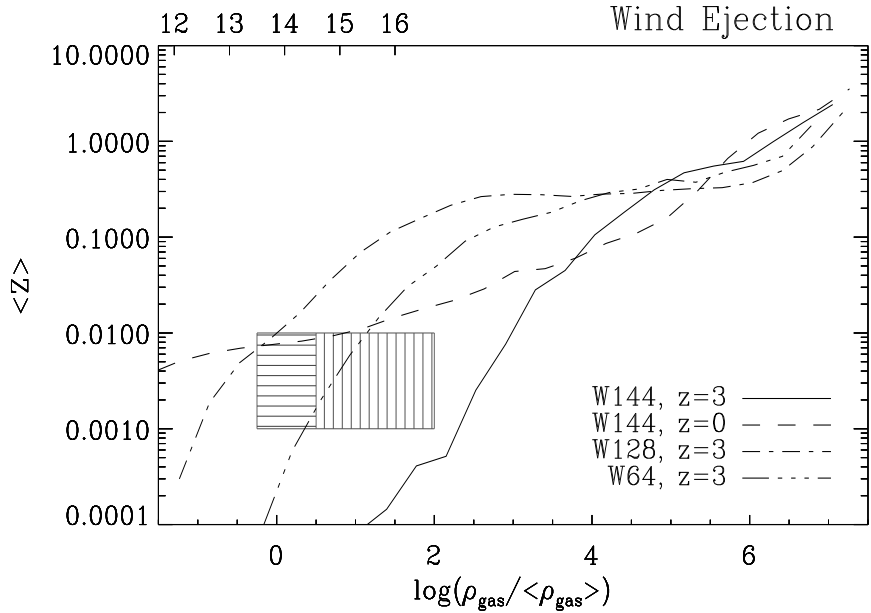}}
\caption{
Mean metallicity vs. overdensity for three different simulations at $z = 3$ and for a
simulation at $z = 0$ (see details in Aguirre et al. 2001c).
The top axis is $\log N_{HI}$ of an absorber corresponding to the bottom-axis
overdensity at $z = 3$. The vertically striped box outlines the approximate current
constraints from the Ly$\alpha$ forest at $z \approx 3$, while the horizontally
striped box shows an extension of these constraints to lower densities.  }
\label{aguirre}
\end{figure}

As metal factories (stars) are almost always associated with high density regions, in
the absence of any efficient
diffusion/transport mechanism, one would then expect a very strong density/metallicity 
correlation. Although such positive correlation is seen, it is not as strong as 
expected on the basis of the above naive assumption. However, a relation is predicted 
by simulations, as is clearly seen in Fig.~\ref{aguirre} (Aguirre et al. 2001c).
In any case, the observed large-scale clustering of metal absorbers encodes valuable
information about the masses of the objects from which they were ejected. Likewise,
as the maximal extent of each enriched region is directly dependent on the velocity
at which the metals were dispersed, measurements of the small clustering of these
absorbers is likely to constrain the energetics of their sources.

Different mechanisms have been suggested to remove metals from the galaxies in
which they are produced, ranging from dynamical encounters between
galaxies, to ram-pressure stripping, supernova-driven winds or radiation-pressure
driven dust efflux. While dynamical removal undoubtedly occurs at some level,
it is not clear if it can account for the observed metallicity in the
$z \approx 3$ IGM (Aguirre et al. 2001c; but see Gnedin 1998, who finds merging
events as a dominant mechanism for transporting heavy elements into the IGM). 
The same argument applies to the ram-pressure stripping, which has
a minor impact on the overall metallicity and filling factor as it
occurs only in the densest and most polluted regions of space (Scannapieco,
Ferrara \& Madau 2002). 

\begin{figure}
\centerline{\includegraphics[width=22pc]{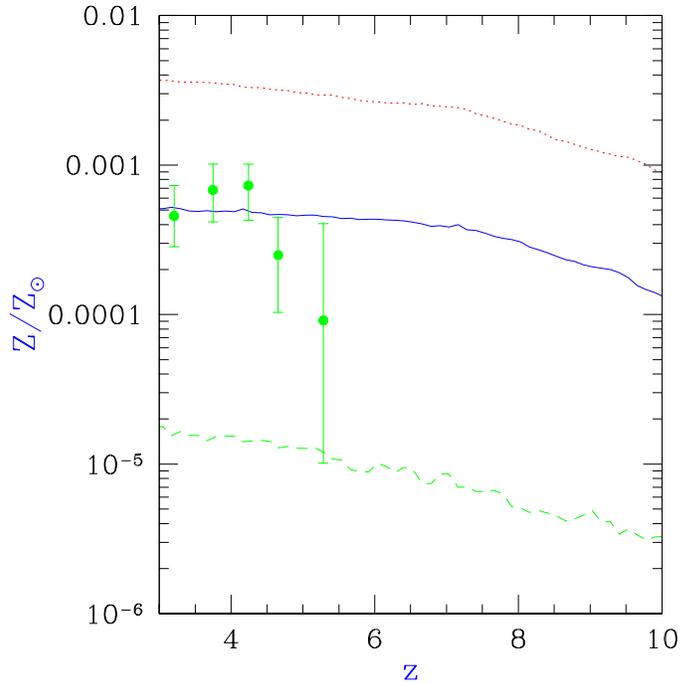}}
\caption{ Metallicity evolution for low-density ($12  < \log N_{HI} < 15$ ) IGM. 
Points are the data of Songaila (2001); dotted, solid, and dashed lines are the 
predicted values for star formation
efficiencies 0.5, 0.1, and 0.01, respectively (see Scannapieco, Ferrara \& Madau
2002 for details).}
\label{smf3}
\end{figure}

{\it Metal ejection by galactic winds} is one of the most popular mechanisms for 
the IGM pollution (e.g. Cen \& Ostriker 1999; Aguirre et al. 2001b; 
Madau, Ferrara \& Rees 2001; Scannapieco, Ferrara \& Madau 2002; Qian \& Wasserburg 2005;
Scannapieco 2005; Pieri, Martel \& Grenon 2006; Tornatore, Ferrara \& Schneider 2007).
Madau, Ferrara \& Rees (2001) find that
pre-galactic outflows from the same primordial halos that reionize the IGM,
could also pollute it to a mean metallicity of $Z \gsim 10^{-3}~Z_\odot$.
The advantage of metals produced in these low-mass halos is that they
can more easily escape from their shallow potential wells than those at lower
redshift. In addition, the enriched gas had to travel much shorter distances
between neighboring halos at these early times, and it might therefore have
been easier to obtain non-negligible cosmic metal filling factors (see also
Scannapieco 2005).
Moreover, the velocities associated with these pre-galactic outflows are small
enough to leave the thermal and structural properties of the IGM unperturbed,
in contrast with stronger galaxy winds (with velocities $\gsim 200-300$~km~s$^{-1}$),
which, although capable of enriching the IGM to the mean level observed,
are likely to overly disturb it (e.g. Cen \& Ostriker 1999; Aguirre et al. 2001b).
Rayleigh-Taylor instabilities at the interfaces
between the dense shell which contains the swept-up material and the hot,
metal-enriched low-density bubble may contribute to the mixing and diffusion of
heavy elements. The volume filling factor of the ejecta can be up to 30\% or higher,
depending on the star formation efficiency, with the majority of the enrichment
occurring relatively early, $5 \lsim z \lsim 14$ (Madau, Ferrara \& Rees 2001;
Scannapieco, Ferrara \& Madau 2002; Thacker, Scannapieco \& Davis 2002), and
a possibly large contribution from very massive stars (Qian \& Wasserburg 2005).
Such early enrichment model can explain quite well the observed redshift 
evolution of elemental abundances derived from QSO absorption line experiments, as
seen from Fig. \ref{smf3}, where model predictions are compared with available data. 
Nevertheless, enrichment by outflowing galaxies is likely
to have been incomplete and inhomogeneous.
Oppenheimer \& Dav\'e (2006), via hydrodynamic simulations, find that momentum-driven
winds from galaxies at $z<5$ are able to reproduce the measured \CIV over the
range $z\sim 2-5$. Thus it seems that a very early generation of stars is not
required to explain these observations. 
The worry remains that the hydrodynamical motions induced by the outflow may spoil
the nice agreement of Ly$\alpha$ absorption line statistics derived from studies in
which these phenomena were not included. This problem has been convincingly washed 
out by the results of Bertone \& White (2006), who, by combining N-body simulations 
and semi-analytical descriptions of the winds, find that winds do not sensibly alter 
the statistical properties of the Ly$\alpha$ forest.

But if indeed, as discussed in Sec.~\ref{firststars}, the first stars were
in the range for PISN and were
produced, in the absence of feedback effects, in halos of masses of $M \approx
10^5-10^6$~M$_\odot$, the more energetic explosions would expel $\gsim 90\%$ of the
stellar metals into a region of $\approx 1$~kpc (Bromm, Yoshida \& Hernquist 2003).
Due to this burst-like initial
SF episode, a large fraction of the universe could have been endowed with
a metallicity floor $Z \gsim 10^{-4}~Z_\odot$ already at $z\gsim 15$.
Similarly, Matteucci \& Calura (2005) find that $\sim$~110 PISN are enough to
pollute a cubic comoving Mpc to such metallicity floor. The authors find that,
although it is easy for PISN to enrich the gas at high redshift, a contribution
from Pop~II stars is needed to reproduce the metallicity patterns observed in
Ly$\alpha$ forest and in damped Ly$\alpha$ systems.

An alternative mechanism for the enrichment of the IGM is based on the {\it ejection
of dust grains}, within which heavy elements are locked, 
by radiation pressure, along the lines
suggested by Ferrara et al. (1991). Aguirre et al. (2001a) and Bianchi \& Ferrara (2005), 
have investigated the
IGM pollution caused by such a mechanism. An attractive feature of this model is 
that, unlike winds, enrichment by dust would not
impact the thermal/structural properties of the IGM, as no shock waves are involved
in the transport. According to this study,
dust pollution
can account for the observed level of C and Si enrichment of the $z \approx 3$ IGM,
but not for all the chemical species
observed in clusters at low redshift. Thus, the authors suggest a possible hybrid
scenario in which winds expel gas and dust into galaxy halos but radiation
pressure distributes the dust uniformly through the IGM. Of course, it is still to
be demonstrated that grains are efficiently sputtered in the IGM and therefore
can release their metals.

As discussed in Sec.~\ref{firststars}, the presence of dust, as well as
metals, can deeply affect the primordial star formation process. Thus, it
is important to determine the dust production in the early universe.
Recently, this issue as been discussed by Nozawa et al. (2003) and
Schneider, Ferrara \& Salvaterra (2004), which investigated the formation
of dust grains in the ejecta of PISN and core-collapse SNe. 
The fraction of mass locked into dust grains is $2-5$\% of the progenitor mass
for SN~II and $15-30$\% for PISN; however, PISN
dust depletion factors (fraction of produced metals locked into dust grains)
are smaller ($<40\%$) than SN~II ones. These
conclusions depend very weakly on the thermodynamics of the ejecta, which instead
affect considerably the grain formation epoch, composition and size distribution
(Nozawa et al. 2003). Thus, if the first generation of stars were 
very massive, a {\it large amount of dust grains might be produced in the early universe}.

An additional problem is that our cosmic inventory of the cosmic metals associated
with the observed star formation history is lacking a consistent fraction of heavy
elements. This discrepancy is often referred to as the {\it missing metals} problem (see
Pettini 2006 for the most recent review on the subject). 
One possibility is that missing metals escaped detection because they reside in the (hot) 
halos of the parent galaxies, as suggested by Ferrara, Scannapieco \& Bergeron (2005) who 
found that up to 95\% of the produced metals could be hidden in such a hot gas phase. 
Consistently with this hypothesis, Scannapieco et al. (2006) find that the correlation 
function of \CIV absorbers favor a metal enrichment mode in which metals are confined
within bubbles with a typical radius of 2 Mpc around galaxies of mass $\approx 10^{12} M_\odot$. Note
that this is not in contradiction with the fact that these metals could have been injected
by early progenitors of these galaxies, as predicted by the pre-enrichment scenario.

\subsection{Key Observations}

\subsubsection{Escape Fraction}
Observationally, a wide range of values for the escape fraction has been
deduced, but it appears that  -- broadly speaking -- relatively low values of $f_{esc}$
are favored. 
For example, most of the detections of starburst galaxies with 
{\tt HUT}\footnote{http://praxis.pha.jhu.edu/hut.html} (Hopkins Ultraviolet Telescope) and 
{\tt FUSE}\footnote{http://fuse.pha.jhu.edu} (Far Ultraviolet
Spectroscopic Explorer) (Leitherer et al. 1995; Hurwitz,
Jelinsky \& Dixon 1997; Heckman et al. 2001) are consistent with $f_{esc}<10\%$,
although objects have been observed with  $f_{esc}<57\%$ (Hurwitz,
Jelinsky \& Dixon 1997). Tumlinson et al. (1999), based on H$\alpha$ images,
derive a value of the escape fraction lower than 2\%.
However, absorption from undetected interstellar
components could allow the true escape fractions to exceed these upper limits. 
Values of $f_{esc}\le 10 \%$ are found from the detection of flux beyond
the Lyman limit in a composite spectrum of 29 LBGs at $z = 3.4$
(Steidel, Pettini \& Adelberger 2001; Haehnelt et al. 2001).
An $f_{esc} \simlt 20\%$ would be compatible also with the observed UVB (Bianchi,
Cristiani \& Kim 2001).
Fern\'andez-Soto, Lanzetta \& Chen (2003) found that, in a sample
of 27 spectroscopically identified galaxies of redshift $1.9<z<3.5$ in the
Hubble Deep Field, on average no more than 4\% of the ionizing photons
escape galaxies and even when all systematic effects are included, the data
could not accommodate any escape fraction larger than 15\%.
A high value of $f_{esc}$ has been measured only by
Bland-Hawthorn \& Maloney (1999; see Erratum 2001), who used optical line emission
data for the Magellanic Stream to derive $f_{esc} \approx 45$\%.
For a compilation of the existing measurements see Inoue, Iwata 
\& Deharveng (2006).                     
In conclusion, {\it observations of the escape fraction, with few exceptions,
give $f_{esc} \le 15\%$}.

\subsubsection{Gunn-Peterson Test}     
\begin{figure}
\centerline{\includegraphics[width=25pc]{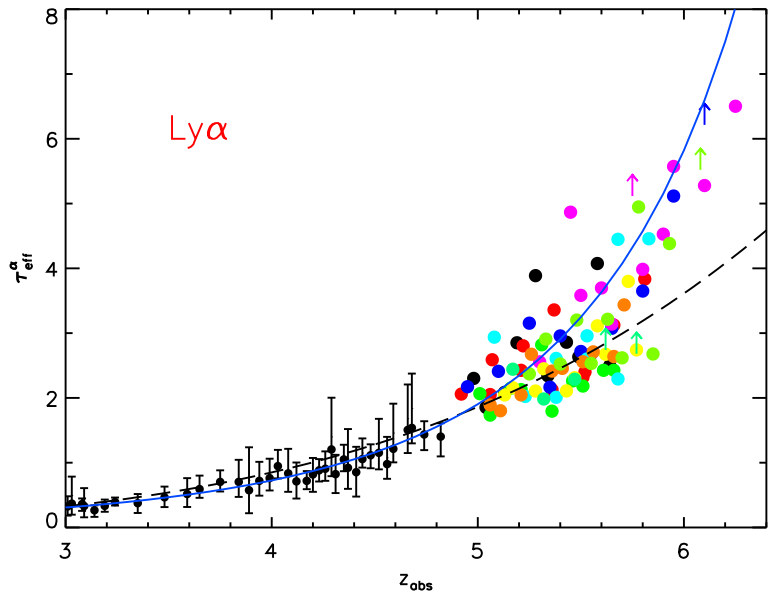}}
\caption{
Evolution of the effective Ly$\alpha$ optical depth. Points correspond to data.
The dashed line shows the best-fit at $z<5.5$ (Fan et al. 2006), the solid line
is calculated from the lognormal distribution of Ly$\alpha$ optical depths.
For details see Becker et al. (2007). }
\label{tau}
\end{figure}

The discovery of quasars at $z>5.8$ (e.g. Fan et al. 2000, 2001, 2003, 2004;
Goto 2006) are
finally allowing quantitative studies of the high-redshift
IGM and its reionization history. In particular,
the detection of a Gunn-Peterson trough (Gunn \& Peterson 1965) in
the {\tt Keck}\footnote{http://www2.keck.hawaii.edu} (Becker et al. 2001) and 
{\tt VLT}\footnote{http://www.eso.org/projects/vlt} (Very Large Telescope)
(Pentericci et al. 2002) spectra of the
Sloan Digital Sky Survey quasars SDSS J1030-0524 at $z=6.28$, and
SDSS J1148+5251 at $z=6.37$ (White et al. 2003) might indicate
that the universe is approaching the reionization epoch at
$z\approx 6$, consistent also with the dramatic increase in the neutral hydrogen
optical depth at $z>5.2$ observed by Djorgovski et al. (2001). 
However, while the theoretical optical depth distribution commonly used to
measure the \HI ionization rate gives a very poor fit to the observations,
more accurate models of the density distribution (i.e. an inverted 
temperature-density relation, see Sec. 2) provide a better agreement
and suggest that the measured dramatic increase might be explained other than
with a reionization at $z\sim 6$ (Becker, Rauch \& Sargent 2007).
Analytical fits to the measured effective optical depth are shown in Fig.~\ref{tau}.

These observations suggest that the IGM is in the post-overlap stage
(Barkana 2002). It should be noted, though, that an analysis of the same high
redshift quasars which takes into account the observed dimension of the ionized
region surrounding them, implies a neutral fraction of $\simgt 0.1$,
much higher than previous lower limits (Wyithe \& Loeb 2004; Mesinger \&
Haiman 2004; Wyithe, Loeb \& Carilli 2005; but see Oh \&
Furlanetto 2005 for a discordant view). The main uncertainty in the above
estimates is the quasars lifetime. Also analyzing the QSOs and modeling their
spectra separately, Mesinger \& Haiman (2007) find a lower limit of 0.033.
Maselli et al. (2007), using a
combination of SPH simulations and radiative transfer and analyzing mock
spectra along lines of sight through the simulated QSO environment, show
that the observed dimension of the ionized region is typically 30\% smaller
than the physical one. This induces an overestimate of the neutral fraction,
which they find to be $<0.06$. This value is consistent with that found by
Yu \& Lu (2005) using a semi-analytic model to describe the QSO's environment
(including a clumping factor and the contribution to ionization from stars).
A similar result has been found by Bolton \& Haehnelt (2007a) for a variety of
quasars lifetime and luminosity and including the treatment of He in their
radiative transfer simulation. The bottom line is that, according to these
studies, {\it the size of the observed ionized regions cannot put stringent limits 
on IGM ionized fraction}.  Stronger
constraints can be obtained considering the relative size of the ionized
regions in the Ly$\alpha$ and Ly$\beta$ (Bolton \& Haehnelt 2007b). 
An additional complication follows from the location of QSOs in biased
regions. In fact, as the extent of the \HII regions produced by nearby galaxies
can be substantial, their contribution should be included when using the \HII regions 
from QSO to determine the abundance of \HI in the IGM (e.g. Alvarez \& Abel 2007; Lidz et al. 2007).

By itself, the detection of a Gunn-Peterson trough
does not uniquely establish the fact that the
source is located beyond the reionization epoch, since the lack of any
observed flux could be
equally caused by ionized regions with some residual neutral fraction,
individual damped Ly$\alpha$ absorbers or line blanketing from lower
column density Ly$\alpha$ forest absorbers. On the other hand,
for a source located at a
redshift beyond but close to reionization, the Gunn-Peterson trough
splits into disjoint Lyman-$\alpha$, -$\beta$ and possibly higher Lyman
series troughs, with some transmitted flux in between them, that can be                   
detected for sufficiently bright sources and used to infer the
reionization redshift (Haiman \& Loeb 1999).
Alternatively, metal absorption lines, rather than H ones, have been
proposed to probe the pre-reionization epoch (Oh 2002). To this goal, one should
select absorption-line probes which are still unsaturated when the
IGM is predominantly neutral (this is possible for ions that are
less abundant than \HI or have smaller oscillator strengths), have ionization
potentials similar to that of H, and that should lie redward of the Ly$\alpha$ 
wavelength (to avoid confusion with lower-redshift Ly$\alpha$ forest). This
suggested experiment remains speculative as it is not clear how it would be
possible to have a highly neutral, but metal polluted IGM. 

A recently suggested technique to study reionization is through the statistical
analysis of QSO absorption spectra. In particular, the dark gap width distribution,
together with the complementary leaks distribution (Fan et al. 2006; 
Gallerani, Choudhury \& Ferrara 2006; Gallerani et al. 2007; Feng et al. 2008; 
Liu, Bi \& Fang 2008)
could help in discriminating between a late and an early reionization history,
as for the former $>$30\% of the line of sights to QSOs in the redshift range 
$5.7<z<6.3$ are expected to have dark gaps with widths $>50 \AA$, while none
is expected for the latter (Gallerani, Choudhury \& Ferrara 2006). In order
to discriminate between different reionization histories though, a sample at
higher redshift is required (Gallerani et al. 2007).

\subsubsection{Ly$\alpha$ Emitters}
Ly$\alpha$ emission has been crucial
in determining the redshift of distant galaxies (e.g. Hu et al. 2002; 
Kodaira et al. 2003; Rhoads et al. 2003; Hu et al. 2004; Pell\'o et al. 2004;
Kurk et al. 2004; Nagao et al. 2004; 
Stern et al. 2005; Wang, Malhotra \& Rhoads 2005; Willis \& Courbin 2005;
Kashikawa et al. 2006), with the highest $z$ Ly$\alpha$ emitter being at 
$z=6.96$ (Iye et al. 2006; Ota et al. 2007;
Stark et al. 2007b have found two possible gravitationally
lensed candidates at $z\approx 10$) and
it represents an invaluable tool to probe the high redshift universe and,
possibly, the sources responsible for the IGM reionization. 
As Ly$\alpha$ is a resonant transition with a large cross section, even a
small hydrogen neutral fraction in the intervening absorber can cause large
opacity to Ly$\alpha$ photons from high-$z$ sources. 
In particular, only the blue side of the Ly$\alpha$ line would be scattered 
away, while the red side would be observed, with a profile that depends
on the ionization state of the IGM surrounding
the galaxy or the presence of galactic winds (Miralda-Escud\'e 1998; Santos 2004).
As a result of the absorption, the Ly$\alpha$ photons are scattered until
they redshift out of resonance, resulting in a compact halo of
Ly$\alpha$ light surrounding the source of the Ly$\alpha$ emission 
(Loeb \& Rybicki 1999). Although the
brightness of such diffuse Ly$\alpha$ component is very weak,
this radiation is highly polarized and thus in principle detectable 
(Rybicki \& Loeb 1999).  If, on the contrary, the source is
surrounded by a large \HII region, the Ly$\alpha$ photons can avoid absorption,
provided they redshift out of resonance before they reach the boundary of the
\HII region. Madau \& Rees (2000) and
Cen \& Haiman (2000) (see also Cen 2003c) show that a sufficiently bright quasar placed
at $z>6$ can produce a Str\"omgren sphere large enough to render its
Ly$\alpha$ emission line detectable, e.g. by {\tt JWST} or {\tt Keck}. 
Haiman (2002) has revisited
this problem for a fainter source, showing that even for a galaxy as the one
discovered by Hu et al. (2002) a significant fraction of the emission line can
remain observable. Moreover, if one considers the contribution to the formation
of the \HII region by neighboring galaxies, the prospect for observing a high-$z$
galaxy are even better (Furlanetto, Hernquist \& Zaldarriaga 2004), also prior
to complete reionization (Wyithe \& Loeb 2005).
By means of numerical simulations of cosmological reionization,
Gnedin \& Prada (2004) show that it is possible to find galaxies at $z\approx 9$
(and possibly at $z>10$)                      
that are barely affected by the dumping wing of the Gunn-Peterson absorption.
Later on, Mesinger, Haiman \& Cen (2004) have shown that it should be possible to 
statistically extract relevant parameters, including the mean neutral fraction in the 
IGM and the radius of the local cosmological Str\"omgren region, from the flux 
distribution in the observed spectra of distant sources. 
Malhotra \& Rhoads (2006) have estimated a lower limit for the volume
average ionization fraction of 20-50\% by assigning to each observed Ly$\alpha$ 
emitter a minimum \HII region.
The evolution in the luminosity function of Ly$\alpha$ emitters observed by
Kashikawa et al. (2006) from $z=5.7$ to $z=6.5$ has been interpreted as due to
a rapid evolution in the IGM ionization state. Dijkstra, Wyithe \& Haiman (2007),
though, argue that the observations are consistent with a mild or no evolution in
the IGM transmission and can be explained with the evolution in the mass function
of dark matter halos hosting Ly$\alpha$ emitters. Following an approach similar
to that of Malhotra \& Rhoads (2006) they estimate a volume averaged ionization
fraction $> 80\%$.

\subsubsection{\HI 21 cm Line Emission/Absorption}
It has long been known (e.g. Field 1959) that neutral hydrogen in the IGM and
gravitationally collapsed systems may be directly detectable in emission or absorption
against the CMB at the frequency corresponding to the redshifted \HI 21~cm line
(associated with the spin-flip transition from the triplet to the singlet ground
state). Madau, Meiksin, \& Rees (1997) first showed that 21~cm tomography could
provide a direct probe of the era of cosmological reionization and reheating. In general,
21~cm spectral features will display angular structure as well as structure
in redshift space due to inhomogeneities in the gas density field,
hydrogen ionized fraction, and spin temperature. Several different
signatures have been
investigated in the recent literature: {\it (i)} fluctuations in the
21 cm line emission induced by the ``cosmic web'' (Tozzi et al. 2000),
by the neutral hydrogen surviving reionization (Ciardi \& Madau 2003;
Furlanetto, Sokasian \& Hernquist 2004; Zaldarriaga, Furlanetto \& Hernquist 2004;
Furlanetto, Zaldarriaga \& Hernquist 2004b; He et al. 2004; Mellema et al. 2006b;
Santos et al. 2007),
and by minihalos with virial temperatures below $10^4\,$K
(Iliev et al. 2002, 2003; Furlanetto \& Oh 2006; Shapiro et al. 2006); 
{\it (ii)} a global feature (``reionization step'') in the
continuum spectrum of the radio sky that may mark the abrupt overlapping
phase of individual intergalactic \HII regions (Shaver et al. 1999);
{\it (iii)} and the 21 cm narrow lines generated in absorption against very
high-redshift radio sources
by the neutral IGM (Carilli, Gnedin, \& Owen 2002; Furlanetto 2006) and by intervening
minihalos  and protogalactic disks (Furlanetto \& Loeb 2002; Furlanetto 2006).
While an absorption signal would be preferable, being of a higher intensity,
it relies on the existence of powerful radio sources at high redshift, which
have not yet been found. A possibility would be the radio afterglow of GRBs or
hypernovae (which could in principle be visible up to $z\approx 20-30$), but
the estimated absorption lines are very difficult to detect also with the
next generation of radio telescopes (Ioka \& Meszaros 2004).
On the other hand, a signal in emission, although weaker
than the one in absorption, would always be present, as long as the IGM is
not completely ionized and its temperature is above the one of the CMB.

\begin{figure*}
\centerline{\includegraphics[width=28pc]{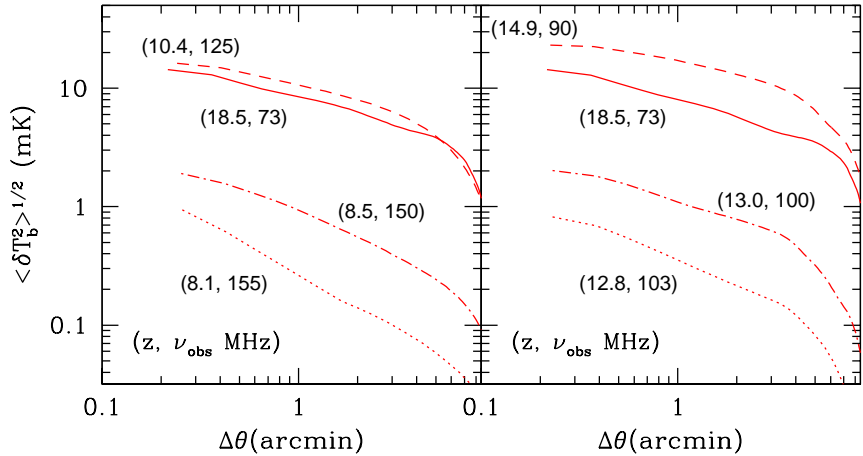}}
\caption{
Expected rms brightness temperature fluctuations as a function of beam size
for a model with reionization
at $z \approx 7.5$ (left panel) and at $z \approx 12.5$ (right panel). A fixed
bandwidth $\Delta \nu=1$~MHz has been assumed. Each curve corresponds to a
different emission redshift or, equivalently, observed frequency
(see Ciardi \& Madau 2003).}
\label{deltaTb}
\end{figure*}
Ciardi \& Madau (2003) were the first to use large scale
 numerical simulations of hydrogen reionization by stellar
sources to investigate the 21 cm signal expected from the diffuse, neutral IGM. They
found that the predicted brightness temperature fluctuations on arcmin scales have
typical values in the range $5-20$~mK on scales below 5 arcmin, with the maximum
corresponding to the epoch when roughly 50\% of the IGM is ionized, in agreement with
subsequent studies (see Fig.~\ref{deltaTb}). 
Iliev et al. (2002, 2003) find that the signal expected from minihalos is comparable
to the signal from the IGM if feedback effects are assumed to be negligible. 
Shapiro et al. (2006) use numerical simulations to
calculate the relative importance of IGM and minihalos, finding that the latter
dominate the emission at $z\simlt 18$, while they are negligible at higher redshifts. 
In their analysis though they neglect the effect of Ly$\alpha$ pumping and feedback,
which can be crucial once the first sources of radiation start to shine.
The inclusion of such mechanisms shows that the contribution from minihalos is
almost always lower than that from the diffuse IGM (Furlanetto \& Oh 2006).

More specific applications of the 21 cm signal allow to track the physical properties
of individual \HII regions as they form and grow around reionization sources, in
particular Pop~III stars and QSOs (e.g.
Wyithe, Loeb \& Barnes 2005; Zaroubi \& Silk 2005; Chen \& Miralda-Escud\'e 2006;
Cen 2006; Rhook \& Haehnelt 2006; Liu et al. 2007; Thomas \& Zaroubi 2008). 
As the expected angular resolution is
sensitive to scales that correspond to the transition between ionized and neutral
regions for hard ionizing photons sources, additional information on their nature
could come form observations of single ionized regions (Zaroubi \& Silk 2005),
also in the form of spectral dips in frequency space along single lines of sight
(Kohler et al. 2005).
This technique though would be compromised once ionized regions start to overlap.
In addition, the 21 cm radiation emitted from 
the partially neutral IGM outside \HII regions, if detected, could provide a novel
way to probe the growth of BHs (Rhook \& Haehnelt 2006).

A key physical mechanism to ensure a detectable 21 cm signal is the decoupling of
the CMB temperature from the spin temperature, as the latter regulates the 21~cm line 
absorption/emission.
It has been shown (Chuzhoy \& Shapiro 2006) that the exact value of the spin temperature
depends on physical processes that usually are not included in the calculations, the
most important being the backreaction of resonant scattering on the pumping radiation
and the scattering by resonant photons other than Ly$\alpha$. 
In addition, it is usually assumed that the spin temperature is decoupled from the
CMB temperature due to scattering with Ly$\alpha$ photons (the Wouthuysen-Field
effect; Wouthuysen 1952; Field 1959) but its strength can be overestimated if a
proper treatment of the Lyman series cascades is not included (Hirata 2006; 
Pritchard \& Furlanetto 2006). 
Moreover, before a Ly$\alpha$ background is established, the bias in the galaxy
distribution can induce fluctuations in the Ly$\alpha$, and thus 21~cm, flux (Chen 
\& Miralda-Escud\'e 2006).  
These fluctuations could be used to probe the distribution of the very
first galaxies (Barkana \& Loeb 2005b; Wyithe \& Loeb 2007).
Additional fluctuations arise if the natural broadening
of the line is considered. In this case in fact the optical depth 
in the wings causes scattering to happen nearer to the source, with a distribution
function that depends on the optical depth and the frequency of the photon
(Chuzhoy \& Zheng 2007).
These fluctuations though should affect the 21~cm signal only during a
narrow range of redshift near the beginning of reionization, as a later times
there are so many UV photons that these corrections are minor.

Kuhlen, Madau \& Montgomery (2006) use hydrodynamic simulations of structure formation
to study the effect of X-rays on 21~cm line. 
In a simulation without sources they find that, while in regions with
$\delta \simlt 1$ decoupling is not effective, for higher density contrasts H-H collisions
are efficient at decoupling and most of the gas is shock heated, producing a signal in
emission. When the effect of mini-quasars is included, the X-ray radiation preheats
the IGM to a few thousand kelvins and the increased electron fraction boosts both the
H-H and the H-e collisions (see also Nusser 2005a for the efficiency of H-e collisions
in the decoupling during the very early stages of reionization).
 Thus also low density regions in the IGM can be seen in emission. It should be noted
though that, although X-ray are usually considered to uniformly penetrate the IGM,
this is not true at all times and might affect the 21~cm line signature (Pritchard \&
Furlanetto 2007).

Also the mechanism that can heat the gas above the CMB temperature is crucial for
the observation of the line in emission. In addition to Ly$\alpha$ and X-ray photons
(e.g. Madau, Meiksin \& Rees 1997; Chen \& Miralda-Escud\`e 2004; Pelupessy, Di Matteo
\& Ciardi 2007; Ciardi \& Salvaterra 2007; Ripamonti, Mapelli \& Zaroubi 2008), also UHECRs and
decaying dark matter particles can heat the gas (Shchekinov \& Vasiliev 2007). But
it's been shown that Ly$\alpha$ scattering is not an efficient mechanism to heat
the IGM (Chen \& Miralda-Escud\`e 2004; Chuzhoy \& Shapiro 2006; Furlanetto \&
Pritchard 2006), while X-rays are better, although they can uniformly heat the
IGM only sometime after the first structure formation (Pritchard \&
Furlanetto 2007; Zaroubi et al. 2007).

Another signal detectable in the 21~cm radiation is its polarization. 
Although intrinsically unpolarized,
in the presence of a magnetic field, the 21~cm line would show a left- and 
right-handed polarized component due to the Zeeman effect (Cooray \&
Furlanetto 2005). In addition, scattering between electrons produced
during reionization and the 21~cm quadrupole would induce polarization
(Babich \& Loeb 2005). Such observations
would probe the magnetic field on scale of \HII regions around bright quasars
or the intergalactic magnetic fields. An alternative method to probe such
magnetic fields would be through the imprint they leave on the brightness
temperature fluctuations (Tashiro \& Sugiyama 2006b).           

The signatures in the brightness temperature discussed above can possibly be detected
by the planned facilities {\tt SKA}, {\tt LOFAR}, {\tt 21cmA/PAST} and {\tt MWA}
(see e.g. Valdes et al. 2006; Mellema et al. 2006b for the observability of its fluctuations).
The same telescopes should also be able to distinguish between different reionization
models and discriminate, e.g., a history with discrete \HII regions from one
with partial uniform ionization or from a double reionization (Furlanetto,
Zaldarriaga \& Hernquist 2004b).
In principle, if peculiar velocities are associated with \HI, they could induce
non linear distortions
in the redshift space and reduce the power spectrum of the brightness temperature.
Wang \& Hu (2006) though find that the next generation of radio telescopes are unlikely
to be affected by such distortions.

\begin{figure}
\centerline{\includegraphics[width=30pc]{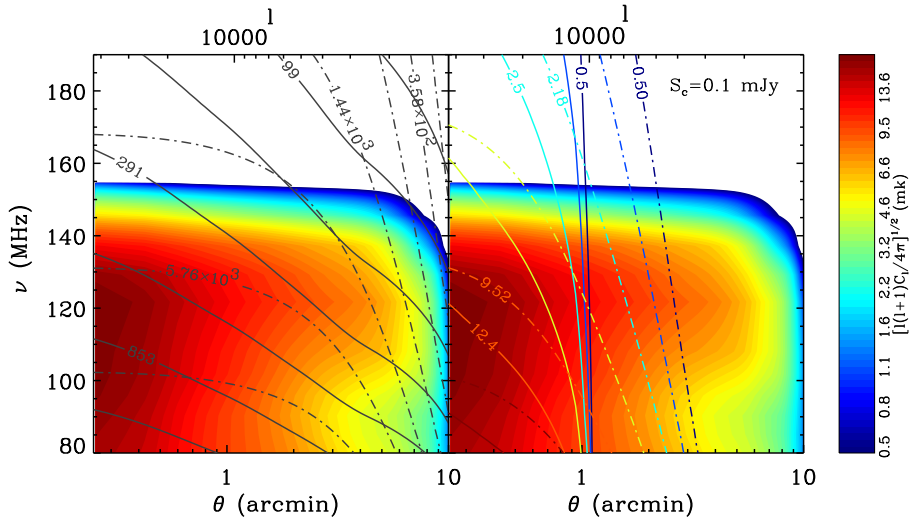}}
\caption{
Contours of the angular power spectrum (in mK) of the the 21~cm fluctuations
(shaded area) and the angular power spectra due to intensity
fluctuations of radio galaxies (solid contours) and radio halos
(dot-dashed contours) as a function of angle ($l$) and frequency.
The lines in the right panel show the contours for the total
foreground signals. In the left panel, foreground sources above a
flux limit $S_{c} \approx 0.1$ mJy have been removed (see Ciardi \& Madau 2003;
Di Matteo, Ciardi \& Miniati 2004).  }
\label{forgr}
\end{figure}
However, these experiments are extremely challenging due to
foreground contamination from unresolved extragalactic radio sources (e.g. Di Matteo
et al. 2002), free-free emission from the same reionizing halos
(e.g. Oh \& Mack 2003), synchrotron emission from cluster radio halos and relics
(Di Matteo, Ciardi \& Miniati 2004) and the Galactic free-free and synchrotron
emission (Shaver et al. 1999). As the foregrounds are slowly varying as a 
function of frequency, it appears that the best 
strategy for measuring the 21~cm signal would be to search for frequency
rather than angular fluctuations (Di Matteo et al. 2002; Gnedin
\& Shaver 2004; Zaldarriaga, Furlanetto \& Hernquist 2004; Santos, Cooray \& Knox
2005; Wang et al. 2006; Gleser, Nusser \& Benson 2008).
Di Matteo, Ciardi \& Miniati (2004) have used cosmological
simulations to predict the impact of extragalactic foreground produced by both
extended (e.g. cluster radio halos and relics) and point-like (e.g. radio galaxies
and free-free emission from ISM) sources in a self-consistent way. 
They find that the contribution to the
angular fluctuations at scales $\theta\simgt 1$~arcmin is dominated by the spatial
clustering of bright sources. Hence, efficient removal of such sources may be
sufficient to allow the detection of angular fluctuations in the 21~cm emission
free of extragalactic foregrounds (see Fig.~\ref{forgr}).

\subsubsection{\it CMB Footprints}         
\begin{figure}
\centerline{\includegraphics[width=20pc]{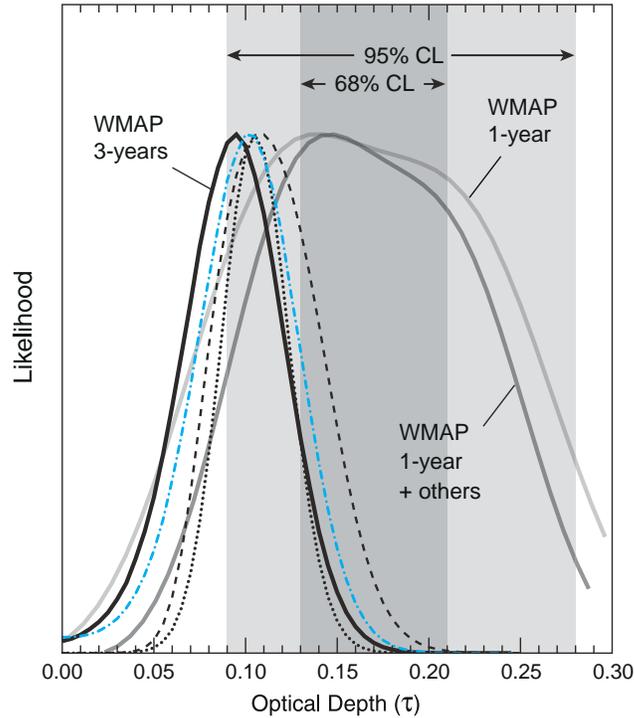}}
\caption{Comparison between the likelihood curves for $\tau$ in the {\tt WMAP}-1yr and
{\tt WMAP}-3yr data. Different curves refer to different analysis (see Page et al. 2006
for details).}
\label{wmap}     
\end{figure}

In addition to the temperature anisotropies discussed in Sec.~2, reionization 
introduces also a polarization
signal in the CMB spectrum. It is well known that the primordial
polarization is generated at the recombination epoch through Thomson scattering
of the quadrupole of the temperature anisotropy. The same mechanism operates
during the reionization epoch and again distorts the shape of the polarization
power spectrum.
The {\tt WMAP} telescope during the first year of operation detected a correlation between CMB temperature and 
polarization on large angular scales (Kogut et al. 2003; 
Spergel et al. 2003), whose amplitude is related to the total optical
depth of CMB photons to Thomson scattering, giving $\tau_e \approx 0.16\pm 0.04$
(the uncertainty quoted for this number depends on the analysis technique 
employed). This high value strongly favored an early reionization; e.g. 
$0.12<\tau_e < 0.20$ translates into a reionization redshift of $13< z
< 19$ for a model of instantaneous reionization. The limits shift at lower
redshift for a more realistic reionization, their precise value
depending on the adopted history. In the light of the first three years of 
the {\tt WMAP} experiment, the above results have been revised to
$\tau_e = 0.09\pm 0.03$ (see Fig.~\ref{wmap}; Page et al. 2006; Spergel et al. 2006). 
Such value has been only marginally updated by the five years {\tt WMAP} results, 
having now converged to $\tau_e = 0.087\pm 0.017$ (Dunkley et al. 2008); 
note, however, the considerably smaller errorbar.  
The updated Thomson scattering optical depth can be better interpreted by a reionization redshift
$z\approx 10$ (Spergel et al. 2006; Dunkley et al. 2008), substantially lower than the
first year value.  
An interesting analysis by Hansen et al. (2004) shows that
the estimate of $\tau_e$ varies substantially
when inferred from power spectra computed separately on the northern and the southern
hemispheres, and that the full sky estimate by the {\tt WMAP} team could in large
part originate in structures associated with the southern hemispheres.
In addition, the interpretation of the Thomson scattering optical depth could change
if magnetic fields were present during the epoch of reionization. In this case, the
observed anisotropies at large scales could be partially explained with the existence
of magnetic fields (Gopal \& Sethi 2005).

Another valuable tool to study the reionization process is to detect
the imprint it should leave on the CMB on small angular scales.
When the IGM gets reionized, Thomson scattering off the newly formed electrons
would, on one hand, blend photons initially coming from different directions
thus damping the primary anisotropies, and on the other hand, create small-scale
secondary anisotropies. Several groups have studied the CMB temperature anisotropies
produced by an inhomogeneous reionization (Knox, Scoccimarro \& Dodelson 1998;
Bruscoli et al. 2000; Benson et al.
2001; Gnedin \& Jaffe 2001; Santos et al. 2003; McQuinn et al. 2006; Salvaterra
et al. 2005; Zahn et al. 2005; Iliev et al. 2006c). 
In particular, calculations of anisotropies from simulations of patchy reionization
give peak values of the power spectrum between $\approx 10^{-14}$ (e.g.
Gnedin \& Jaffe 2000) and $\approx$ few $10^{-13}$ (e.g. Salvaterra
et al. 2005; Iliev et al. 2006c), which might 
be probed by future interferometers like {\tt ALMA}, 
{\tt ACT}\footnote{http://www.hep.upenn.edu/angelica/act/act.html} or
{\tt CQ}\footnote{http://brown.nord.nw.ru/CG/CG.htm.}.  
The detection of such anisotropies would be an invaluable tool to discriminate
between different sources of ionizing photons (e.g. McQuinn et al. 2006), 
especially in combination with
measurements of 21~cm line emission (Salvaterra et al. 2005; Alvarez et al. 2006a;
Holder, Iliev \& Mellema 2007; Adshead \& Furlanetto 2007; Tashiro et al. 2008;
but see also Cooray 2004), 
taking advantage of the correlation between the two signals.    

Future missions, like {\tt PLANCK}\footnote
{http://astro.estec.esa.nl/SA-general/Projects/Planck}, 
should also be able to distinguish between
different reionization models (e.g. single, double reionization) based on the
temperature-polarization power spectrum (e.g. Holder et al. 2003; Hu \& Holder
2003; Kaplinghat et al. 2003; Naselsky \& Chiang 2004; Zhang, Pen \& Trac 2004;
Colombo et al. 2005).
The overall differences in the temperature (polarization) angular power spectra
between prompt and extended reionization models with equal optical depths are less than
1\% (10\%) (Bruscoli, Ferrara \& Scannapieco 2002).
The same observations should also allow to discriminate a contribution
to reionization from decaying particles (e.g. Naselsky \& Chiang 2004).

Analogously to free electrons, heavy elements leave an imprint on the CMB
due to scattering in their fine-structure lines, partially erasing original
temperature anisotropies and generating new fluctuations. The main difference
with the Thomson scattering is that the latter gives equal contribution over
the whole CMB spectrum, while the line scattering gives different
contributions to different spectral regions. Since every line affects
only a given range of redshift, having observations at different
wavelengths can provide an upper limit to abundances of different species at
various epochs (Basu, Hern\'andez-Monteagudo \& Sunyaev 2004; Hern\'andez-Monteagudo,
Verde \& Jimenez 2006; Hern\'andez-Monteagudo et al. 2007, 2008).

It should be noted that the time and duration of recombination directly influence
the characteristics of CMB anisotropies. 
Several authors (e.g. Leung et al. 2004; Dubrovich \&
Grachev 2005; Chluba \& Sunyaev 2006) have found that the detailed treatment of
processes such as the physics of the H$^+$ and He$^+$ populations, the softening
of the matter equation of state due to the transition from ionized to neutral
matter, the induced emission of the  softer photons which change the two-photon
decay rate of the 2s level of H, affect the value of the anisotropies at large
multipoles at a percent level. Thus, in this epoch of precision cosmology,
also those processes that can affect the anisotropies at a percentage level, should
be included in the analysis. 

\subsubsection{Gamma Ray Bursts as Cosmic Lighthouses} 

\begin{figure*}
\centerline{\includegraphics[width=27pc]{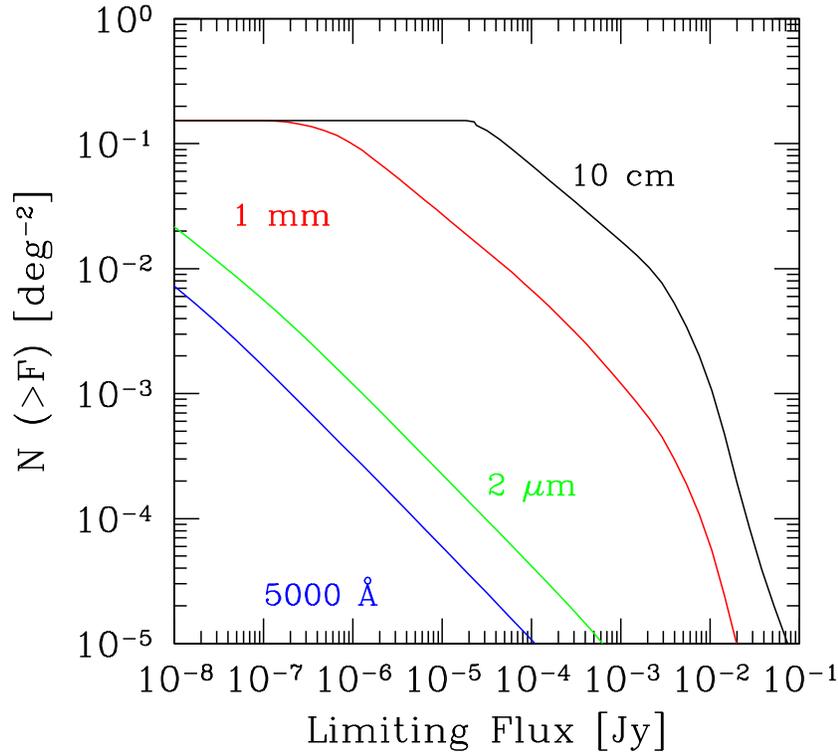}}
\caption{Predicted number of GRBs with observed flux greater than a limiting flux
$F$, at different observed wavelengths (see Ciardi \& Loeb 2000 for more details).} 
\label{grb}
\end{figure*}

The observation of the high-$z$ universe, beyond the epoch of reionization,
could be feasible directly if bright enough sources are present. If indeed GRBs
are related to the SFR of the host galaxy and occur at high redshift
(the highest GRB has been detected by {\tt SWIFT}
at $z=6.29$, Watson et al. 2006), 
their afterglow emission can be used to probe the ionization history of the IGM
(Totani et al. 2006).
In contrast to other sources, such as galaxies or quasars, the observed
infrared flux from a GRB afterglow at a fixed observed age is only a weak function of 
its redshift (e.g. Ciardi \& Loeb 2000). 
Thus, the afterglow of these sources could
be detected by the next generation of telescopes, e.g. {\tt JWST}, and used to probe
the SF and ionization history of the universe (see Fig.~\ref{grb};
e.g. Ciardi \& Loeb 2000; Blain \&
Natarajan 2000; Lamb \& Reichart 2000; Porciani \& Madau 2001; Bromm \& Loeb 2002; 
Choudhury \& Srianand 2002; Gallerani et al. 2008; McQuinn et al. 2008), 
as well as the characteristics of the circumburst
medium (e.g. Frail et al. 2006). 
A proof that very high-redshift GRBs could be observed is GRB 080319B, that 
would have been easily detected in $K$ band with meter-class telescopes up to $z \approx 17$ 
(Bloom et al. 2008).
In particular, it has been noted (Inoue, Yamazaki \& Nakamura 2004) that as the 
detection of a GRB afterglow through a NIR filter $i$ beyond the Ly$\alpha$ drop-out
redshift, $z^i_{{\rm Ly}\alpha,{\rm out}}$, probes the ionization state of the
IGM around $z^i_{{\rm Ly}\alpha,{\rm out}}$, the $I$, $J$, $H$ and $K$ filters
would be suitable to probe the ionization state at $z\approx 5-8$, $8-11$, $11-15$ and
$15-20$, respectively. Thus, the null detection of the GRB afterglows at $z\approx 11$
in $J$-band would imply a high neutral fraction at $z\approx 8-11$. The 
{\tt Subaru}\footnote{http://subarutelescope.org/} telescope has sufficient resolution to do
the photometry. To better constrain the redshift, spectroscopy is required and 
it will be only feasible with {\tt SWIFT} or future missions as {\tt ASTRO-F} or {\tt JWST}.

Free electrons along the light path cause distortions in the spectrum and light curve
of an early radio afterglow. This distortion can be quantified through the
Dispersion Measure (DM).  At $z>6$ the DM due to the
ionized IGM probably dominates the contribution from the Galactic plasma and the plasma
in the host galaxy. Thus, DM of GRBs at various redshifts and 
directions would help in determining the reionization history and possibly the 
topology of the \HII regions. Both {\tt SKA} and {\tt LOFAR} should be able to
detect bright afterglows and measure the DM towards high redshift GRBs (Ioka 2003;
Inoue 2004).

\subsubsection{Helium Reionization}

The Gunn-Peterson trough technique to infer the abundance of neutral hydrogen
in the IGM can be extended to the determination of \HeII abundance. The first
evidence of such a trough has been found in the spectrum of the quasar
Q0302-003 ($z=3.286$, Jakobsen et al. 1994), followed by more determinations 
in the range $2.4 < z < 3.5$ (e.g. Davidsen, Kriss \& Zheng 1996; Hogan, 
Anderson \& Rugers 1997; Anderson et al. 1999; Smette et al. 2002; 
Jakobsen et al. 2003; Zheng et al. 2004). In particular,
high resolution spectra of the quasar HE 2347-4342 at $z=2.885$ 
(and more recently of QSO 1157+3143, Reimers et al. 2005a) have shown a largely 
fluctuating $\eta$, in the range $0.1-1000$, with an
average of $80$ (e.g. Reimers et al. 1997; Kriss et al. 2001; 
Smette et al. 2002; Shull et al. 2004). This implies that the metagalactic
radiation field exhibits inhomogeneities at the scale of $\sim 1$~Mpc and
has been interpreted
as evidence for a patchy \HeII ionization in the IGM. However, it is unclear
if the fluctuations could be caused by fluctuations of
the IGM density or of the ionizing background flux (Miralda-Escud\'e, Haehnelt
\& Rees 2000; Kriss et al. 2001; Bolton et al. 2006), a wide range in the QSO spectral indexes
(Telfer et al. 2002), finite QSO lifetimes (Reimers et al. 2005b) or the
filtering of QSO radiation by radiative transfer effects (Shull et al. 2004;
Maselli \& Ferrara 2005).
In fact, also QSOs that are believed to be in the post-reionization phase (e.g.
HS 1700+6416, that, together with HE 2347-4342, is the only line of sight with
\HeII absorption resolved in the Ly$\alpha$ forest structure) show a large
variation in $\eta$ (Fechner et al. 2006).
Particular care should be taken when estimating the $\eta$ value. 
In fact, if the same line width is assumed for \HeII and \HI
absorption features without including thermal broadening, as done in most analysis, 
this leads to a spurious correlation of the inferred $\eta$ value with \HI column
density (Fechner \& Reimers 2007).

\begin{figure}
\centerline{\includegraphics[width=18pc]{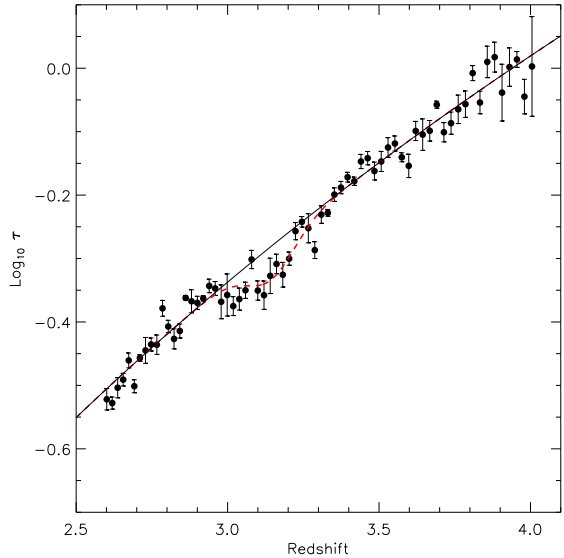}}
\caption{Effective optical depth as a function of redshift as measured from a
sample of 1061 QSO spectra from the {\tt SDSS} database (Bernardi et al. 2003).}
\label{taueff}
\end{figure}

An \HeII reionization at $z\approx 3$ was initially favored by a claim of an
abrupt change at that redshift in the \SiIV/\CIV
ratio in a sample of 13 QSO spectra, indicating that, 
while at $z\simlt 3$ the ionization balance is consistent with a pure 
power-law ionizing spectrum, at higher redshift
the spectrum must be softer. An analogous result is obtained by Heap et al.
(2000), who analyzed the Gunn-Peterson \HeII absorption trough at $z=3.05$, 
finding that the spectrum is much softer than the one deduced at $z=2.87$.
However, a more recent {\tt VLT/UVES} study (Kim, Cristiani
\& D'Odorico 2002) 
using 7 QSOs finds no strong discontinuity for the quantity \SiIV/\CIV
around $z \approx 3$ and suggests that it might not be a good indicator of \HeII reionization.
Also the analysis of 19 high-quality quasar absorption spectra by Aguirre et al.
(2004) seems to indicate no evolution in the \SiIV/\CIV ratio. 
Finally, Boksenberg, Sargent \& Rauch (2003) 
find no jump at $z \approx 3$ in the median column density ratio \SiIV/\CIV, and also
other ionic ratios vary continuously with redshift. 
On the other hand, the presence of a significant amount of O~$\scriptstyle\rm VI\ $
at $z\simgt 3$ suggests either a considerable volume of \HeIII bubbles embedded
in the more general region in which the ionizing flux is heavily broken, or the
addition of collisional ionization to the simple photoionization models (Songaila
1998). 

As usually $\eta$=N(\HeII)/N(\HI)$\gg 1$ and $\tau_{\rm He \scriptstyle\rm II}=
(\eta/4) \tau_{\rm H \scriptstyle\rm I}$, \HeII can
be tracked into much lower density regions than \HI, making the \HeII absorption
a good probe of low-density regions or hot, collisionally ionized regions in the
IGM. Shull et al. (2004) find  that $\eta$ is systematically larger
in voids ($\tau_{\rm H \scriptstyle\rm I}<0.05$), compared to filaments, possibly 
implying a softer ionizing radiation field.

\begin{figure}
\centerline{\includegraphics[width=30pc]{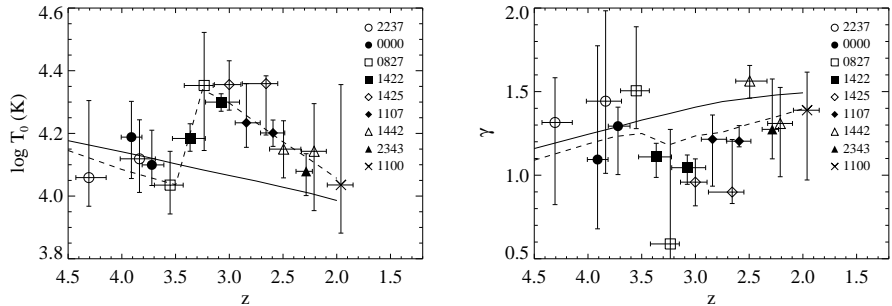}}
\caption{Temperature at the mean density (left panel) and slope of the effective
equation of state (right panel) as a function of redshift. The dashed lines are
from a simulation designed to fit the data. For more details see Schaye et al.
(2000b).}
\label{tempIGM}
\end{figure}
There have also been suggestions (Ricotti, Gnedin \& Shull 2000;
Schaye et al. 2000b; Theuns, Schaye \& Haehnelt 2000) for a relatively sudden
increase in the line widths between $z=3.5$ and 3.0, which could be associated
with entropy injection resulting from the reionization of \HeII. In particular,
Theuns et al. (2002b) use hydrodynamical simulations to show that the observed
temperature increase leads to a relatively sudden decrease in the
hydrogen Ly$\alpha$ effective optical depth. Bernardi et al. (2003)
find clear evidence for such
a feature in the evolution of the effective optical depth determined from a
sample of about 1100 quasars obtained from the {\tt SDSS} (Fig.~\ref{taueff}). 
However, the existence of 
the peak in the IGM temperature evolution is still
quite uncertain, as the data are also consistent with no feature at $z=3$.
From Fig.~\ref{tempIGM} it is clear that the temperature peaks at $z \approx 3$
and that the gas becomes close to isothermal (adiabatic index $\gamma \approx 1$), but the present
constraints are not sufficient to distinguish between a sharp rise (as indicated
by the dashed line) and a more gradual increase.
In addition, several different reionization histories and radiative transfer
effects can fit the observed temperature data equally well, 
as is discussed in detail by Miniati et al. (2004).

The IGM temperature might be affected also by the energy deposited by dark
matter particles if these are either decaying or annihilating (the most interesting
candidates being sterile neutrinos with masses of a few keV, light dark matter 
with mass $\approx 100$~MeV
 and neutralinos). If so, the IGM temperature might deviate from adiabatic evolution
before the heat input from luminous sources comes into play
(Mapelli, Ferrara \& Pierpaoli 2006; Ripamonti, Mapelli \& Ferrara 2007a; 2007b; 
Valdes et al. 2007).

\subsubsection{Metal Enrichment}

Observationally, our best view of the IGM is still provided by the
Ly$\alpha$ forest in the spectra of distant QSOs 
and the detection of metals in the forest ranks as one of the most
significant discoveries initially made possible by the {\tt Keck}
telescopes (Cowie et al. 1995; Burles \& Tytler 1996) and now also 
exploiting the power of {\tt VLT/UVES}.  From
an analysis of these data, Hellsten et al. (1997) and
Rauch, Haehnelt, \& Steinmetz (1997)
concluded that the measured column density ratios $N$(\CIV)/$N$(\HI)
imply a typical [C/H]~$ \simeq -2.5$ at $z \simeq 3$,
with a one order of magnitude dispersion
in the metallicity of different clouds about this mean value.
These measurements, however, still refer to overdense regions
of the universe, traced by Ly$\alpha$ clouds with column densities in
excess of log~$N$(\HI) = 14.5\,.

The situation is far less clear-cut when we turn to regions 
with log~$N$(\HI)$ < 14.5$  and density increasingly closer to the mean 
(a good fitting formula is given by $\rho_b/\overline\rho_b\simeq 
0.8 \left[\,N_{\rm \HI}/10^{13}{\rm cm}^{-2}\right]^{0.7}$), 
where the observations are very challenging
even with a 10-m telescope.
Two early studies addressed this problem, with conflicting results.
Lu et al. (1998) applied a stacking technique
to nearly 300 \CIV regions in QSO spectra, but still found no composite signal.
They interpreted this non-detection as evidence
for a highly non-uniform degree of metal enrichment at $z \simeq 3$,
with the voids having [C/H]~$\simlt -3.5$\,.
On the other hand, Cowie \& Songaila (1998) used a pixel-to-pixel optical depth technique
to conclude that the average \CIV/\HI ratio remains essentially constant
over the full range of neutral hydrogen column densities tested,
down to log~$N$(\HI)~$\simeq 13.5$\,.
If this is indeed the case, the ejection and transport of metals away
from galaxies must have been much more efficient than envisaged.
Ellison et al. (1999, 2000) re-examined the problem and found
that both approaches suffer from limitations which had not been properly
taken into account in previous analysis. 

To date, the number of direct metal line
detections associated with log~$N$(\HI)$ < 14.5$ lines is small. For this reason,
searches of warm photoionized \OVI have been used.
Schaye et al. (2000a) reported the first (statistical) detection of \OVI
in the low-density IGM
at high redshift ($2< z < 4.5$). By performing a pixel-by-pixel search for \OVI
absorption in eight quasar spectra, they detected \OVI in the form of a
positive correlation between the \HI Ly$\alpha$  optical depth and the optical
depth in the corresponding \OVI pixel, down to $\tau_{\rm \HI} \approx 0.1$.
However, the interpretation of the \OVI data is not straightforward
because the ionization mechanism is often ambiguous. Simcoe et al. (2002)
found evidence that the \OVI corresponding to strong \HI absorbers
(log~$N$(\HI)$ > 15$) closely resemble hot, collisionally ionized gas found
near regions of significant overdensity ($\delta > 300$) with a tiny filling
factor of about $10^{-4}$; furthermore, they seem to be clustered on velocity
scales of $100-300$ km~s$^{-1}$ and show weaker clustering out to 750 km~s$^{-1}$.
The \OVI corresponding to low \HI absorbers ($\log N(\HI) < 15$) on the other hand,
is more consistent with warm, photo-ionized gas (Carswell et al. 2002; Bergeron et
al. 2002; Simcoe et al. 2004).
The presence of O~$\scriptstyle\rm VI\ $ has been detected down to $\tau_{\rm \HI}
\approx 0.2$ also by Aracil et al. (2004), but associated only with gas located
within $\approx 400$~km~s$^{-1}$ from strong H~$\scriptstyle\rm I\ $ lines, i.e.
possibly associated with galactic winds.
As a caveat on these studies one should keep in mind that such difficult
experiments could be affected by a number of errors. For example, the apparent
O~$\scriptstyle\rm VI\ $ absorptions could be due to spurious
coincidences with  H~$\scriptstyle\rm I\ $ absorptions at other redshifts and
thus the O~$\scriptstyle\rm VI\ $ optical depth would be independent on that of the
H~$\scriptstyle\rm I\ $ (Pieri \& Haehnelt 2004).

Considerable progress has been made concerning the distribution of
predominantly photoionized species, as \CIV and \SiIV. Schaye et al. (2003) have
measured the distribution of carbon in the IGM as a function of redshift $z$  and 
overdensity $\delta$, obtaining the following fit to its median abundance:
\begin{equation} 
[{\rm C/H}]= -3.47^{+0.07}_{-0.06} + 0.08^{+0.09}_{-0.10} (z-3) + 0.65^{+0.10}_{-0.14} (\log \delta- 0.5),
\end{equation} 
which is valid in the range $1.8 < z < 4.1$ and $-0.5 < \log \delta < 1.8$
(the quoted errors are statistical; systematic errors
due to the uncertainty in the spectral shape of the UV background
radiation can be substantially greater, see Schaye et al. 2003). Two points
are worth noticing about this formula. The first is that even in underdense regions
($\delta = -0.5$) carbon is detected with an abundance [C/H]$\approx -4$. Secondly,
it appears that very little evolution has occurred in such a long cosmic time span
(2.16 Gyr). The slow evolution with column density and redshift is observed 
up to $z=5.7$ (Ryan-Weber, Pettini \& Madau 2006).

The former evidence implies that metal transport must have occurred efficiently 
and at early times to simultaneously account for the cool temperatures and widespread
distribution of elements like C and Si.   
Thus, observations of metals can possibly allow testing of the mechanisms that distributed
them or establish when/where they were produced. For example, if metals are carried by
galactic winds, they should be seen in absorption against bright
background sources, such as quasars or GRBs, in narrow lines, with
characteristic equivalent widths 0.5~\AA $\lsim W \lsim$ 5~\AA ~(Furlanetto
\& Loeb 2003).

The  ``no evolution'' trend seems to
extend to even higher redshifts, as already pointed out by Songaila (2001) 
(and further supported by Pettini et al. 2003),
who concluded as well that the \CIV column density distribution function is 
consistent with being invariant throughout the redshift range $z=1.5-5.5$.
This range has been extended to $z\approx 6$ by Simcoe (2006).
This is quite a puzzling result as, even if the observed elements were produced by a 
high-$z$ pre-enrichment episode, large variations of the ionizing radiation
field would be expected anyway during such extended time intervals.     
Alternatively, the \CIV systems may be associated with outflows from 
massive star-forming galaxies at later times, while the truly intergalactic metals 
may reside in regions of the Ly$\alpha$ forest of density lower than those probed up to now.
Songaila (2006) finds that, in the redshift range $z=2-3.5$, of a total sample of 53
absorbers, only half of those with N$\CIV > 2 \times 10^{13}$~cm$^{-2}$ could be associated
with galactic outflows, while all the rest lie in the IGM. 

Only recently, since a larger number of observations has been available, long-duration
GRBs (which are thought to be associated with the death of massive metal-poor stars)
have been used to probe the metallicity of the host galaxies. The available sample
is still to small to draw conclusive results, but it looks like they can probe more
metal-rich galaxies compared to QSOs absorption lines. Although these studies are
still in their infancy, they look extremely promising in probing metal-enrichment
(Savaglio 2006).

\section{Conclusions}  

The investigation of the first cosmic structures has started to attract 
a huge collective effort, as the distant universe is progressively disclosed by the most
advanced observing facilities. Compared to just a few years ago, 
the progresses made in the understanding of several of the 
fundamental aspects in the field are indeed impressive.  It is then not inappropriate to 
claim that a qualitatively clear scenario of these early epochs, 
whose current status is broadly summarized in this review, is becoming rapidly established.  
However, as for any new frontier, we do not yet fully know its extent and 
the territories beyond it. 
The uncertainties in the basic cosmological constants  are pinned down 
to the point that ``precision cosmology'' has become a fashionable way 
to present the data. Nevertheless, the shape of the primordial fluctuation spectrum is still
uncertain and needs to be constrained, along with the degree of non-gaussianity
of the density field. The nature of dark matter is likely to remain an 
extremely challenging problem, which may have intimate connections with
the highly debated halo density profiles. Both problems affect the way 
in which galaxies build up and their observable properties; hence 
their importance extends beyond  the realm of purely theoretical speculations. 

Aside from these general issues, the formation of the first structures 
exhibits its complexity in the large number of different aspects, ranging
from galaxy formation to stellar evolution. In principle, the advantage 
of having to deal with the simplest native environment of the first stars (no 
heavy elements, dust, dynamically significant magnetic fields, 
UV radiation background) is rapidly dissolved within the lack of similar counterparts 
in the local universe that can be used as working laboratories. Although the
formation of small mass stars in a primordial environment is possible, we tend to
believe that the first stars were very massive, but as yet, no ultimate study       
has shown the formation of such stars starting from realistic conditions,
including all the relevant physics (turbulence, radiative transfer, rotation) in detail. 
For this reason, we have essentially no clue regarding what the mass distribution, i.e. the 
IMF, of such stars could be; even indirect observations are of limited help due to 
the poor number statistics. Other constraints have been proposed, but they 
remain very weak. Also the evolution of massive stars   is
under work: for example, the combined effects of metallicity, rotation, and mass loss 
still need to be thoroughly studied and understood. 

These uncertainties must be added to the one related to the physical mechanism driving
the characteristic stellar mass to lower (and more common) values with time. Metallicity
increase is probably the most accepted explanation, but one can suspect that other 
agents may drive the transition: turbulence in the protostellar clumps, strengthening
of the UV radiation field, or heating due to cosmic rays. A sudden change in the properties
of the first stars would noticeably affect the evolution of reionization
history, as the ionizing power would drop dramatically with increasing (decreasing) 
stellar metallicity (mass). An equally uncertain quantity is the fraction of 
emitted ionizing photons able to escape from galaxies. 
Theoretical approaches to this problem have not yet clarified the dependence 
of such crucial parameter on redshift, galactic mass, and internal structure, and it 
is likely  that the next major advance will come from a dedicated observational campaign. 

Despite the wealth of literature on the formation and evolution of
low mass galaxies and on the variety of feedback effects which may hinder their formation, 
these remain controversial topics in the study of the first structures. Also, no general
consensus has been reached on their contribution to the global star formation
rate, IGM metal enrichment and reionization. For this reason, a stronger 
effort will be required in the future to clarify these issues.     

The widespread interest stimulated by reionization has produced,
as a side effect, a rapid growth of numerical radiative transfer techniques.
At the same time additional physical ingredients, like the plethora of feedback effects
which are equally important as radiative transfer in shaping the reionization history,
should be properly included in simulations.
On the observational side, a major cornerstone to map        
reionization history will be provided by 21 cm HI line measurements,
which hold the promise to open a paved road toward the distant ``gray ages''.
   
Similar arguments should also apply to the study of the intermediate-low redshift 
IGM, i.e. the Ly$\alpha$ forest. Despite its simplicity, which allowed early 
successes in matching the observed properties in the framework of hierarchical 
cosmological models, few issues remain unsolved. 
A long standing puzzle involves the observed temperature
distribution, which seems to indicate an IGM warmer than predicted by
numerical simulations. Also, an alleged temperature jump at $z\approx 3$ has not yet been 
either fully interpreted theoretically or confirmed experimentally. For this reason,
it is crucial to devote more attention to the study of \HeII reionization. 
In addition, the IGM temperature deserves further investigation as it
is often used to set a maximum redshift for hydrogen reionization.

Finally, the importance of winds from galaxies, both at intermediate and high 
redshifts, has only recently 
started to be appreciated, both theoretically and observationally. Do all galaxies 
go through a wind phase? What fraction of their mass did they lose in such events? 
What fraction of the IGM has been inside galaxies, and possibly polluted by 
nucleosynthetic products of massive stars, as a function of redshift? What is the 
cosmic filling factor of metals and do we understand its evolution? The answers we 
currently possess on these and many other
questions concerning the cosmic chemical evolution are very partial. 
The hope is that the pace we have kept so far will bring us close to the final answers 
during our life time.

\acknowledgements We thank D.R. Flower, F. Haardt, J. Le Bourlot, 
C. Larmour, F. Primas, M. Ricotti and  R. Schneider for comments,
materials, discussions and reading of the original version 
(in 2005) of the manuscript.

\end{article}
\end{document}